\documentclass[twocolumn, 10pt]{IEEEtran}
\usepackage{hyperref} % create hyperlinks
\usepackage{graphicx}
\usepackage{amssymb}
\usepackage{amsmath}
\usepackage{mathtools}
\usepackage{cite}
\usepackage{stfloats}
\usepackage{subfigure}
\usepackage{psfrag}
\usepackage[mathscr]{euscript}
\usepackage{acronym}  % make an acronym
\usepackage{booktabs} 
\usepackage[table]{xcolor}
\acrodef{CCDF}{complementary cumulative distribution function}
\acrodef{CF}{characteristic function}
\acrodef{PPP}{Poisson point process}
\acrodef{CSI}{channel state information}
\acrodef{OFDM}{orthogonal frequency division multiplexing}
\acrodef{OFDMA}{orthogonal frequency division multiple access}

\acrodef{RV}{random variable}
\acrodef{i.i.d.}{independent, identically distributed}
\acrodef{PMF}{probability mass function}
\acrodef{PDF}{probability distribution function}
\acrodef{CDF}{cumulative distribution function}
\acrodef{ch.f.}{characteristic function}
\acrodef{AWGN}{additive white Gaussian noise}
\acrodef{SNR}{signal-to-noise ratio}
\acrodef{LRT}{likelihood ratio test}
\acrodef{DRT}{distance ratio test}
\acrodef{GLRT}{generalized likelihood ratio test}
\acrodef{CRLB}{Cram\'{e}r-Rao lower bound}
\acrodef{CRB}{Cram\'{e}r-Rao bound}
\acrodef{ZZLB}{Ziv-Zakai lower bound}
\acrodef{ZZB}{Ziv-Zakai bound}
\acrodef{LOS}{line-of-sight}
\acrodef{ToF}{time-of-flight}
\acrodef{NLOS}{non-line-of-sight}
\acrodef{GDOP}{geometric dilution of precision}
\acrodef{GPS}{Global Positioning System}
\acrodef{FIM}{Fisher information matrix}
\acrodef{PEB}{position error bound}
\acrodef{SPEB}{squared position error bound}
\acrodef{TOA}{time-of-arrival}
\acrodef{TOF}{time-of-flight}
\acrodef{WSN}{wireless sensor network}
\acrodef{MAC}{medium access control}
\acrodef{RSS}{received signal strength}
\acrodef{WAF}{wall attenuation factor}
\acrodef{TDOA}{time difference-of-arrival}
\acrodef{RF}{radiofrequency}
\acrodef{RTT}{round-trip time}
\acrodef{AOA}{angle-of-arrival}
\acrodef{MF}{matched filter}
\acrodef{ED}{energy detector}
\acrodef{ML}{maximum likelihood}
\acrodef{MSE}{mean-square error}
\acrodef{RMSE}{root-mean-square error}
\acrodef{LEO}{localization error outage}
\acrodef{ppm}{part-per-million}
\acrodef{ACK}{acknowledge}
\acrodef{UWB}{Ultrawide bandwidth}
\acrodef{TNR}{threshold-to-noise ratio}
\acrodef{LS}{least squares}
\acrodef{IR-UWB}{impulse radio UWB}
\acrodef{FCC}{Federal Communications Commission}
\acrodef{TH}{time-hopping}
\acrodef{PPM}{pulse position modulation}
\acrodef{MUI}{multi-user interference}
\acrodef{PDP}{power delay profile}
\acrodef{BPZF}{band-pass zonal filter}
\acrodef{SIR}{signal-to-interference ratio}
\acrodef{RFID}{radio frequency identification}
\acrodef{WPAN}{wireless personal area network}
\acrodef{WWB}{Weiss-Weinstein bound}
\acrodef{DP}{direct path}
\acrodef{MF}{matched filter}
\acrodef{MMSE}{minimum-mean-square-error}
\acrodef{SBS}{serial backward search}
\acrodef{SBSMC}{serial backward search for multiple clusters}
\acrodef{NBI}{narrowband interference}
\acrodef{WBI}{wideband interference}
\acrodef{INR}{interference-to-noise ratio}
\acrodef{CR}{channel response}
\acrodef{CIR}{channel impulse response}
\acrodef{CR}{channel  response}
\acrodef{RADAR}{radar}
\acrodef{MUR}{Multistatic radar}
\acrodef{JBSF}{jump back and search forward}
\acrodef{HDSA}{high-definition situation-aware}
\acrodef{RRC}{root raised cosine}
\acrodef{ST}{simple thresholding}
\acrodef{BTB}{Bellini-Tartara bound}
\acrodef{P-Max}{$P$-Max}  %suggestion, use with \acl{P-Max}
\acrodef{MIMO}{multiple-input multiple-output}
\acrodef{MAP}{maximum a posteriori}
\acrodef{FG}{factor graph}
\acrodef{OP}{outage probability}
\acrodef{WED}{wall extra delay}
\acrodef{RMS}{root mean square}
\acrodef{SPAWN}{sum-product algorithm over a wireless network}
\acrodef{MDD}{minimum distance distribution}
\acrodef{MAP}{maximum a posteriori probability}
\acrodef{PAR}{probabilistic association rule}

\usepackage{color}
\usepackage{dsfont}
\usepackage{bbm}

\newcommand{\msgStyleBlock}[1]{%
\noindent\textcolor{red}{$\blacksquare$ #1 $\blacksquare$}}
\newcommand{\todoStyleDefault}[1]{\msgStyleBlock{#1}}

\newcommand{\ChG}[1]{  \RV{H}_{#1} }

\newcommand{\IndF}[2]{\mathbbm{1}_{#1}\!\!\left(#2 \right)}

\def\lo{\text{o}}
\def\lx{\bold{x}}
\def\x{\bold{X}}
\def\ly{\bold{y}}
\def\y{\bold{Y}}
\def\z{\bold{Z}}
\def\lz{\bold{z}}
\def\rr{r}

\def\RI{\mathsf{R}}

\newcommand{\AU}[1]{\mathcal{U}(#1)}

\newcommand{\FD}{\text{FD}}
\newcommand{\HD}{\text{HD}}
\def\BS{\text{a}}
\def\US{\text{u}}

\def\trm{\zeta}

\def\PPP{\mathsf{\bold{\Pi}}}
\newcommand{\PPPs}[1]{\PPP_{\BS, #1}}
\newcommand{\PPPn}{\PPP_{\US}}
\newcommand{\PPPm}[2]{\PPP_{\BS,#1}^{#2}}
\newcommand{\Con}[2]{c_{#1}\!\left(\!#2\!\right)}
\def\Pwr{P}

\newcommand{\PL}[1]{\alpha_{#1}} % pathloss
\newcommand{\Pbs}[1]{\Pwr_{\BS,#1}}
\newcommand{\Pus}[1]{\Pwr_{\US,#1}}
\newcommand{\Prx}{\Pwr_{\mathsf{r}}}
\newcommand{\Ptx}{\Pwr_{\mathsf{t}}}

\newcommand{\Ws}[2]{{W_{}^{}}} % Symbol bandwidth
\newcommand{\Ts}{T_\text{s}} % Symbol duration

\newcommand{\Int}[2]{I_{#1}^{#2}}

\newcommand{\LT}{\lambda_{\text{t}}}
\newcommand{\LBt}[1]{\lambda_{#1}}
\newcommand{\LB}[2]{\lambda_{#1}^{#2}}
\newcommand{\LUt}{\mu}

\newcommand{\LRatio}[1]{\mathcal{R}_{\lambda, #1}}

\newcommand{\FnCk}[3]{\mathcal{C}^{#1}_{#3}\!(#2)}
\newcommand{\SICdB}[1]{L_{\text{dB},#1}}
\newcommand{\FnI}[2]{\mathcal{I}_{#1}\!\left( #2\right)}
\newcommand{\FnM}[3]{\trm_{#1}^{}\!\left(#3 \right)}
\newcommand{\FnMA}[3]{\sigma_{#1}^{}\!\left(#3 \right)}
\newcommand{\FnMB}[3]{{\tilde{\sigma}}_{#1}^{}\!\left(#3 \right)}
\newcommand{\FnMa}[3]{{\delta}_{#1}^{}\!\left(#3 \right)}
\newcommand{\FnMb}[3]{{\tilde{\delta}}_{#1}^{}\!\left(#3 \right)}
\def\SetI{\mathcal{K}}
\newcommand{\SetT}[2]{\mathcal{T}_{#1}^{#2}}

\def\rDis{\RV{D}}

\newcommand{\Bratio}[1]{\mathcal{B}_{#1}} % Bias ratio
\newcommand{\Bias}[1]{\mathcal{W}_{#1}} % Bias factor
\newcommand{\Biasfactor}[1]{\mathcal{U}_{#1}}

\newcommand{\SIR}[2]{\gamma_{#1}^{#2}}

\newcommand{\TSIR}[2]{{\tau_{}^{}}}
\newcommand{\tSIR}[1]{{\tau_{#1}}}

\newcommand{\tRate}[1]{\mathsf{R}_{#1}}
\def\TRate{\mathsf{R}_{\text{t}}}
\newcommand{\Pa}[2]{p_{\text{A},#1}^{#2}}

\newcommand{\Psk}[2]{p_{\text{S}, #1}^{#2}}
\newcommand{\Ps}[3]{p_{\text{S}, #1}^{#2}\! \left(#3\right)}

\newcommand{\Pfd}[1]{p_{#1}^{\FD}}
\newcommand{\PfdO}[1]{\hat{p}_{#1}^{\FD}}
\newcommand{\SE}{\mathcal{S}}
\newcommand{\CSE}{\mathcal{S}^{\text{c}}}
\newcommand{\NSE}{\mathcal{S}^{\text{n}}}
\DeclareMathAlphabet{\mathsf}{OML}{cmbr}{m}{it}

\newcommand{\redR}[1]{{#1}}

\newcommand{\blue}[1]{{#1}} % {{\textcolor[rgb]{0,0,1}{#1}}} % <-- This is real blue

\newtheorem{definition}{Definition}
\newtheorem{theorem}{Theorem}
\newtheorem{lemma}{Lemma}
\newtheorem{corollary}{Corollary}

\newtheorem{remark}{Remark}

\newcommand{\R}{\mathbbmss{R}}

\newcommand{\RV}[1]{\mathsf{#1}}
\newcommand{\PX}[1] {{\mathbb{P}}\left\{{#1}\right\}}
\newcommand{\EX}[1] {{\mathbb{E}}\left\{{#1}\right\}}
\newcommand{\EXs}[2] {{\mathbb{E}}_{{#1}}\!\!\left\{{#2}\right\}}
\newcommand{\PDF}[2]{p_{{#1}}\left({#2}\right)}

\newcommand{\Erfc}[1]{\textsf{Erfc}\left(#1\right)}

\newcommand{\GF}[1]{\Gamma\!\left(#1\right)}
\newcommand{\HGF}[3]{{}_{#1}F_{#2}\!\left(#3\right)}
\newcommand{\sNorm}[1]{\left|{#1}\right|}
\newcommand{\vNorm}[1]{\left\|{#1}\right\|}
\newcommand{\Lap}[2]{\mathcal{L}_{#1}\left(#2 \right)}

\newcommand{\bd}{\begin{description}}
\newcommand{\ed}{\end{description}}
\newcommand{\be}{\begin{enumerate}}
\newcommand{\ee}{\end{enumerate}}
\newcommand{\bi}{\begin{itemize}}
\newcommand{\ei}{\end{itemize}}
\newcommand{\bl}{\begin{list}}
\newcommand{\el}{\end{list}}
\newcommand{\bt}{\begin{tabbing}}
\newcommand{\et}{\end{tabbing}}

\newcounter{eqncnt}

\newcounter{eqnback}

\acrodef{BS}{base station}
\acrodef{AP}{access point}
\acrodef{HD}{half-duplex}
\acrodef{FD}{full-duplex}
\acrodef{IC}{interference cancellation}
\acrodef{HDHN}{hybrid-duplex heterogeneous network}
\acrodef{TDD}{time-division duplexing}
\acrodef{FDD}{frequency-division duplexing}
\begin{document}

\markboth{IEEE Transactions on Wireless Communications, accepted for publication}{Lee \& Quek:Hybrid Full-/Half-Duplex System Analysis in Heterogeneous Wireless Networks}

\newcommand{\paperTitle}{Hybrid Full-/Half-Duplex System Analysis\\ 
in Heterogeneous Wireless Networks}
%
%\newcommand{\paperTitle}{
%Throughput Analysis for Heterogeneous Network \\
%with Hybrid Full-/Half-Duplex Systems}
%
%\newcommand{\paperTitle}{Heterogeneous Network Throughput \\
% with Hybrid Full-/Half-Duplex Systems}
 
%---------------------------------------------------------------------------%
%                     title, title footnote, header                         %
%---------------------------------------------------------------------------%
 
%\twocolumn

\title{\paperTitle}
% \title{Distributed Secrecy in \\Multilevel Wireless Networks}

% author names, IEEE memberships, corresponding address, title footnote %
\author{
%WGroup
%%%%%%%% uncomment this section for a 2-column formt %%%%%%%
%%%%%%%% [begin] %%%%%%%%
	\vspace{0.2cm}
        Jemin~Lee, \textit{Member, IEEE} and 
        Tony~Q.~S.~Quek, \textit{Senior Member, IEEE}
%[-0.5em]
%
%
% \IEEEauthorblockA{\IEEEauthorrefmark{1}ENDIF, University of Ferrara, Email: \texttt{a.conti@ieee.org}}
%
%
   \thanks{Manuscript received April 07, 2014; revised August 26, 2014 and November 5, 2014.} 
    \thanks{
        J.\ Lee and T.\ Q.\ S.\ Quek are with    
       the Singapore University of Technology and Design,  
       8 Somapah Road, Singapore 487372      
       (e-mail:\texttt{jmnlee@ieee.org}, \texttt{tonyquek@sutd.edu.sg}). 
       The material in this paper was presented, in part, at the Global Communications Conference, Austin, TX, Dec. 2014.
}
    \thanks{
       This research was supported, in part, by Temasek Research Fellowship, the SUTD-MIT International
Design Centre under Grant IDSF1200106OH, and the A$*$STAR SERC under Grant
1224104048.
       }
}

%% make the title area
%% Don't write page number 0 to the cover page.
\maketitle %% make the title area
%% Don't write page number 0 to the cover page.

%
% \markboth{Submitted to IEEE Journal on Selected Areas in Communications}{\title}

%
%%%%%%%%% uncomment this section for a 2-column formt %%%%%%%
%%%%%%%%% [begin] %%%%%%%%
%\thispagestyle{empty}
%  \textcolor{blue}{\framebox{\textsf{\small{Today: \today}}}}\\

%
%\newpage
%%%%%%%%% [end] %%%%%%%%
\setcounter{page}{1}
\acresetall
%%---------------------------------------------------------------------------%
%%                           abstract and key words                          %
%%---------------------------------------------------------------------------%
\begin{abstract}
Full-duplex (FD) radio has been introduced for bidirectional communications on the same temporal and spectral resources
so as to maximize spectral efficiency. 
% , and this figure can be utilized for heterogeneous networks for enhanced link capacity, spectrum usage flexibility, and communication security.
%
In this paper, motivated by the recent advances in {FD} radios, 
%we consider hybrid-duplex cell networks for future heterogeneous cellular networks. 
% 
%Different from conventional heterogeneous networks, 
\blue{we provide a foundation for \acp{HDHN}, composed of multi-tier networks 
with a mixture of \acp{AP}, operating either in bidirectional {FD} mode or downlink \ac{HD} mode.}
Specifically, we characterize the network interference from {FD}-mode cells, and derive the \ac{HDHN} throughput 
by accounting for % network parameters including 
\ac{AP} spatial density, self-\ac{IC} capability, and transmission power of \acp{AP} and users. 
By quantifying the \ac{HDHN} throughput, 
we present the effect of network parameters and the self-\ac{IC} capability on the \ac{HDHN} throughput,
and show the superiority of {FD} mode 
for larger \ac{AP} densities (i.e., larger network interference and shorter communication distance) or higher self-\ac{IC} capability.
% and show how to optimally determine the duplex modes. 
% in terms of the \ac{HDHN} throughput in the presence of the network interference.
% as well as the self-interference. 
Furthermore, our results show operating all \acp{AP} in {FD} or \ac{HD} achieves higher throughput 
compared to the mixture of two mode \acp{AP} in each tier network, 
and introducing hybrid-duplex for different tier networks improves the heterogenous network throughput. 
\end{abstract}


%\clearpage
\acresetall
%%%%%%%%%%%%%%%%%%%%%%%%%%%%%%%%%%%%

%%---------------------------------------------------------------------------%
%%                           Sec: Introduction                                    %
%%---------------------------------------------------------------------------%

\section{Introduction}
%
%Motivated by the recent advances in full-duplex radios, we propose the novel idea of hybrid-duplex cell networks for future heterogeneous cellular networks. Different from conventional heterogeneous networks, we consider a network with a mixture of \acp{AP} with full-duplex and half-duplex capabilities. In this work, our main contributions are as follows:
%\begin{itemize}
%\item Based on the mathematically tractable two-tier network model in \cite{CheQueKou:12}, we incorporate the concept of hybrid-duplex small cells within the cellular networks.
%\item Characterize the optimal density of SAP in \ac{FD} mode and \ac{HD} mode, respectively.
%\end{itemize}
%
%Objectives: 
%\bi
%\item To provide a framework for the efficient design of hybrid-duplex heterogeneous networks; 
%\item To investigate the effects of the self-\ac{IC} performance and the network interference
%on the network performance; 
%\item To obtain the optimal density of \ac{FD}-mode cells that maximizes the heterogeneous network throughput.
%\ei

%\mynote{General:} \mysubnote{Why full duplex becomes more meaningful for heterogenous networks}
% 
Conventional communication systems operate 
in \ac{HD} such as time-division or frequency-division approaches, 
which require different orthogonal resources in either temporal or spectral domain for bidirectional communications. 
As a way of enhancing the spectral efficiency of communication systems, 
\ac{FD} has been introduced to perform bidirectional communications 
on the same temporal and spectral resources. 
Thus, \ac{FD} radios can potentially be employed in heterogeneous networks 
%can be utilized for heterogeneous networks, especially for small cells supporting simultaneous connections between \acp{AP} and users, in multiple ways
% such as 
for increased link capacity, more flexibility in spectrum usage, 
and improved communication security\cite{SabSchGuoBliRanWic:14}. %\cite{Tap:13}.  
%

%This feature offers the potential to complement and enforce solutions 
% for high data rate and ultra reliable communications, which will be essential for the sustainability of future communication. 

% \mynote{Prior works: Full duplex} 
Different \ac{FD} systems have been studied
considering 
the asynchronous transmission and reception\cite{SahSab:12}, 
the one-way relay transmission\cite{JuLimKimPooHon:12,LiLiTeh:11,DayMarBliSch:12b}, 
the two-way relay transmission\cite{JuOhHon:09,JuLimKimPooHon:12}, 
the imperfect channel estimation and limited dynamic range in a \ac{MIMO} system\cite{DayMarBliSch:12a}, and 
relays with different self-\ac{IC} capabilities in multi-hop transmission\cite{WeeCodLatEph:10}.
%
%The \ac{MAC} protocol design of \ac{FD} mode nodes\cite{SinGunProRadBal:11,MiuBan:12}
%and  
The achievable rates of \ac{HD} and \ac{FD} in MIMO systems have been compared in \cite{BarRan:12},
\blue{and the degree-of-freedom of the system with a \ac{FD} \acp{BS} with \ac{HD} users has been analyzed 
by considering the intra-cell inter-node interference\cite{SahDigSab:13}.} 
The \ac{FD} radios has also been used in jamming techniques for communication secrecy\cite{ZheKriLiPetOtt:13}
and 
bidirectional broadcast communications by implementing rapid on-off-division duplex\cite{ZhaGuo:14}. 
\blue{The hybrid of \ac{FD}- and \ac{HD}-relaying schemes have also been presented,
which allows a relay to opportunistically switch between two modes
based on instantaneous\cite{YamHanMurYos:11,NgLoSch:12,RiiWerWic:11a} or statistical \ac{CSI}\cite{RiiWerWic:11a}.}

The key challenging in implementing a \ac{FD} radio is the presence of self-interference, 
received at a node from its own transmission while transmitting and receiving at the same time. 
Self-\ac{IC} techniques for \ac{FD} systems with multiple antennas 
have been proposed by exploiting the following domains: 
\blue{1) propagation-domain schemes including
antenna separation % , to isolate the transmitting and receiving signals via physical pathloss
\cite{DuaSab:10,SnoFulCha:11}
% cross-polarizarion\cite{}, 
and
directional transmit/receive antennas (e.g., beamforming-based techniques)
\cite{LioVibColAth:10,RiiWerWic:11,LeeSimChaKan:14};}
2) analog circuit-domain including
channel-unaware schemes\cite{DuaSab:10,DuaDicSab:12} and
channel-aware schemes\cite{DuaSab:10,ChoJaiSriLevKat:10,JaiChoKimBhaSir:11,BhaMcmKat:13};
3) digital circuit-domain \cite{RiiWerWic:11a,NgLoSch:12}; 
and 
4) hybrid of analog and digital domains\cite{RiiWic:12,HuaMaLiaCir:13}. 
% 
%2) active circuit-domain cancellation 
%including analog and digital domains % cancellation
%\cite{DuaDicSab:12,ChoJaiSriLevKat:10,JaiChoKimBhaSir:11,AhmEltSab:13,
%RiiWerWic:11a,NgLoSch:12},
%and the hybrid of two domains\cite{RiiWic:12,HuaMaLiaCir:13}.  
%% analog-digital hybrid method\cite{HuaMaLiaCir:13}. 
%\blue{Furthermore, beamforming-based self-\ac{IC} techniques 
%have been proposed to exploit the antenna array for the cancellation\cite{LioVibColAth:10,RiiWerWic:11,LeeSimChaKan:14}.}
%% dynamic rande reduction in 
%% signal processing in analog and digital domains 
%% \cite{JaiChoKimBhaSir:11,ChoJaiSriLevKat:10}. 
%%the analog domain cancellation by modifying transmitted signal that can be canceled by cancellation vector
%
%
{The self-\ac{IC} techniques are being researched actively as
the current self-\ac{IC} capability is still challengeable.}
Recently, the feasibility of single (shared)-array \ac{FD} transceivers 
% the feasibility of \ac{FD} system with a single antenna 
has also been presented in \cite{CoxAck:13,Kno:12,BhaMcmKat:13}. 
%
%-------------
% However, there is no work that characterizes and considers the network interference generated from randomly distributed \ac{FD}-nodes
% for the performance evaluation of \ac{FD} systems.
However, there is no work that considers the self-interference together with 
the network interference generated from randomly distributed \ac{FD}-mode nodes
for the performance evaluation of \ac{FD} systems.
% although it is paramount on the design of \ac{FD} wireless networks.  
%However, there is no work considering the network interference along with imperfect self-\ac{IC}.
% As we increases nodes operating in \ac{FD}, 
If more nodes operate in \ac{FD}, 
the number of communicating nodes in the network increases,
% we have more communicating nodes in the network,
but network interference also increases, % due to the simultaneous transmission and receoption,
which can degrade the communication reliability between nodes. 
%the network interference increases since receivers are also transmitting, 
%which can degrade the communication reliability between nodes in the network. 
%\redR{The network interference from \ac{FD}-mode nodes has also been presented in \cite{GoyLiuHuaPan:13,TonHae:14},
%but the perfect cancellation of self-interference are assumed in both works, 
%and the intra-cell interference issue has not been considered in \cite{GoyLiuHuaPan:13} and 
%the random link distance between a user to its associated \ac{BS} has not been considered in \cite{TonHae:14}. 
%}
\redR{The network interference from \ac{FD}-mode nodes has been presented in \cite{GoyLiuHuaPan:13,TonHae:14}, 
but one tier network was considered with the perfect self-\ac{IC} assumption.}\footnote{\redR{Furthermore, 
	simpler network model was used in \cite{GoyLiuHuaPan:13} 
	by ignoring the intra-cell interference, which can be generated by users accessing the same resource in a cell, 
	and the fixed link distance between a user and its communicating \ac{AP}
	was used in  \cite{TonHae:14}.}     
}  
 
%\redR{The network interference from \ac{FD}-mode nodes has also been presented for a tier network in \cite{GoyLiuHuaPan:13,TonHae:14}. However, both assumed the perfect self-\ac{IC}, and 
%%with the perfect self-interference cancellation for a tier network, and 
%% but the perfect cancellation of self-interference are assumed,
%the intra-cell interference, generated by users accessing the same resource in a cell, 
%and the link distance distribution, determined by a cell association policy,  
%has not been considered in \cite{GoyLiuHuaPan:13} and \cite{TonHae:14}, respectively.} 
% and the intra-cell interference issue has not been considered in \cite{GoyLiuHuaPan:13} and 
% the random link distance between a user to its associated \ac{BS} has not been considered in \cite{TonHae:14}. 

% \mynote{Heterogenous networks}
% \mysubnote{Prior works: HetSNet}
The performance of heterogeneous networks has been studied
\cite{QueRocGuvKou:13,CheQueKou:12,DhiGanBacAnd:12,NovDhiAnd:13,DhiGanAnd:13,
LopGuvRocKouQueZha:11,LeeAndHon:13,
HuaLauChe:09,LeeAndHon:11,WilQueSluRab:13,SohQueKouShi:13,
JoSanXiaAnd:12,SinDhiAnd:13,
SohQueKouCai:13} 
%by taking into account the network interference 
by taking into account the spatial node distribution using the \ac{PPP},
which is widely used
in wireless networks\cite{WinPinShe:J09,PinWin:J10,PinWin:J10a,LeeConRabWin:J13,HeaKouBai:13,WinRabLeeCon:14,BanLeeKimHon:15}.\footnote{\blue{Recently, 
the \ac{PPP} has also been used to evaluate some advanced techniques such as 
the coordinated multiple-point (CoMP) with \ac{BS}-centric\cite{AkoHea:13} and user-centric\cite{GarZhoShi:14} clustering,
and the self-powered transmitters using energy harvesting techniques\cite{DhiLiNugPiAnd:14}.}}
%\cite{WinPinShe:J09,PinWin:J10,PinWin:J10a,HeaKouBai:13}.
% 
% \cite{Win:T07CTW,WinPinShe:J09,HaeAndBacDouFra:09,OrrBar:03,SalZan:09,RabQueShiWin:J11,Sou:92,GhaSou:08a,PinWin:J10,PinWin:J10a,IloHatVen:98,YanPet:03,GovBliSta:07,LeeAndHon:11}.
%
The heterogeneous network throughput has been presented
by considering 
the $K$-tier spectrum sharing network in downlink\cite{DhiGanBacAnd:12} and in uplink\cite{NovDhiAnd:13},
the \ac{BS} loads of different tier networks\cite{DhiGanAnd:13},
% the biased \ac{BS} association rules\cite{JoSanXiaAnd:12}, 
the interference cancellation capability\cite{LopGuvRocKouQueZha:11,LeeAndHon:13}, 
the spectrum sharing methods\cite{HuaLauChe:09,LeeAndHon:11},
%the spectrum allocation\cite{CheQueKou:12},
and
the trade-off between traffic offloading and energy consumption of small cells\cite{WilQueSluRab:13,SohQueKouShi:13}. 
\blue{To solve the load balancing problem in heterogeneous networks with \ac{HD} systems,
the concept of cell range expansion has also been considered
to offload users to less loaded networks 
using a biased cell association rule\cite{JoSanXiaAnd:12,SinDhiAnd:13}. }
%For load balancing among different tier networks, 
%the cell range expansion has also used to offload   
%The cell range expansion has been also 
%the biased \ac{BS} association rules\cite{JoSanXiaAnd:12}, 
% 
The design of duplex communication modes is also presented by
considering 
the coordinated \ac{TDD} underlay structure in two-tier networks\cite{LimBenGhaLat:10}, and
the hybrid division duplexing in the network  
composed of macro-cells in \ac{FDD} and cognitive femto-cells in \ac{TDD}\cite{SohQueKouCai:13}. 
% CheQueKou:12
%  
% [HeaKouBai:13] Modeling HetNet Interference using PPP
%--- [K Tier Networks] 
% [DhiGanBacAnd:12] Modeling and Analysis of K-Tier Downlink Heterogeneous cellular networks 
% [DhiGanAnd:13] Load-Award Modeling and ANalysis of Heterogeneous 
% [] Analytical Modeling of Uplink Cellular Networks
%
%--- [Associataion]
%
%---[TDD FDD]
% [LimBenGhaLat:10] Interference management for self-organized femtocells towards free networks 
However, most of these works is based on \ac{HD} and 
does not explore the effect of \ac{FD} on network throughput, % heterogeneous network throughput,
impeding the efficient duplex mode design for heterogeneous networks. 

%According to the different network parameters (e.g., transmission power) and the interference, each tier network may need to have different duplex schemes

 %\/\/\/\/\/\/\/\/\/\/\/\/\/\/\/\/\/\/\/\/\/\/\/\/\/\/\/\/\/\/\/\/\/\/\/\/\/\/\/\/\/\/\/\/\/\/\/\/\/\/\/\/\/\/\/\/\/\/\/\/\/\/\/\/\/\/\/\/\/\/\/\/\/\/\/\/\/\/\/\/\/\/\/\/\/\/\/
%***** x-axis:  [tc][bc][0.7] y-axis: [bc][tc][0.7], legend: [Bl][Bl][0.59]
\begin{figure}[t!]
    \begin{center}   
    { 
	 \includegraphics[width=1.00\columnwidth]{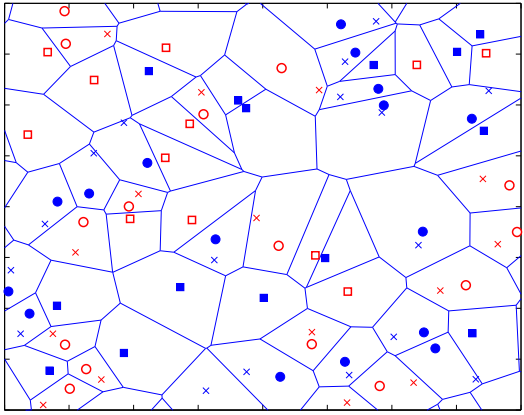}
	\vspace{-10mm}
    }
    \end{center}
    \caption{
    		An example of downlink two-tier \acp{HDHN} 
		(squares, circles, crosses are 
		\ac{HD}-mode \acp{AP}, \ac{FD}-mode \acp{AP}, and \ac{FD}-mode users sharing the same channel
		in the first-tier network (empty red) and in the second-tier network (filled blue)). 
%		(empty red squares, circles, crosses are 
%		\ac{HD}-mode \acp{AP}, \ac{FD}-mode \acp{AP}, and transmitting users in the first-tier network, and 
%		filled blue squares, circles, crosses are 
%		\ac{HD}-mode \acp{AP}, \ac{FD}-mode \acp{AP}, and transmitting users in the second-tier network.)		
		 }
   \label{fig:NetworkModel}
\end{figure}
%\/\/\/\/\/\/\/\/\/\/\/\/\/\/\/\/\/\/\/\/\/\/\/\/\/\/\/\/\/\/\/\/\/\/\/\/\/\/\/\/\/\/\/\/\/\/\/\/\/\/\/\/\/\/\/\/\/\/\/\/\/\/\/\/\/\/\/\/\/\/\/\/\/\/\/\/\/\/\/\/\/\/\/\/\/\/\/

\begin{table}
\caption{Notations used throughout the paper.} \label{table:notation}
\begin{center}
\rowcolors{2}%{green!10!yellow}{}
{cyan!15!}{}
\renewcommand{\arraystretch}{1.3}
\begin{tabular}{c  p{7cm} }
\hline 
 {\bf Notation} & {\hspace{2.5cm}}{\bf Definition}
\\
\midrule
\hline
$\PPPs{k}$ & PPP for \ac{AP} distribution of network $k$ \\ \addlinespace
$\PPPm{k}{m}$ & PPP for $m$-mode \ac{AP} distribution in network $k$ \\ \addlinespace
$\LB{k}{}$ 	& Spatial density of \acp{AP} of network $k$\\ \addlinespace
$\LB{k}{m}$ 	& Spatial density of $m$-mode \acp{AP} in network $k$\\ \addlinespace
$\Ts$		& Symbol time\\ \addlinespace % [sec]
$\Ws{k}{m}$ 		& Communication bandwidth %  of a $m$-mode \ac{AP} in network $k$ 
					\\ \addlinespace %[Hz]
% $\Pwr$ 		& Transmission power of a user  \\ \addlinespace
$\ChG{\lx,\ly}$ 		& Fading level of the link between nodes at $\lx$ and $\ly$\\ \addlinespace
$\rDis_{\lx,\ly}$ 		& Distance of the link between nodes at $\lx$ and $\ly$\\ \addlinespace
$\rDis_k^{m}$		& Distance of the link between an user and its associated $m$-mode \ac{AP} in network $k$ \\ \addlinespace
$\PL{k}$ 			& Pathloss exponent in network $k$\\ \addlinespace
$\Bias{k}$ & Weighting factor of network $k$\\ \addlinespace 
$\Bratio{ik}$		& Ratio between association factors, $\Bias{i}/\Bias{k}$\\ \addlinespace
$\Pbs{k}$	& Transmission power of a \ac{AP} in network $k$\\ \addlinespace
$\Pus{k}$	& Transmission power of a user in network $k$\\ \addlinespace
$\Prx$	& Transmission power at a receiving node\\ \addlinespace
$\Int{k}{m}$	& Interference from $m$-mode \acp{AP} in network $k$\\ \addlinespace
$\SIR{k}{m}$	& {SIR} at $m$-mode node in network $k$\\ \addlinespace
$\TSIR{k}{m}$	& Target {SIR} \\ \addlinespace % of $m$-mode node in network $k$
$\FnCk{m}{\Prx}{k}$		& Self-\ac{IC} capability of $m$-mode node in network $k$\\ \addlinespace
% $\FnCk{m}{\Prx}{k}$		& Self-\ac{IC} capability of $m$-mode node in network $k$\\ \addlinespace
% 
$\Pfd{k}$		& Portion of \ac{FD}-mode \acp{AP} in network $k$\\ \addlinespace
% $\Psk{k}{m}$	& Successful transmission probability of $m$-mode node in network $k$\\ \addlinespace
$\Psk{k}{m}$	& Successful transmission probability of $m$-mode node\\ \addlinespace
$\SE$		& \ac{HDHN} throughput [bits/sec/Hz/m$^2$]\\ \addlinespace
$\SE_k$		& \ac{HDHN} throughput of network $k$ [bits/sec/Hz/m$^2$]\\ \addlinespace
$\CSE$		& Cell throughput of \acp{HDHN} [bits/sec/Hz/cell]\\ \addlinespace
%$\Pfd{k}$		& Portion of \acp{AP} in $k$th-tier network\\ \addlinespace
% $\SIR{k}{m}$	& \ac{SIR} at $m$-mode node in $k$th-tier network\\ \addlinespace
\hline 
\end{tabular}
\end{center}\vspace{-0.63cm}
\end{table}%

Motivated by the recent advances in \ac{FD} radios, 
we propose the novel idea of hybrid-duplex cell networks for future heterogeneous cellular networks. 
%Different from conventional heterogeneous networks, 
We consider \acp{HDHN}, composed of multi-tier networks with a mixture of \acp{AP} operating either in 
\blue{ bidirectional \ac{FD}  mode or downlink \ac{HD} mode}. 
% \ac{FD} or \ac{HD}. 
%
We develop a framework for \acp{HDHN} in the presence of self-interference and network interference. 
Specifically, after characterizing the network interference of \acp{HDHN}, 
we define a performance metric, namely the \ac{HDHN} throughput, %to determine the overall throughput in the network aspect.
to measure the average data rate achieved by \acp{AP} and users successfully communicating in this network.  
Based on this metric, we present the effect of network parameters such as the \ac{AP} spatial density, the network interference, 
and the self-interference on the \ac{HDHN} throughput.
We then determine the portion of \ac{FD}-mode \acp{AP} that maximizes the \ac{HDHN} throughput
based on the system parameters including \ac{AP} spatial density. 
\blue{Note that this is different from the opportunistic mode switching based on \ac{CSI} in \cite{YamHanMurYos:11,NgLoSch:12,RiiWerWic:11a}.}
%, especially, we present the optimal spatial density for the case of \ac{FD}-mode \acp{AP} with perfect self-\ac{IC} capability. 
% 
The main contributions of this paper can be summarized as follows:
\bi
\item we characterize the network interference generating from distributed \acp{AP} and users in \ac{FD}-mode cells; 
\item we introduce and derive the \ac{HDHN} throughput that accounts 
for self-interference, spatial \ac{AP} densities, and transmission power of \acp{AP} and users; and 
%the \ac{AP} association rule; and 
%
\item we quantify the \ac{HDHN} throughput and present the optimal portion of \ac{FD}-mode \acp{AP} 
to maximize the \ac{HDHN} throughput 
according to the self-\ac{IC} capability and network parameters. % of the \ac{HDHN}. 
\ei

The remainder of this paper is organized as follows: 
Section~\ref{sec:models} describes the \ac{HDHN} model and provides the statistical characterization of interference from \ac{FD}-mode cells. 
%
% Section~\ref{sec:metrics} introduces and analyzes the \ac{D2D-CNS} throughput. 
% Section~\ref{sec:STP} analyzes the successful transmission probability in the \ac{HDHN}. 
Section~\ref{sec:throughput} analyzes the successful transmission probability of \acp{HDHN}, 
and introduces and analyzes the \ac{HDHN} throughput. 
Section~\ref{sec:numerical} quantifies the effects of network parameters and self-interference capability on the \ac{HDHN} throughput and
determines the optimal portion of \ac{FD}-mode \acp{AP} in \acp{HDHN}.
Finally, conclusions are given in Section~\ref{sec:conclusion}.

{\it Notation:} The notation used throughout the paper is reported in Table~\ref{table:notation}.

%%---------------------------------------------------------------------------%
%%                           Sec: Network Model                               %
%%---------------------------------------------------------------------------%
%  \textcolor{blue}{\framebox{\textsf{\small{Today: \today}}}}

\section{Hybrid-Duplex Heterogeneous Network Model}\label{sec:models}
%
%
%\bi
%\item Netowrk structure
%\item Node distribution - PPP
%\item FD and HD modes -- Some portion randomly choose -- independent PPP
%\item Transmission power
%\item The function of self-interference cancellation: $\FnC{m}{\Prx} > 0$ if $m = \FD$, otherwise $\FnC{m}{\Prx} = 0$.
%\item AP association 
%\ei
In this section, we describe the \ac{HDHN} model
and characterize the network interference of the \ac{HDHN}. 
%and provide the definition of \ac{SIR} in this network. 

%
\subsection{Network Model}
We consider \acp{HDHN} composed of $K$-tier wireless networks. 
The $k$th-tier network consists of \acp{AP} distributed in space according to a homogeneous \ac{PPP} $\PPPs{k}$ 
with spatial density $\LBt{k}$. 
Each \ac{AP} forms a cell and communicates to nodes
in either \blue{downlink \ac{HD} mode or bidirectional \ac{FD} mode}. 
The portion of \ac{FD}-mode \acp{AP} in the $k$th-tier network is $\Pfd{k}$, 
and the distributions of \ac{HD}-mode and \ac{FD}-mode \acp{AP} also follow \acp{PPP}, 
$\PPPm{k}{\HD}$ and $\PPPm{k}{\FD}$, 
with spatial densities $\LB{k}{\HD}= \LBt{k}(1-\Pfd{k})$ and $\LB{k}{\FD}= \LBt{k}\Pfd{k}$, respectively. 
\blue{All \ac{HD}-mode cells are in downlink while all \ac{FD}-mode cells have both uplink and downlink communications.}
% due to the independent thinning property of a \ac{PPP}\cite{Kin:B93}. 
%due to the independence between two mode \ac{AP} distribution.
\blue{In the \acp{HDHN}, 
users in \ac{FD}-mode cells and all \acp{AP} of the $k$th-tier network 
transmit with power $\Pus{k}$ and $\Pbs{k}$, respectively,
and generally $\Pbs{k}\geq\Pus{k}$.}
%In downlink, 
%all \acp{AP} and nodes in the $k$th-tier network transmit with power $\Pbs{k}$ and $\Pus{k}$, respectively,
%and generally $\Pbs{k}\geq\Pus{k}$.
% {All $K$-tier networks share the same spectrum, but in a cell, 
Each channel is used by 
one user in a cell to avoid intra-cell interference, and the whole spectrum is utilized in each cell.\footnote{Note
that if channels are not always used in every cells,
it only affects the spatial density of interfering nodes in the same framework of this paper. 
% Hence,  which we can easily extend without changing the framework of this paper. 
% which we can  $\LI$ are not used in some cells, 
}
%We assume that each cell has one user that uses the same channel at the same time.\footnote{The node
%	can be one of open-access nodes or close-access nodes.
%}
An example of downlink \acp{HDHN} is presented in Fig.~\ref{fig:NetworkModel}.

Users are scattered in \acp{HDHN} according to a homogeneous \ac{PPP} $\PPPn$
with spatial density $\LUt$, and a node located at $\lx_\text{o}$ connects to an \ac{AP} % located at $\hat{\lx}_k$ 
in the $k$-tier networks  
based on the association rule, 
presented by\cite{SinDhiAnd:13}
\begin{align}\label{eq:association}	
	k = \arg \max_{i\in \SetI}
		%\underbrace{ 
		\left\{ 
				\max_{\x_i  \in \PPPs{i}}\Bias{i} \rDis_{\x_i, \lx_\text{o}}^{-\PL{i}}
		\right\}
		% }_{\hat{\lx}_i}
\end{align}
where $\SetI = \{1,\,2, \cdots, K \}$ is the index set of $K$ tier networks;
$\Bias{i}$ is the weighting factor for the $i$th-tier network; 
$\rDis_{\ly,\lx}$ is the distance between nodes at $\ly$ and $\lx$; and 
$\PL{i}>2$ is the pathloss exponent in the $i$th-tier network. 
This association rule can be extended to special cases such as 
the rule that makes nodes associate to the nearest \ac{AP}, i.e., $\Bias{i}=1$, 
or to the \ac{AP} providing the maximum average received power, i.e., $\Bias{i} = \Pbs{i} \Biasfactor{i}$
where $\Biasfactor{i}$ is the association bias of the $i$th-tier network. 
\blue{Let us denote $\rDis_i^{m}$ as the distance to the $m$-mode \ac{AP} with 
the maximum $\Bias{i} \rDis_{\x_i, \lx_\text{o}}^{-\PL{i}}$ for all $\x_i  \in \PPPm{k}{i}$. % in the $i$th-tier network. 
Using the association rule and $\rDis_i^{m}$, 
the probability that a user is associated to an $m$-mode \ac{AP} in the $k$th-tier network is given by 
\begin{align}
	\Pa{k}{m}
	& =
		\PX{
			\bigcup_{i \in \SetI, m_\lo \in \{\FD, \HD \}}
			\!\!\!\!\!\!
			\Bias{k} \left({\rDis_k^{m}}\right)^{-\PL{k}}
			> \Bias{i} \left({\rDis_i^{m_\lo}}\right)^{-\PL{i}}
		}
	\nonumber \\
& \mathop = \limits^{\mathrm{(a)}} 
		2 \pi \LB{k}{m} 
		\int_0^\infty 
		\!\!\!
			x 
			\exp\left\{ 
					- \pi 
					\!\!\!\!\!\!
					\sum_{i \in \SetI, m_\lo \in \{\FD, \HD \} } 
					\!\!\!\!\!\!
					\LB{i}{m_\lo} \Bratio{ik}^{2/\PL{i}} 
					x^{2 \PL{k}/\PL{i}}
				\right\}
			dx
	\nonumber \\
	& =
		2 \pi \LB{k}{m} 
		\int_0^\infty 
			x 
			\exp\left\{ 
					- \pi \sum_{i \in \SetI} \LB{i}{} \Bratio{ik}^{2/\PL{i}} 
					x^{2 \PL{k}/\PL{i}}
				\right\}
			dx
%		\frac{
%			\LB{k}{m}
%		}{
%			\sum_{i \in \SetI} 
%			\LB{i}{} \Bratio{ik}^{2/\PL{i}}
%		}
\label{eq:Paa}	
\end{align}
where (a) is from Lemma~4 in \cite{SinDhiAnd:13}, and 
$\Bratio{ik} = \Bias{i}/\Bias{k}$ is the ratio between the association factor.} 
%The association at a node is independent to those at other nodes, so the distribution of nodes associated to a $m$-node \ac{AP} in the $i$th-tier network also follows a \ac{PPP} with the spatial density $\LU{k}{m}= \LU{k}{}\Pa{k}{m}$
% due to the thinning property\cite{Kin:B93}. 
%
\blue{Using \eqref{eq:Paa}, for a node associated to $m$-mode \ac{AP} in the $k$th-tier network, 
the \ac{PDF} of the link distance to the associated \ac{AP}, $\rDis_k^{m}$, is given by\cite{SinDhiAnd:13}}
%In addition, the \ac{PDF} of the link distance from a typical node to its associated $m$-mode \ac{AP} 
% in the $k$th-tier network is given by  \cite{SinDhiAnd:13}
%
%
\begin{align}
	f_{\rDis_k^{m}}(x)
	& =
		\frac{2 \pi \LB{k}{m} x}{\Pa{k}{m}}
		\exp
		\left\{ 
			- \pi \sum_{i \in \SetI} \LB{i}{} \Bratio{ik}^{2/\PL{i}} x^{2 \PL{k}/\PL{i}}
		\right\}\,
	 \nonumber\\
	& =
		\frac
		{
			x
			\exp \left\{ 
			- \pi \sum_{i \in \SetI} \LB{i}{} \Bratio{ik}^{2/\PL{i}} x^{2 \PL{k}/\PL{i}}
			\right\}
		}
		{	\int_0^\infty 
			y
			\exp\left\{ 
					- \pi \sum_{i \in \SetI} \LB{i}{} \Bratio{ik}^{2/\PL{i}} 
					y^{2 \PL{k}/\PL{i}}
				\right\}
			dy
		}\,.
		\label{eq:Pa}
\end{align}
Note that $\rDis_k^{m}$ depends not on the spatial density $\LB{k}{m}$, but on 
the sum of scaled spatial densities, i.e., 
$
\sum_{i \in \SetI} \LB{i}{} 
{(\Bias{i}/\Bias{k})}^{2/\PL{i}}
% \Bratio{ik}^{2/\PL{i}} 
x^{2 \PL{k}/\PL{i}}
$. 

%
%=============Beta ==========================================
\setcounter{eqnback}{\value{equation}}
\setcounter{equation}{12}
\begin{figure*}[t!]
%\centering \rule[0pt]{18cm}{0.3pt}
\begin{align}
& \Lap{\Int{i}{\FD}}{s}
= 
\nonumber\\
	& \left\{
		\begin{aligned}
		%\Lap{\Int{i}{\FD}}{s}
		& 
		\exp
		\left\{
			- \pi \LB{i}{\FD} 
			% \frac{2}{\PL{i}}
			\left[
			-	%\frac{\PL{i}}{2} 
				\hat{\rDis}^{2}
			+ 	%\frac{\PL{i}}{2}
				s^{2/\PL{i}} 
				%\EXs{\RV{G}_i}{\RV{G}_i^{2/\PL{i}}}
				\frac{2}{\PL{i} }
				\frac{1}{\Pus{i} - \Pbs{i}}
			\left\{	
				\pi \csc\left(\frac{2\pi}{\PL{i}} \right)
				% \GF{\frac{2}{\PL{i}}}
				% \GF{1-\frac{2}{\PL{i}}}
				%
				\left(
					\Pus{i}^{\frac{2}{\PL{i}} + 1}  - \Pbs{i}^{\frac{2}{\PL{i}} + 1}
				\right)
%				\frac
%				{
%					\Pus{i}^{\frac{2}{\PL{i}} + 1}  - \Pbs{i}^{\frac{2}{\PL{i}} + 1}
%				}
%				{
%					\Pus{i} - \Pbs{i}
%				}
		\right.\right.\right.
		% \label{eq:LapDiff}
		\\ 
		& \left.\left.\left.
		\quad\quad\quad\quad\quad\quad\quad\quad\quad\quad\quad\quad
			+ 	
%			\frac{2}{\PL{i}}
%			s^{2/\PL{i}}
%			\frac{1}{\Pus{i} - \Pbs{i}}
%			\left\{
			\FnI{1}{
				\frac{2}{\PL{i}}+1,
				\frac{1}{\Pus{i}},
				\frac{-2}{\PL{i}},
				\frac{s}{\hat{\rDis}^{\PL{i}}}
%				 \frac{\TSIR{k}{m}}{\Ptx \Bratio{ik}}
				 }
			- 
			\FnI{1}{
				\frac{2}{\PL{i}}+1,
				\frac{1}{\Pbs{i}},
				\frac{-2}{\PL{i}},
				\frac{s}{\hat{\rDis}^{\PL{i}}}
				 %\frac{\TSIR{k}{m}}{\Ptx \Bratio{ik}}
				}
			\right\} 
			%\EXs{\RV{G}_i}{\RV{G}_i^{2/\PL{i}} \GF{-\frac{2}{\PL{i}}, \frac{s\RV{G}_i}{\hat{\rDis}_{\x,i}^{\PL{i}}}}}
			\right]
		\right\},
		\,\, \text{if } \Pbs{i} \neq \Pus{i},
	\\
			& \exp
		\left\{
			- \pi \LB{i}{\FD} 
			% \frac{2}{\PL{i}}
			\!
			\left[
			-	%\frac{\PL{i}}{2} 
				\hat{\rDis}^{2}
			\! +\! 	%\frac{\PL{i}}{2}
				\frac{2 s^{2/\PL{i}} }{ \PL{i}  \Pbs{i}^2}
				%\EXs{\RV{G}_i}{\RV{G}_i^{2/\PL{i}}}
				% \frac{2}{\PL{i} }
				\left\{
					 \Pbs{i}^{\frac{2}{\PL{i}}+2} \!
					\left(1 \! +\! 	 \frac{2}{\PL{i}} \right) \!
					\pi \csc\left(\frac{2\pi}{\PL{i}} \right)
%		\right.\right.\right.
	%\label{eq:LapEqual}
%	\\
%	& \quad\quad\quad\quad\quad\quad\quad\quad\quad\quad
%	    \quad\quad\quad\quad\quad\quad
%	\left.\left.\left.
					\! +\! 	
					\FnI{1}{
						\frac{2}{\PL{i}}\! +\! 	2,
						\frac{1}{  \Pbs{i}},
						\frac{-2}{\PL{i}},
						\frac{s}{\hat{\rDis}^{\PL{i}} }
					}
				\right\}
			\right]
		\right\}
		,\, \text{otherwise}\,.
		\end{aligned}
	\right.
	 \label{eq:LapDiff}
\end{align}

\setcounter{eqncnt}{13}
\centering \rule[0pt]{18cm}{0.3pt}
\end{figure*}
\setcounter{equation}{\value{eqnback}}
%
%======================================================= 
%
%

All users and \acp{AP} have a single antenna,
%\footnote{\blue{Note that 
%our framework can be easily extended to the case that \acp{AP} 
%are equipped with multiple antennas, each of which is operating in orthogonal frequency channel.
%In this case, the \ac{HDHN} throughput will be 
%the sum of the throughput (derived in this paper) obtained at each antenna.}  
%} 
and 
they are transmitting and receiving at the same time in \ac{FD} mode\cite{CoxAck:13,Kno:12}. 
%It is assumed that each node has a single antenna, 
%so nodes or \acp{AP} in \ac{FD}-mode cell are transmitting and receiving data at the same time. 
A node in \ac{FD} mode receives self-interference from its transmitted signal,
and performs \ac{IC} for the self-interference. 
%and cancel the interference received from its transmitted signal, which is called as {\em self-interference}. 
%For reliable \ac{FD}-mode transmission, a node should be able to cancel its self-interference,
%of which amount depends on its own transmitting power\cite{CoxAck:13}.  
% and the amount of the self-interference after cancellation depends on the transmitting power at the receiver\cite{CoxAck:13}. 
%
Since the amount of self-interference
depends on the transmission power at the receiver $\Prx$\cite{CoxAck:13},   
we define the residual self-interference power after performing cancellation 
as\cite{RiiWerWic:11a,RiiWerWic:11,NgLoSch:12} 
% we define the self-\ac{IC} capability of nodes in the $k$th-tier network as % $\FnCk{\FD}{\Prx}{k}$,
%we present the amount of received self-interference 
% as a function of the transmitting power such as
%
%
\begin{align}\label{eq:SelfIC}	
	 \FnCk{m}{\Prx}{k}  = \Prx \ChG{\RI,k} % \sNorm{\RV{h}_{\text{s},k}}^2
	%0 \leq \FnCk{m}{\Prx}{k} \leq \Prx %, \,\, \forall k \in \SetI,\, \forall m\in\{\text{FD},\text{HD}\}
\end{align}
for $ \forall k \in \SetI$ and $\forall m\in\{\text{FD},\text{HD}\}$.  
% where $\Prx$ is the transmission power at the receiver.
% and $\RV{h}_{\text{L},k}$ is the self-interfering channel of  a node in the $k$th-tier network. 
\blue{Here, $\ChG{\RI,k} = \sNorm{\RV{h}_{\RI,k}}^2$ shows the {\em self-\ac{IC} capability} of nodes
where $\RV{h}_{\RI,k}$ is the residual self-interfering channel 
of a node in the $k$th-tier network. 
In \eqref{eq:SelfIC}, $\FnCk{\FD}{\Prx}{k}=0$ denotes perfect self-\ac{IC}, 
and $\FnCk{\HD}{\Prx}{k}=0$ since \ac{HD}-mode nodes are not transmitting while receiving data.
}

\blue{
The residual self-interfering channel gain $\ChG{\RI,k}$ in \eqref{eq:SelfIC} needs to be characterized 
according to cancellation algorithms. % \footnote{
% \blue{
	For instance, after a digital-domain cancellation,
	$\RV{h}_{\RI,k}$ can be presented as 
	$\RV{h}_{\RI,k}= \RV{h}_{\mathsf{S},k} - \hat{\RV{h}}_{\mathsf{S},k}$
	where $\RV{h}_{\mathsf{S},k}$ and $\hat{\RV{h}}_{\mathsf{S},k}$ are the 
	self-interfering channel and its estimate
	as the self-interference is subtracted using its estimate\cite{RiiWerWic:11,NgLoSch:12,KimJuParHon:13,CirRonHua:14}. 
	Then,  
	$\ChG{\RI,k}$ can be modeled as a constant value such as 
	$\ChG{\RI,k} = \sigma_\text{e}^2$
	for the estimation error variance $\sigma_\text{e}^2$\cite{NgLoSch:12,KimJuParHon:13,CirRonHua:14}.   
	%Our parameterization of the self-\ac{IC} capability in \eqref{eq:SelfIC}
	% can make the analysis more generic. 
	However, for other cancellation techniques such as 
	analog-domain schemes\cite{DuaSab:10,DuaDicSab:12,ChoJaiSriLevKat:10,JaiChoKimBhaSir:11,BhaMcmKat:13},  
	propagation-domain schemes\cite{SnoFulCha:11,LioVibColAth:10,RiiWerWic:11,LeeSimChaKan:14}, 
	and combined schemes of different domains\cite{RiiWic:12,HuaMaLiaCir:13}, 
	the modeling of $\ChG{\RI,k}$ is a still challenging problem. 
Hence, the parameterization of the self-\ac{IC} capability in \eqref{eq:SelfIC} can make the analysis more generic. 
We consider $\ChG{\RI,k}$ as a constant value in this paper,
but note that the analysis can be easily extended for the case of random $\ChG{\RI,k}$
within our framework.}\footnote{For instance, 
	 once the \ac{PDF} of $\ChG{\RI,k}$ is available for a certain self-\ac{IC} algorithm, 
	% although it can be difficult to present in close forms, 
	by averaging analytic results of the paper 
	over the distribution of $\ChG{\RI,k}$, 
	the results for the random $\ChG{\RI,k}$ can be obtained. }

%  our framework and analytical result can be easily extended to the case of random $\ChG{\RI,k}$. 
% 
%We used the constant value for $\ChG{\RI,k}$, but our framework and analytical result can be easily extended to the case of % random $\ChG{\RI,k}$. 

% can be modelled as a constant value or random variables, 
% but we use the constant parameter 

%  in the $k$th-tier network. 
%the amount of the remained self-interference after the cancellation, 
% where $\Prx$ is the transmission power at the receiver. 
% In \eqref{eq:SelfIC}, $\FnCk{\FD}{\Prx}{k}=0$ and $\FnCk{\FD}{\Prx}{k}=\Prx$ denote perfect self-\ac{IC}, and $\FnCk{\HD}{\Prx}{k}=0$ since nodes are not transmitting while receiving in \ac{HD} mode.
%  and no self-\ac{IC}, respectively. 
% \mynote{Will add some examples of $\FnC{\FD}{\Prx}$ if any.} 
% Note that $\FnCk{\HD}{\Prx}{k}=0$ since nodes are not transmitting while receiving in \ac{HD} mode.
% \mynote{Add why self interference is static.}

\subsection{Network Interference Characterization}

\blue{
In the $k$th-tier \ac{HDHN}, 
the \ac{SIR} received by a node at $\lx_\lo$ from a transmitter at $\ly_\lo$
for a propagation channel model with pathloss and Rayleigh fading 
is defined as
\begin{align}\label{eq:SIR}	
	\SIR{k}{m}
	=
		\frac{
			\Ptx \ChG{\lx_\lo,\ly_\lo} \rDis_{\lx_\lo,\ly_\lo}^{-\PL{k}}
		}{
			\FnCk{m}{\Prx}{k} 
			+ \sum_{i \in \SetI}
				\left( \Int{i}{\HD} + \Int{\text{o},i}{\FD} \right)
		}
\end{align}
where $\Ptx$ is the transmission power at the transmitter,
$\Prx$ is the transmission power at the receiver, and 
$\ChG{\lx_\lo, \ly_\lo}$ is the i.i.d. fading channel gain of the link, i.e., $\ChG{\lx_\lo, \ly_\lo} \sim \exp(1)$. 
In~\eqref{eq:SIR}, when a user at $\lx$ associates to an \ac{AP} at $\ly$, 
$\Int{i}{\HD}$ and $\Int{\text{o},i}{\FD}$ are the aggregate interference received 
from \ac{HD}-mode cells and \ac{FD}-mode cells in the $i$th-tier network, given by 
\begin{align}	
	\Int{i}{\HD} 
	& =\!\!
		\sum_{\z \in \PPPm{i}{\HD} / \{ \ly\}}\!\!
		\Pbs{i} \ChG{\lx, \z} \rDis_{\lx, \z}^{- \PL{i}}
		% \vNorm{{\RV{Z}}}^{-\PL{i}}
		% \rDis_{\RV{Z}}^{-\PL{i}}
%		+
%		\sum_{\RV{Z} \in \PPPm{i}{\HD}}
%		\Pbs{i} \ChG{\RV{Z}} \rDis_{\RV{Z}}^{-\PL{i}}
	\label{eq:HDint} \\
	\Int{\text{o},i}{\FD} 
	& =\!\!	
		\sum_{\z \in \PPPm{i}{\FD}/ \{ \ly\}}\!\!
		\Pbs{i} \ChG{\lx,\z} \rDis_{\lx,\z}^{-\PL{i}}
		\!+\!
		\Pus{i} \ChG{\lx, \z+N(\z)} \rDis_{\lx, \z + N(\z)}^{-\PL{i}}	
	\label{eq:FDint}
\end{align}
where 
$N(\lz)$ is the relative location of a user to its associated \ac{AP} at $\lz$. }
% communicating in \ac{HD} mode. 
%$\rDis_{\x, \y} = \vNorm{\x - \y}$ is the distance between the transmitter and receiver; and
% $\ChG{\x, \y}$ is the fading channel gain of the link. 
Note that in \eqref{eq:FDint}, interference from a \ac{FD}-mode cell 
consists of the interference from an \ac{AP} and a user.  
% since there is only one transmitting node using the same resource with a node at $\x$. 
%
%

\blue{Let us consider a user at $\lx$, its associating \ac{AP} at $\ly$, the other \acp{AP} at $\lz \in \PPPm{i}{\FD}/\{ \ly\}$,
and their associated users at $\lz+N(\lz)$. 
Generally, the distance between $\lx$ and $\lz$ is greater than
the distance between $\lz$ and $\lz+N(\lz)$, i.e., $\vNorm{\lx-\lz} \gg \vNorm{N(\lz)}$. 
}
% Generally, the distance between a user at $\lx$ and an unassociated \ac{AP} at $\lz$ is greater than
% the distance between the \ac{AP} and its associating user at $\lz+N(\lz)$, i.e., $\vNorm{\lx-\lz} \gg \vNorm{N(\lz)}$. 
%
Hence, due to the difficulty in obtaining the exact characteristics of $\Int{i}{\FD}$, 
we assume that the distance between a user at $\lx$ and a user at $\lz+N(\lz)$ 
can be approximated 
to the distance between a user at $\lx$ and the unassociated \ac{AP} at $\lz$ as\footnote{\blue{From 
	the law of cosines, we have $\vNorm{\lx- \left(\lz +N(\lz) \right)}^2 \approx \vNorm{\lx-\lz}^2 $ for $\vNorm{\lx-\lz} \gg \vNorm{N(\lz)}$. 
%	we have 
%	%
%	%
%	\begin{align}\nonumber	
%		\vNorm{\lx- \left(\lz +N(\lz) \right)}^2 = \vNorm{\lx-\lz}^2 + \vNorm{N(\lz)}^2 - (\lx-\lz) \circ N(\lz)
%	\end{align}
%	%
%	where `$\circ$' means the inner product, and here, $\vNorm{\lx- \left(\lz +N(\lz) \right)}^2 $ 
%	approximates to $\vNorm{\lx-\lz}^2 $ for $\vNorm{\lx-\lz} \gg \vNorm{N(\lz)}$. 
	}}
\begin{align}\label{eq:approximation}	
	% \vNorm{\lx - (\lz+N(\lz))} \approx \vNorm{ \lx - \lz}\,. 
	\rDis_{\lx,\lz+N(\lz)} \approx \rDis_{\lx, \lz}\,.
\end{align}
Using the approximation in \eqref{eq:approximation}, 
the interference received from \ac{FD}-mode cells can be presented as
\begin{align}\label{eq:IntFD_app}	
	\Int{i}{\FD} 
	& =	
		\sum_{\z \in \PPPm{i}{\FD}/ \{ \ly\}}
		\RV{G}_{i} \rDis_{\lx,\z}^{-\PL{i}}
\end{align}
%\Pwr_{\text{o},i}^{-1}
%
where $\RV{G}_{i}$ is given by
\begin{align} \label{eq:GG}	
	\RV{G}_{i} = \Pbs{i} \ChG{\lx, \z} + \Pus{i} \ChG{\lx, \z+N(\lz)}\,.
\end{align}
%
%
% In \eqref{eq:GG}, 
% if $\Pbs{i} = \Pus{i}$, $\RV{G}_{i} $ follows the Erlang distribution with a rate $ \Pbs{i}^{-1}$, 
% otherwise, it is a hypo-exponential random variable with rates ${\Pbs{i}}^{-1}$ and ${\Pus{i}}^{-1}$. 
\blue{In \eqref{eq:GG}, 
if $\Pbs{i} = \Pus{i}$, $\RV{G}_{i} $ is the sum of two exponential random variables all with same rate $ \Pbs{i}^{-1}$ 
and it follows the Erlang distribution\cite{BolGreMeeTri:06}. % with a rate $ \Pbs{i}^{-1}$.  
On the other hand, if $\Pbs{i} \neq \Pus{i}$, 
$\RV{G}_{i}$ is the sum of two exponential random variables with different rates ${\Pbs{i}}^{-1}$ and ${\Pus{i}}^{-1}$ and  
it follows the hypo-exponential distribution\cite{BolGreMeeTri:06}. }% with rates ${\Pbs{i}}^{-1}$ and ${\Pus{i}}^{-1}$. 
Hence, the \ac{PDF} of $\RV{G}_{i}$ is given by\cite{BolGreMeeTri:06}
\begin{align}\label{eq:HypoPDF}	
	\PDF{\RV{G}_{i}}{x}
	=
	\left\{
		\begin{aligned}
		& 
		\frac{1}{  \Pbs{i}^2}  x e^{-x/  \Pbs{i}}, % \quad\quad\quad\quad 
%		\left(\frac{1}{  \Pbs{i}} \right)^2 x e^{-x/  \Pbs{i}}, % \quad\quad\quad\quad 
		\, \text{if } \Pbs{i} = \Pus{i}, 
		\\
		& \frac{e^{- x/\Pus{i}} - e^{- x/\Pbs{i}}}{\Pus{i} - \Pbs{i}}, \, \text{otherwise}
		\end{aligned}
	\right.
\end{align}
and we have\footnote{\blue{From \eqref{eq:HypoPDF}, 
	$\EXs{\RV{G}_{i}}{\RV{G}_{i}^\delta} = \int_0^{\infty} x^\delta \PDF{\RV{G}_{i}}{x} dx$
	can be obtained
	by substituting $x/  \Pbs{i}$ (or $x/  \Pbs{i}$) to $y$ and using the Gamma function $\GF{t} = \int_0^{\infty} y^{t-1} e^{-y} dy$.}}  
% its $n$th moment is given by 
%
%
\begin{align}\label{eq:HypoMoment}	
	\EXs{\RV{G}_{i}}{\RV{G}_{i}^\delta}
	=
	\left\{
		\begin{aligned}
		&  \Pbs{i}^{\delta} \GF{2+ \delta},% \quad\quad\quad\quad\quad\quad 
		\quad \, \text{if } \Pbs{i} = \Pus{i},
		\\
		&\frac
		 {
			\GF{1+\delta} \left( \Pus{i}^{\delta + 1}  - \Pbs{i}^{\delta + 1} \right)
		}
		{
			\Pus{i} - \Pbs{i}
		}, %\quad 
		\, \text{otherwise}
		\end{aligned}
	\right.
\end{align}
for $\delta > -1$ and the gamma function $\GF{\cdot}$. 
With the approximation in \eqref{eq:approximation}, we now obtain the Laplace transform of $\Int{i}{\FD}$ as follows. 
%
%\mynote{Will add a simulation result to show the effect of this assumption}

%\/\/\/\/\/\/\/\/\/\/\/\/\/\/\/\/\/\/\/\/\/\/\/\/\/\/\/\/\/\/\/\/\/\/\/\/\/\/\/\/\/\/\/\/\/\/\/\/\/\/\/\/\/\/\/\/\/\/\/\/\/\/\/\/\/\/\/\/\/\/\/\/\/\/\/\/\/\/\/\/\/\/\/\/\/\/\/
\begin{figure}[t!]
    \begin{center}   
    { 
	\psfrag{Simulation}[Bl][Bl][0.59]   {Simulation}
	\psfrag{SimulationAssumptionAssump}[Bl][Bl][0.59]   {Simulation w/ approx. in \eqref{eq:approximation}}
	\psfrag{ApproxOneTheory}[Bl][Bl][0.59]   {Analysis w/ approx. in \eqref{eq:approximation}}
	%\psfrag{RatioThirtyZFive}[Bl][Bl][0.59]   {$\RatioE_1  = 30$, $\tslot =5$}
	%
%	\psfrag{LIFiveFour}[Bl][Bl][0.7]   {$\LB{i}{\FD} = 5\cdot 10^{-4}$}
%	\psfrag{LIThreeFour}[Bl][Bl][0.7]   {$\LB{i}{\FD}  = 1\cdot 10^{-3}$}
	\psfrag{DFifty}[Bl][Bl][0.7]   {$\LB{i}{\FD} \!=\! 10^{-3}$, $\hat{\rDis}\!=\!50$}
	\psfrag{DThirty}[Bl][Bl][0.7]   {$\LB{i}{\FD} \!=\! 10^{-3}$, $\hat{\rDis}\!=\!30$}
	\psfrag{DOne}[Bl][Bl][0.7]   {$\LB{i}{\FD} \!=\! 10^{-3}$, $\hat{\rDis}\!=\!10$}
	\psfrag{DThirty2303}[Bl][Bl][0.7]   {$\LB{i}{\FD} \!=\! 2 \!\cdot\!10^{-3}$, $\hat{\rDis}\!=\!30$}
	\psfrag{SS}[tc][bc][0.7] {$s$}
	\psfrag{Laplace}[bc][tc][0.7] {$\Lap{\Int{i}{\FD}}{s}$}
	 \includegraphics[width=1.00\columnwidth]{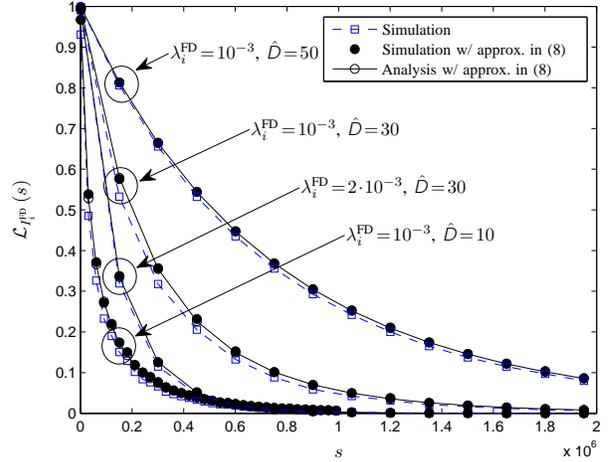}
    }
    \end{center}
    \caption{
    		An example of Laplace transform of network interference from \ac{FD}-mode cells
		in the $i$th-tier network. 
		% where $\LT$ is the density of 		
		 }
   \label{fig:Laplace}
\end{figure}
%\/\/\/\/\/\/\/\/\/\/\/\/\/\/\/\/\/\/\/\/\/\/\/\/\/\/\/\/\/\/\/\/\/\/\/\/\/\/\/\/\/\/\/\/\/\/\/\/\/\/\/\/\/\/\/\/\/\/\/\/\/\/\/\/\/\/\/\/\/\/\/\/\/\/\/\/\/\/\/\/\/\/\/\/\/\/\/

%
%
\begin{lemma}\label{lem:Laplace}
The Laplace transform of the {approximated} interference received from \ac{FD}-mode cells in the $i$th-tier network, $\Int{i}{\FD}$ in \eqref{eq:IntFD_app}, 
is given by \eqref{eq:LapDiff} ({on top of the page}) 
where $\hat{\rDis}$ is the minimum distance to an interfering node and $\FnI{1}{x, y, z, \nu}$ is defined by\setcounter{equation}{13}
\begin{align}	
	& \FnI{1}{x, y, z, \nu}
	= 
		\int_{0}^{\infty}
			t^{x-1}e^{-y t} \GF{z, \nu t} dt
	\nonumber 
	 \\
	& \quad\quad\quad
	=
		\frac{
			\nu^z \GF{x+z}
		}{
			x(y+\nu)^{x+z}
		}
		\HGF{2}{1}{1, x+z; x+1; \frac{y}{y+\nu}}\,
	\label{eq:Integral1} 
\end{align}
for all constants $\nu+y>0$, $y>0$, and $x+z>0$, $\GF{\cdot, \cdot}$ is the upper incomplete function, and $\HGF{2}{1}{\cdot, \cdot; \cdot; \cdot}$ is the hypergeometric function.
\end{lemma}
\begin{IEEEproof}
See Appendix~\ref{app:lem1}. 
\end{IEEEproof}
%
%
%An example of the Laplace transform of $\Int{i}{\FD}$ is presented in Fig.~\ref{fig:Laplace} 
%for the cases with and without the approximation in \eqref{eq:approximation},
%which shows a good match between two cases. 
%This Laplace transform will be used to obtain the successful transmission probability of \acp{HDHN}
%in the following section. 

%
%=============Beta ==========================================
\setcounter{eqnback}{\value{equation}}
\setcounter{equation}{16}
\begin{figure*}[t!]
%\centering \rule[0pt]{18cm}{0.3pt}
\begin{align} \label{eq:term11} 
	\FnMB{ik}{m}{\PL{}, \Ptx} 	
	= 
%	\\
%	&\quad  
	\left\{
		\begin{aligned}
			& % 2\pi \LB{i}{}
					%\left\{
%						\frac{ \Bratio{ik}^{2/\PL{}}}{2}
%					+
%						%\LB{i}{\HD} 
%						(1-\Pfd{i})
%						\FnI{0}{1, \frac{\Ptx}{\Pbs{i} \TSIR{k}{m}}, \Bratio{ik}, \PL{}}
%					+
%						\Pfd{i}
%						\left\{
							\frac{-\Bratio{ik}^{2/\PL{}}}{2}
							+\frac{\Ptx^{-2/\PL{}} \TSIR{k}{m}^{2/\PL{}}  }{ \PL{} \left( \Pus{i} - \Pbs{i} \right)}
%						\right.
					%\right.
		%\right.
%		 	\\ 
%			&\quad\quad\quad	
%				\times
							\left[ \pi \csc\left( \frac{2\pi}{\PL{}}\right)
							\right.
							\left( \Pus{i}^{2/\PL{}+1} -  \Pbs{i}^{2/\PL{}+1} \right)
			\\
			& \quad \quad \quad\quad\quad	\quad\quad	
				 % \left. 
				 \left. 
							+ \FnI{1}{\frac{2}{\PL{}}+1, \frac{1}{\Pus{i}}, \frac{-2}{\PL{}}, \frac{\TSIR{k}{m}}{\Ptx \Bratio{ik}} }
							- \FnI{1}{\frac{2}{\PL{}}+1, \frac{1}{\Pbs{i}}, \frac{-2}{\PL{}}, \frac{\TSIR{k}{m}}{\Ptx \Bratio{ik}} }
						\right],
					% \right\}\,,
					% \quad \quad\quad\quad\quad \quad\quad\quad \quad\quad\quad 
					\, \text{if } \Pus{i}\neq  \Pbs{i}, 
			%\right\}
		\\
			& %2\pi \LB{i}{}
				%	\left\{
%						\frac{ \Bratio{ik}^{2/\PL{}}}{2}
%					+
%						%\LB{i}{\HD} 
%						(1-\Pfd{i})
%						\FnI{0}{1, \frac{\Ptx}{\Pbs{i} \TSIR{k}{m}}, \Bratio{ik}, \PL{}}
%					+
%						\Pfd{i}
%						\left\{
							\frac{-\Bratio{ik}^{2/\PL{}}}{2}
							\!+\! 
							\frac{\Ptx^{-2/\PL{}} \TSIR{k}{m}^{2/\PL{}}  }{ \PL{}  \Pbs{i}^{2}} \!
							\left[ 
								\Pbs{i}^{2/\PL{}+2} 
								\left(1\!+\! \frac{2}{\PL{}}\right) \pi \csc\left( \frac{2\pi}{\PL{}}\right)
%							\right.	
%		 	\\ 
%			&\quad\quad\quad\quad\quad\quad\quad\quad\quad\quad\quad\quad	
%				 \left. 	
							\!+\!  \FnI{1}{\frac{2}{\PL{}}\!+\! 2, \frac{1}{\Pbs{i}}, \frac{-2}{\PL{}}, \frac{\TSIR{k}{m}}{\Ptx \Bratio{ik}} }
						\right]\!
%					\right\}\,,
			%\quad 
			, \,\, \text{otherwise}\,. 
			%\right\}
		\end{aligned}
	\right.
	% \nonumber
\end{align}

\setcounter{eqncnt}{\value{equation}}
\centering \rule[0pt]{18cm}{0.3pt}
\end{figure*}
\setcounter{equation}{\value{eqnback}}
%
%======================================================= 

An example of the Laplace transform of $\Int{i}{\FD}$ is presented in Fig.~\ref{fig:Laplace} 
when the association policy in \eqref{eq:association} is applied. 
For other parameters, the values presented in Table~\ref{table:parameter} are used.  
Fig.~\ref{fig:Laplace} shows a good match between the cases with and without the approximation in \eqref{eq:approximation},
especially for dense networks, i.e., large $\LB{i}{\FD}$.\footnote{Note that $\Lap{\Int{i}{\FD}}{s}$ 
	can be extended to the case with transmission power control 
	for $\Pus{i}$ or $\Pbs{i}$ such as \cite{NovDhiAnd:13} 
	by taking expectation to the exponential exponent in \eqref{eq:Pro4}
	according to the power distribution.
%	For instance, 
%	if a power control for users is applied such as \cite{NovDhiAnd:13}, 
%	$\Int{i}{\FD}$ can be obtained from \eqref{eq:Pro4} by taking expectation 
%	according to the distribution of $\Pus{i}$ in both \eqref{eq:ProG2} and \eqref{eq:HypoMoment}. 
	%This gives in non-tractable analysis results, so this paper does not consider the transmission power control 
	%to provide more insights on the full-duplex network interference. 
}

\blue{Note that the Laplace transform of interference from \ac{FD}-mode cells in Lemma~\ref{lem:Laplace}  
and the following analytical results related to \ac{FD} mode 
can also be used 
when each \ac{FD}-mode node has two antennas, one for transmitting and the other for receiving. 
In this case, the self-\ac{IC} capability will be determined differently to the case of single antenna.}

%%---------------------------------------------------------------------------%
%%                       Sec: Successful Trans. Prob.                     %
%%---------------------------------------------------------------------------%
\section{Hybrid-Duplex Heterogeneous Network Throughput}\label{sec:throughput}

In this section, we analyze the successful transmission probability of \acp{HDHN}, and 
define and derive the \ac{HDHN} throughput as a new performance measurement for \acp{HDHN}. 

\subsection{Successful Transmission Probability}\label{sec:STP}

In this subsection, we analyze the successful transmission probability of \acp{HDHN}. 
% 
% The achievable communication throughput in \acp{HDHN} is affected by 
% the spatial node distribution and the self-\ac{IC} capability. 
%
We present the successful transmission probability of a $m$-mode node in the $k$th-tier network 
as $\Ps{k}{m}{\TSIR{k}{m}} = \PX{\SIR{k}{m} \ge \TSIR{k}{m}}$, 
where $\TSIR{k}{m}$ is the target \ac{SIR} value. % of a $m$-mode node in the $k$th-tier network. 
Users and \acp{AP} may have different target data rates such as
$\tRate{\US}$ and $\tRate{\BS}$, respectively. In this case, 
the target \acp{SIR} of user and \ac{AP} can be set to 
$\tSIR{\US} = 2^{ \tRate{\US}/\Ws{k}{m}}-1$ and $\tSIR{\BS} = 2^{ \tRate{\BS}/\Ws{k}{m}}-1$, respectively, 
where $\Ws{k}{m}$ is 
a communication bandwidth. 
%
%This metric also presents the portion of nodes 
%satisfying a certain value of target data rate $\TRate$
%%, i.e., $\Ps{k}{m}{\TSIR{k}{m}} = \PX{\Rate{k}{m} \ge \TRate}$ 
%when $\TSIR{k}{m}$ is set to guarantee $\TRate$ such as  
%$\TSIR{k}{m} =  2^{ \TRate/(\Ts\Ws{k}{m})}-1$ 
%for a communication bandwidth $\Ws{k}{m}$
%and a symbol time $\Ts$ 
%%and the achievable data rate $\Rate{k}{m}$ 
%of a $m$-mode node in the $k$th-tier network.
The $\Ps{k}{m}{\TSIR{k}{m}}$ is derived as follows. 
\begin{theorem}\label{trm:STP}
In \acp{HDHN}, 
the successful transmission probability % (i.e., \ac{CCDF} of \ac{SIR}) 
of a $m$-mode node ($m \in \{\HD, \FD\}$) in the $k$th-tier network
is given by
\begin{align}		
	& \Ps{k}{m}{\Ptx, \Prx, \TSIR{k}{m}}
	=
		% \frac{2 \pi \LB{k}{m}}{\Pa{k}{m}}
		2 \pi \sum_{t \in \SetI} \LB{t}{} \Bratio{tk}^{2/\PL{t}} \!\!
	\nonumber\\
		&\!\! \times \!\!
		\int_{0}^\infty
		%\left\{
			 \!\!\!
			r
			\exp	
			\left\{
				\!- \frac{ r^{\PL{k}} \FnCk{m}{\Prx}{k} \TSIR{k}{m} }{ \Ptx }
				\!-\!
					\sum_{i \in \SetI} 
					r^{2\PL{k}/\PL{i}}
					2\pi \LB{i}{}
					\FnM{ik}{m}{\PL{i}, \Ptx}
			\right\}
		% f_{\rDis_k^{m}}(x)
		\!
		dr
	\label{eq:STP} 
\end{align}
where $\Ptx$ and $\Prx$ are the transmission power of the transmitter and the receiver, respectively. 
In \eqref{eq:STP}, $\FnM{ik}{m}{\PL{}, \Ptx}$ is given by 
\begin{align}\label{eq:term1} 	
	\FnM{ik}{m}{\PL{}, \Ptx}
	\! =\!  
		\frac{ \Bratio{ik}^{2/\PL{}}}{2}
		\! + \! 
		(1-\Pfd{i}) \FnMA{ik}{m}{\PL{}, \Ptx}
		\! + \! 
		\Pfd{i} \FnMB{ik}{m}{\PL{}, \Ptx}
\end{align}
where $\FnMB{ik}{m}{\PL{}, \Ptx}$ is given by \eqref{eq:term11} (on top of the page) and 
$\FnMA{ik}{m}{\PL{}, \Ptx}$ is defined as
\setcounter{equation}{17}
% and $\FnMA{ik}{m}{\PL{}, \Ptx} $ and $\FnMB{ik}{m}{\PL{}, \Ptx}$ are defined by 
%
\begin{align}\label{eq:term22} 
	\FnMA{ik}{m}{\PL{}, \Ptx} 
	& =
		\FnI{0}{1, \frac{\Ptx}{\Pbs{i} \TSIR{k}{m}}, \Bratio{ik}, \PL{}}\,.
\end{align}
Here, $\FnI{1}{x, y, z, \nu}$ is defined in \eqref{eq:Integral1}  and $\FnI{0}{x, y, z, \nu}$ is defined as  
\begin{align}	
	\FnI{0}{x, y, z, \nu}
	&  =
		\int_{(x z)^{1/\nu}}^{\infty}
		\frac{t}{x^{2/\nu} (1+y/x t^{\nu})} dt
	\nonumber\\
	& %\quad\quad%\quad
	= 
		\frac{z^{2/\nu-1}}{(\nu - 2)y}
	   	\HGF{2}{1}{1, 1-\frac{2}{\nu}; 2 - \frac{2}{\nu}; \frac{-1}{zy}}
		% , \quad \forall \nu >2\,.
	\label{eq:Integral0}  
\end{align}
for all constants $\nu >2$ and $x,y,z>0$. 
\end{theorem}
\begin{IEEEproof}
See Appendix~\ref{app:trm1}. 
\end{IEEEproof}
The successful transmission probability $\Ps{k}{m}{\Ptx, \Prx, \TSIR{k}{m}}$ in \eqref{eq:STP} 
also presents the portion of $m$-mode nodes or \acp{AP}
satisfying %a certain value of 
target data rate in the $k$th-tier network. 
%, i.e., $\Ps{k}{m}{\TSIR{k}{m}} = \PX{\Rate{k}{m} \ge \TRate}$ 
%when $\TSIR{k}{m}$ is set to guarantee $\TRate$ such as  
%$\TSIR{k}{m} =  2^{ \TRate/(\Ts\Ws{k}{m})}-1$ 
%for a communication bandwidth $\Ws{k}{m}$
%and a symbol time $\Ts$ 
%%and the achievable data rate $\Rate{k}{m}$ 
%of a $m$-mode node in the $k$th-tier network.
%
\blue{The $\Ps{k}{m}{\Ptx, \Prx, \TSIR{k}{m}}$ in \eqref{eq:STP} 
is difficult to be presented in a closed form, 
except for the cases of 
$\PL{i}= 4, \, \forall i \in \SetI$ or $\FnCk{\FD}{\Prx}{k} = 0$.
The successful transmission probabilities are provided in closed forms 
for those special cases in the following corollaries.  }
%
%Due to the difficulty in deriving \eqref{eq:STP} for a general case, 
% we present the successful transmission probabilities for special cases in the following corollaries. 
%  in close-forms for special cases. % as follows. 
%
\begin{corollary}% [Special case for a \ac{FD}-mode cell]
For $\PL{i}= 4, \, \forall i \in \SetI$, 
the successful transmission probability of a \ac{FD}-mode node in the $k$th-tier network
$\Ps{k}{\FD}{\Ptx, \Prx, \TSIR{k}{\FD}}$ is given by 
\begin{align}	
	& \Ps{k}{\FD}{\Ptx, \Prx, \TSIR{k}{\FD}}
	=
		%\frac{2 \pi \LB{k}{\FD}}{\Pa{k}{\FD}}
		% \sqrt{\pi} 
		\frac{
			\pi^{3/2} \sqrt{\Ptx} \sum_{t \in \SetI} \LB{t}{} \Bratio{tk}^{1/2}
		}{
			2 \sqrt{ \FnCk{\FD}{\Prx}{k} \TSIR{k}{\FD}}
		}
		% 2 \pi^{3/2} \sum_{t \in \SetI} \LB{t}{} \Bratio{tk}^{1/2}
		%2 \pi^{3/2} \left( \sum_{i \in \SetI} \LB{i}{} \Bratio{ik}^{2/\PL{}}\right)
	\nonumber \\
	& \quad \quad\quad \quad
	\times	\exp\left\{\left( 
				\frac{ 
					\pi \sqrt{ \Ptx} \sum_{i \in \SetI} \LB{i}{} \FnM{ik}{\FD}{4, \Ptx}
				}{ 
					\sqrt{\FnCk{\FD}{\Prx}{k} \TSIR{k}{\FD} }
				} 
			\right)^2\right\} 
		 \nonumber \\
	& \quad \quad \quad \quad% \quad \quad \quad \quad \quad \quad \quad \quad \quad \quad \quad \quad \quad \quad \quad 
	\times
		\Erfc{				
			\frac{ 
					\pi \sqrt{ \Ptx} \sum_{i \in \SetI} \LB{i}{} \FnM{ik}{\FD}{4, \Ptx}
				}{ 
					\sqrt{\FnCk{\FD}{\Prx}{k} \TSIR{k}{\FD} }
				} 
			}  
	\label{eq:STPb}
\end{align}
where $\FnM{ik}{m}{\PL{}, \Ptx}$ is defined in \eqref{eq:term1} and 
$\Erfc{x} = \frac{2}{\sqrt{\pi}} \int_x^\infty e^{-t^2} dt$ is the complementary error function.
\end{corollary}
\begin{IEEEproof}
When $\PL{i}=\PL{}, \, \forall i \in \SetI$, we can present $\Ps{k}{m}{\Ptx, \Prx, \TSIR{k}{m}}$ in \eqref{eq:STP} as 
\begin{align}		
	& \Ps{k}{m}{\Ptx, \Prx,\TSIR{k}{m}}
	\! =\!
%	\nonumber \\
%		%\frac{2 \pi \LB{k}{m}}{\Pa{k}{m}}
%	&	
	2 \pi \sum_{t \in \SetI} \LB{t}{} \Bratio{tk}^{2/\PL{}} \!\!
		\int_{0}^\infty  \!\! \!\!
		%\left\{
			r
			\exp	
			\left\{
				- r^{\PL{}} \frac{\FnCk{m}{\Prx}{k} \TSIR{k}{m} }{ \Ptx } 
			\right.
	\nonumber \\		
	& \left.
	\quad\quad\quad\quad\quad\quad\quad% \quad
				-
					r^{2}
					\sum_{i \in \SetI} 
					2\pi \LB{i}{}
					\FnM{ik}{m}{\PL{}, \Ptx}
			\right\}
		% f_{\rDis_k^{m}}(x)
		dr\,. 
		\label{eq:STPs}
\end{align}
\blue{For $\PL{}=4$, by substituting $r^2$ with $t$ in \eqref{eq:STPs}, we have
\begin{align}		
	& \Ps{k}{m}{\Ptx, \Prx,\TSIR{k}{m}}
	\! =\!
%	\nonumber \\
%		%\frac{2 \pi \LB{k}{m}}{\Pa{k}{m}}
%	&	
	\pi \sum_{t \in \SetI} \LB{t}{} \Bratio{tk}^{2/\PL{}} \!\!
		\int_{0}^\infty  \!\! \!\!
		%\left\{
			% t
			\exp	
			\left\{
				- t^2 
				\frac{\FnCk{m}{\Prx}{k} \TSIR{k}{m} }{ \Ptx } 
			\right.
	\nonumber \\		
	& \left.
	\quad\quad\quad\quad\quad\quad\quad% \quad
				-
					t
					\sum_{i \in \SetI} 
					2\pi \LB{i}{}
					\FnM{ik}{m}{\PL{}, \Ptx}
			\right\}
		% f_{\rDis_k^{m}}(x)
		dt\,. 
		\label{eq:STPs4}	
\end{align}
which results in \eqref{eq:STPb} by \cite[eq. (3.322)]{GraRyz:B07}.  }
\end{IEEEproof}

\begin{corollary}% [Special case for a \ac{FD}-mode cell]
For $\PL{i}=\PL{} >2, \,\forall i \in \SetI$, 
when the self-\ac{IC} is perfect, i.e., $\FnCk{\FD}{\Prx}{k} = 0$, 
the successful transmission probability of a FD-mode node in the $k$th-tier network
$\Ps{k}{\FD}{\Ptx, \Prx, \TSIR{k}{\FD}}$ is given by 
\begin{align}\label{eq:STPsFD}	
	\Ps{k}{\FD}{\Ptx, \Prx, \TSIR{k}{\FD}}
	=
		\frac{
			\sum_{i \in \SetI} \LB{i}{} \Bratio{ik}^{2/\PL{}}
		}{ 
			2 \sum_{i \in \SetI} \LB{i}{}\FnM{ik}{\FD}{\PL{}, \Ptx}
		}
%		\frac{\pi \LB{k}{\FD}}{\Pa{k}{\FD} \sum_{i \in \SetI} 2\pi \LB{i}{}\FnM{ik}{\FD}{\PL{}}}\,
\end{align}
where $\FnM{ik}{m}{\PL{}, \Ptx}$ is defined in \eqref{eq:term1}.
\end{corollary}
\begin{IEEEproof}
When $\FnCk{\FD}{\Prx}{k} = 0$ and $\PL{i}=\PL{}, \, \forall i \in \SetI$, from \eqref{eq:STPs}, we have 
\begin{align}		
	& \Ps{k}{\FD}{\Ptx, \Prx, \TSIR{k}{\FD}}
	= 2 \pi \sum_{i \in \SetI} \LB{i}{} \Bratio{ik}^{2/\PL{}}
	\nonumber\\
	& \quad	\quad\quad\quad
	%2 \pi \sum_{i \in \SetI} \LB{i}{} \Bratio{ik}^{2/\PL{}}
		%\frac{2 \pi \LB{k}{\FD}}{\Pa{k}{\FD}}
		\times
		\int_{0}^\infty
		%\left\{
			r
			\exp	
			\left\{
				% - x^{\PL{}} \frac{\FnC{m}{\Prx} \TSIR{k} }{ \Ptx } 
				-
					r^{2}
					\sum_{i \in \SetI} 
					2\pi \LB{i}{}
					\FnM{ik}{\FD}{\PL{}, \Ptx}
			\right\}
		% f_{\rDis_k^{m}}(x)
		dr
		\label{eq:STPsFD2} 
\end{align}
%for $\PL{i}=\PL{}, \, \forall i \in \SetI$, 
which results in \eqref{eq:STPsFD}.
\end{IEEEproof}

% \mynote{when the perfect self-interference cancellation is possible}
% \begin{corollary}% [Special case for a \ac{HD}-mode cell]
\blue{In \ac{HD} mode, there is no self-interference (i.e., $\FnCk{\HD}{\Prx}{k}=0$), 
so for $\PL{i}=\alpha> 2, \forall i \in \SetI$, 
the successful transmission probability in $k$th-tier network can also be obtained 
from \eqref{eq:STPsFD2} as %, and it is given by 
% we can readily obtain the successful transmission probability in $k$th-tier network 
% using \eqref{eq:STPsFD} as 
%$\Ps{k}{\HD}{\Pbs{k}, 0, \TSIR{k}{\HD}}$ 
%
%
\begin{align}\label{eq:STPsHD}	
	\Ps{k}{\HD}{\Pbs{k}, 0, \TSIR{k}{\HD}}
	= 
		\Ps{k}{\FD}{\Pbs{k}, 0, \TSIR{k}{\FD}}\,
\end{align}
where $\Ps{k}{\FD}{\Ptx, \Prx, \TSIR{k}{\FD}}$ is defined in \eqref{eq:STPsFD}. 
}
\setcounter{eqnback}{\value{equation}}
\setcounter{equation}{30}
\begin{figure*}[t!]
%\centering \rule[0pt]{18cm}{0.3pt}
\begin{align}
	\SE
	& =
		\frac{1}{\Ws{}{}}%\TRate
		\sum_{k=1}^K
		\LB{k}{}
		 \sum_{t \in \SetI} \LB{t}{} \Bratio{tk}^{1/2}
		\Bigg\{
			\frac{
				\tRate{\BS}\left(1 - \Pfd{k}\right)
			}{
				2  \sum_{i \in \SetI} \LB{i}{} \FnM{ik}{\HD}{4, \Pbs{k}}
			}
			+
			% \frac{\pi^{3/2}\Pfd{k} }{2 \sqrt{\TSIR{k}{\FD}}}
			\frac{\pi^{3/2}\Pfd{k} }{2 }	
		%\right.
	 \nonumber \\
	&		% +
			%2 \pi^{3/2} 
			% \frac{\pi^{3/2}\Pfd{k} }{2 \sqrt{\TSIR{k}{\FD}}}
			\times
			\left[
				\frac{
					\tRate{\BS} \sqrt{\Pbs{k}}
				}{
					\sqrt{\FnCk{\FD}{\Pus{k}}{k} \tSIR{\BS}}
				}
				\exp\left\{\left( 
				\frac{ 
					\sqrt{\Pbs{k}} \pi
					\sum_{i \in \SetI}  \LB{i}{} \FnM{ik}{\FD}{4, \Pbs{k}}
				}{ 
					\sqrt{\FnCk{\FD}{\Pus{k}}{k}\tSIR{\BS}}
				} 
				\right)^2\right\}
				\Erfc{				
				\frac{ 
					\sqrt{ \Pbs{k}}  \pi 
					\sum_{i \in \SetI} \LB{i}{} \FnM{ik}{\FD}{4, \Pbs{k}}
				}{ 
					 \sqrt{\FnCk{\FD}{\Pus{k}}{k} \tSIR{\BS}}
				} 
				} 
		\right. 
		 \nonumber\\
		& \left.\left. %\quad \quad \quad
		% \times
				+ 
				\frac{
					\tRate{\US} \sqrt{\Pus{k}}
				}{
					\sqrt{\FnCk{\FD}{\Pbs{k}}{k} \tSIR{\US}}
				}
				\exp\left\{\left( 
				\frac{ 
					\sqrt{ \Pus{k}} \pi
					\sum_{i \in \SetI}  \LB{i}{} \FnM{ik}{\FD}{4, \Pus{k}}
				}{ 
					\sqrt{\FnCk{\FD}{\Pbs{k}}{k} \tSIR{\US} }
				} 
				\right)^2\right\}
				\Erfc{				
				\frac{ 
					\sqrt{ \Pus{k}}  \pi 
					\sum_{i \in \SetI} \LB{i}{} \FnM{ik}{\FD}{4, \Pus{k}}
				}{ 
					 \sqrt{\FnCk{\FD}{\Pbs{k}}{k} \tSIR{\US} }
				} 
				} 
			\right]
		\right\}\,.
		 \label{eq:SAEFour} 
%		2 \pi^{3/2} \left( \sum_{i \in \SetI} \LB{i}{} \Bratio{ik}^{2/\PL{}}\right)
%		\exp\left\{\left( 
%				\frac{ 
%					\sqrt{ \Ptx} \sum_{i \in \SetI} 2\pi \LB{i}{} \FnM{ik}{\FD}{4, \Ptx}
%				}{ 
%					2 \sqrt{\FnC{\FD}{\Prx} \TSIR{k}{\FD} }
%				} 
%			\right)^2\right\} 
%		\\ \nonumber
%	& \quad \quad \quad \quad \quad \quad \quad \quad \quad \quad \quad \quad \quad \quad \quad 
%	\times
%		\Erfc{				
%			\frac{ 
%					\sqrt{ \Ptx} \sum_{i \in \SetI} 2\pi \LB{i}{} \FnM{ik}{\FD}{4, \Ptx}
%				}{ 
%					2 \sqrt{\FnC{\FD}{\Prx} \TSIR{k}{\FD} }
%				} 
%			}
\end{align}

\setcounter{eqncnt}{\value{equation}}
\centering \rule[0pt]{18cm}{0.3pt}
\end{figure*}
\setcounter{equation}{\value{eqnback}}
%
%======================================================= 
%

%%---------------------------------------------------------------------------%
%%                       Sec: Spectral Area Efficiency                     %
%%---------------------------------------------------------------------------%

\subsection{\ac{HDHN} Throughput Analysis}% \label{sec:throughput}
In this subsection, we derive the \ac{HDHN} throughput for various network settings. 
We first define the \ac{HDHN} throughput as follows. 
\begin{definition}\label{def:SE}
The \ac{HDHN} throughput is defined by 
\begin{align}	
	& \SE 
	=
		% \frac{\TRate}{\sNorm{\mathcal{A}}}
		\frac{1}{\Ws{}{}\sNorm{\mathcal{A}}}
		\mathbb{E} 
		\left\{
			\sum_{k=1}^K
			\left[
				\sum_{\x \in \PPPm{k}{\HD} \cap \mathcal{A}}
				\tRate{\BS}
				\IndF{\SetT{k}{\HD}}{\x, \AU{\x}}
			\right.
		\right.
	\label{eq:SEDef} \\
	%\nonumber \\ 
	& 
	% \quad \quad \quad  \quad 
		\left.
			\left.
				+
				\sum_{\x \in \PPPm{k}{\FD} \cap \mathcal{A}}
				\left(
					\tRate{\BS} \IndF{\SetT{k}{\FD}}{\x, \AU{\x}}
					+
					\tRate{\US} \IndF{\SetT{k}{\FD}}{\AU{\x}, \x}
				\right)
			\right]
		\right\}
	%\label{eq:SEDef} 
	\nonumber
\end{align}
where $\mathcal{A}$ is a bounded space with area $\sNorm{\mathcal{A}}$, 
$\AU{\lx}$ is the associating user to an \ac{AP} at $\lx$, 
and 
\begin{align}	
	\IndF{\SetT{}{}}{\lx, \ly} 
	& \triangleq
		\begin{cases}
		1, & \text{if $(\lx, \ly) \in \SetT{}{}$}
		% $\SIR{t} >\epsilon$, $\SIR{e} \leq \delta$},
		\\
		0, & \text{otherwise}\,.
		\end{cases}
\end{align}
Here, $\SetT{k}{m}$ is a random set of transmitter-receiver pairs $(\lx, \ly)$
that a transmitter at $\lx$ and its corresponding receiver at $\ly$
communicates successfully with higher received \ac{SIR} than a threshold value $\TSIR{k}{m}$,
i.e., $(\lx, \ly) \in \SetT{k}{m}$ when 
%$\SetT{k}{m} = \left\{ (\lx, \ly) \in \R^d: \SIR{k}{m} > \TSIR{k}{m}\right\}$.
%
%
\begin{align}\nonumber	
	\SetT{k}{m} = \left\{ (\lx, \ly) \in \R^d: \SIR{k}{m} \ge \TSIR{k}{m}\right\}\,.
\end{align}
\end{definition}

The \ac{HDHN} throughput measures the average data rate achieved by nodes (e.g., \acp{AP} and users)
communicating successfully in the network, 
and its unit is bits/sec/Hz/m$^2$. 
%By normalizing $\SE$ 
\blue{One can also define the {\em cell \ac{HDHN} throughput} by normalizing the total \ac{HDHN} throughput achieved over the network, $\sNorm{\mathcal{A}} \SE$, with respect to 
the average number of cells in \acp{HDHN}, $\sum_{i\in\SetI} \sNorm{\mathcal{A}} \LB{i}{}$, as
\begin{align}	
	\CSE 
	= 
		\frac{\SE}
		{
			\sum_{i\in\SetI} \LB{i}{}
		}\,.
	\label{eq:CSE}
\end{align}
}This shows the average data rate per cell in this network and its unit is bits/sec/Hz/cell.
% \mynote{hybrid-duplex heterogeneous network (HDHN) throughput }
%
Now, we derive the \ac{HDHN} throughput. 
\begin{lemma}\label{lem:SERay}
The \ac{HDHN} throughput is given by
\begin{align}
	&\SE 
	= 
		\frac{1}{\Ws{}{}}
		% \TRate
		\sum_{k=1}^K
		\LBt{k}
		\Big\{
			\left(1 - \Pfd{k} \right) 
			\tRate{\BS}
			\Ps{k}{\HD}{\Pbs{k}, 0, \tSIR{\BS}}
	%\right.
	\nonumber \\ 
	&% \quad \quad 
	%\left. 
			+
			\Pfd{k} 
			\Big[ 
				 \tRate{\BS} \Ps{k}{\FD}{\Pbs{k}, \Pus{k}, \tSIR{\BS} }
				 +
				 \tRate{\US} \Ps{k}{\FD}{\Pus{k}, \Pbs{k}, \tSIR{\US} }
			\Big]
		\Big\}
	\label{eq:SERay}	
\end{align}
where $\Ps{k}{m}{\Ptx, \Prx, \TSIR{k}{m}}$ is defined in \eqref{eq:STP}.
\end{lemma}
\begin{IEEEproof}
The \ac{HDHN} throughput in \eqref{eq:SEDef} can be represented as
\begin{align}	
	\SE 
	 & =
		%\TRate
		\frac{1}{\Ws{}{}}
		\sum_{k=1}^K
		\LBt{k}
		\left\{
			\left(1 - \Pfd{k} \right) 
			\tRate{\BS} \EX{\IndF{\SetT{k}{\HD}}{\lx, \ly}}
			%\Ps{k}{\HD}{\Pbs{k}, 0, \TSIR{k}{\HD}}
		\right.
	\nonumber\\
	&
	\quad \quad \quad \quad %\quad \quad
		\left.
			+
			\Pfd{k} 
			\left[ 
				 \tRate{\BS} \EX{\IndF{\SetT{k}{\FD}}{\lx, \ly}}
				 %\Ps{k}{\FD}{\Pbs{k}, \Pus{k}, \TSIR{k}{\FD}}
				 +
				 \tRate{\US} \EX{\IndF{\SetT{k}{\FD}}{\ly, \lx}}
				  %\Ps{k}{\FD}{\Pus{k}, \Pbs{k}, \TSIR{k}{\FD}}
			\right]
		\right\}
	\nonumber 
\end{align}
by Campbell's theorem and the stationarity of a homogeneous \ac{PPP}\cite{Kin:B93}.
Since $\EX{\IndF{\SetT{k}{m}}{\lx, \ly}} = \Ps{k}{m}{\Ptx, \Prx, \TSIR{k}{m}}$,
we obtain \eqref{eq:SERay}.
\end{IEEEproof}
We present the \ac{HDHN} throughput in closed-forms for special cases.% in following corollaries. 
\begin{corollary}\label{cor:SEeqPL}
For $\PL{i} = \PL{} >2, \, \forall {i} \in {\SetI}$ and the perfect self-\ac{IC}, i.e., $\FnCk{\FD}{\Prx}{k}=0$, 
the \ac{HDHN} throughput is given by 
\begin{align} \label{eq:SAEIC} 
	\SE
	& = 
		\sum_{k=1}^K \SE_k
\end{align}
where $\SE_k$ is the throughput of the $k$th-tier network, given by
\begin{align}  
	\SE_k
	& = 
		% \TRate
		\frac{1}{2 \Ws{}{}}
		% \sum_{k=1}^K
		\LB{k}{}\!
		\left( \sum_{t \in \SetI} \LB{t}{} \Bratio{tk}^{2/\PL{}}\right)\!
%		\Bigg\{
%			\frac{
%				%1 - \Pfd{k}
%				\tRate{\BS}
%			}{
%				\sum_{i \in \SetI} \LB{i}{} \FnM{ik}{\HD}{\PL, \Pbs{k}}
%			}
	\nonumber \\
	& \quad\quad
%	\times  		
%		\Bigg\{
%			\frac{
%				1 - \Pfd{k}
%			}{
%				\sum_{i \in \SetI} \LB{i}{} \FnM{ik}{\HD}{\PL, \Pbs{k}}
%			}
%		\right.
%		\nonumber \\
%	& \quad\quad\quad\quad\quad\quad\quad\quad 
%		\left.
%			+
%			\frac{
%				\Pfd{k}
%			}{
%				\sum_{i \in \SetI} \LB{i}{} \FnM{ik}{\FD}{\PL, \Pbs{k}}
%			}
		\times
		\Bigg\{
			\frac{
				%1 - \Pfd{k}
				\tRate{\BS}
			}{
				\sum_{i \in \SetI} \LB{i}{} \FnM{ik}{\HD}{\PL, \Pbs{k}}
			}
			+
			\frac{
				\tRate{\US} \,\Pfd{k}
			}{
				\sum_{i \in \SetI} \LB{i}{} \FnM{ik}{\FD}{\PL, \Pus{k}}
			}
		\Bigg\} \,.	
	\nonumber 	
\end{align}
\end{corollary}
\begin{IEEEproof}
It is obtained by substituting \eqref{eq:STPsFD} into \eqref{eq:SERay}.
\end{IEEEproof}
%
%When $\TSIR{k}{\HD} = \TSIR{k}{\FD},\, \forall k\in \SetI$, 
%we have $\FnM{ik}{\HD}{\PL, \Pbs{k}} = \FnM{ik}{\FD}{\PL, \Pbs{k}}$ and 
%the spectral area efficiency can be presented from \eqref{eq:SAEIC} as
%%
%%
%\begin{align}\label{eq:SAEIC2}	
%	\SE
%	& 
%	= 
%		\TRate
%		\sum_{k=1}^K
%		\LB{k}{}\!
%		\left( \sum_{t \in \SetI} \LB{t}{} \Bratio{tk}^{2/\PL{}} \right)\!
%		\Bigg\{
%%			\frac{
%%				1 
%%			}{
%				\sum_{i \in \SetI} \LB{i}{} \!
%				\Bigg[
%					\frac{\Bratio{ik}^{2/\PL{}}}{2}
%			%	\right.
%		%\right.
%				\\
%	& 
%		%\left. 
%				%\left.
%				+
%					\FnMA{ik}{m}{\PL, \Pbs{k}}
%				+ 
%				\left( 
%					 \FnMB{ik}{m}{\PL, \Pbs{k}}
%					- \FnMA{ik}{m}{\PL, \Pbs{k}} 
%				\right) 
%				\Pfd{i}
%				\Bigg]
%% 			}
%		\Bigg\}^{-1}\,.	
%		\nonumber		
%\end{align} 
%%
%%
%
\setcounter{equation}{31}
%We can present the spectral area efficiency in close form for $\PL{i}=4, \, \forall i \in \SetI$ as follows. 
\begin{corollary}% [Special case for $\PL{i}=4, \, \forall i \in \SetI$]
For $\PL{i}=4, \, \forall i \in \SetI$, the $K$-tier \ac{HDHN} throughput is given by
\eqref{eq:SAEFour} (on top of the next page). 
\end{corollary}
\begin{IEEEproof}
It is obtained by substituting \eqref{eq:STPb} and \eqref{eq:STPsHD} into \eqref{eq:SERay}. 
\end{IEEEproof}

\blue{From Lemma~\ref{lem:SERay}, we can see that 
the \ac{HDHN} throughput consists of the densities of \ac{HD}-mode \acp{AP}, \ac{FD}-mode \acp{AP},
and the \ac{FD}-mode users, and their corresponding successful transmission probabilities. 
As $\Pfd{k}$ increases, there exists more transmitting nodes (e.g., \ac{FD}-mode users) in the network, 
but the network interference also increases, % due to the increased number of \ac{FD}-mode users, 
which consequently decreases successful transmission probabilities.
Hence, it is not clear how to determine the portion of \ac{FD}-mode cells to maximize the throughput of each tier network. 
From Corollary~\ref{cor:SEeqPL}, 
we obtain the optimal portion of \ac{FD}-mode cells, $\PfdO{k}$,  for the perfect self-\ac{IC} case as follows.} 
\begin{corollary}\label{cor:PfdOIC}
For $\PL{k} = \PL{} >2$ with $\tRate{\US}=\tRate{\BS}$, % and $\TSIR{k}{\HD} = \TSIR{k}{\FD},\, \forall k\in \SetI$, 
when the self-\ac{IC} in the \ac{FD} mode is perfect, i.e., $\FnCk{\FD}{\Prx}{k}=0$,
the optimal portion of \ac{FD}-mode \acp{AP} in the $k$th-tier network 
that maximizes the throughput of the network 
is $\PfdO{k} = 1, \, \forall {k} \in {\SetI}$. 
\end{corollary}
\begin{IEEEproof}
See Appendix~\ref{app:cor5}. 
 \end{IEEEproof}
\begin{remark}
%
%From \eqref{eq:SAEIC2}, if $\FnMB{ik}{m}{\PL, \Pbs{k}} < \FnMA{ik}{m}{\PL, \Pbs{k}} $, 
%we can see that $\SE$ is an increasing function with $\Pfd{i}$ and the optimal $\Pfd{i}$ is $\PfdO{i}=1$.
%Otherwise, $\SE$ is a decreasing function with $\Pfd{i}$ and $\PfdO{i}=0$.
%As $\Pfd{k}$ increases, 
%the interference from \ac{FD}-mode users increases, which results in lower successful transmission probability.
Corollary~\ref{cor:PfdOIC} shows that, when the self-\ac{IC} is perfect, 
in spite of the degradation of successful transmission probability, 
having more communicating nodes by operating more cells in \ac{FD}
enhances the network throughput. 
Therefore, in this network, the network throughput is maximized 
by operating all \acp{AP} in \ac{FD} mode regardless of network parameters
such as transmission power or \ac{AP} spatial density. 
% 
%the number of transmitting nodes also increases as we have transmitting users as well as \acp{AP}, which results in larger network interference and lower successful transmission probability. . 
% 
\end{remark}

%---------------------------------------------------------------------------%
%                            Sec: Case Study                                 %
%---------------------------------------------------------------------------%

\begin{table}[!t]
\caption{Parameter values if not otherwise specified \label{table:parameter}} 
\begin{center}
\rowcolors{2}%{green!20!yellow}
{cyan!15!}{}
\renewcommand{\arraystretch}{1.5}
\begin{tabular}{l l | l l}
\hline 
 {\bf Parameters} & {\bf Values} & {\bf Parameters} & {\hspace{0.32cm}}{\bf Values} \\
\hline 
\hspace{0.15cm}$\PL{k}$, $\forall k$ & \hspace{0.2cm}$4$ 
			& \hspace{0.12cm}$\LB{1}{}$, $\LB{2}{}$ [nodes/m$^2$] & \hspace{0.2cm}$10^{-3}$ \\ %\addlinespace
\hspace{0.15cm}$\Ts $ [sec] & \hspace{0.2cm}$10^{-4}$ 
			& \hspace{0.12cm}$\tRate{\BS}$ [bits/sec] & \hspace{0.2cm}$10^4$  \\ %\addlinespace
\hspace{0.15cm}$\Ws{k}{m}$ [Hz] & \hspace{0.2cm}$10^4$ 
			& \hspace{0.12cm}$\tRate{\US}$ [bits/sec] & \hspace{0.2cm}$10^4$  \\ %\addlinespace
\hspace{0.15cm}$\SICdB{1}$ [dB] & \hspace{0.2cm}$-40$ 
			& \hspace{0.12cm}$\SICdB{2}$ [dB] & \hspace{0.2cm}$-30$ \\ %\addlinespace
\hspace{0.15cm}$\Pbs{1}$ [W] & \hspace{0.2cm}$30$ 
			& \hspace{0.12cm}$\Pus{1}$ [W] & \hspace{0.2cm}$3$ \\ %\addlinespace
\hspace{0.15cm}$\Pbs{2}$ [W] & \hspace{0.2cm}$30$
			& \hspace{0.12cm}$\Pus{2}$ [W] & \hspace{0.2cm}$6$  \\ %\addlinespace
\hspace{0.15cm}$\Bratio{ij}$, $\forall i,j$ & \hspace{0.2cm}$1$ 
			& \hspace{0.12cm}$\Pfd{2}$ & \hspace{0.2cm}$0$ (\ac{HD} mode) \\ %\addlinespace
\hline
\end{tabular}\vspace{-0.3cm}
\end{center}
\end{table}%

\section{Numerical Results}\label{sec:numerical}

%\/\/\/\/\/\/\/\/\/\/\/\/\/\/\/\/\/\/\/\/\/\/\/\/\/\/\/\/\/\/\/\/\/\/\/\/\/\/\/\/\/\/\/\/\/\/\/\/\/\/\/\/\/\/\/\/\/\/\/\/\/\/\/\/\/\/\/\/\/\/\/\/\/\/\/\/\/\/\/\/\/\/\/\/\/\/\/
\begin{figure}[t!]
    \begin{center}   
    { 
	\psfrag{L2Zero}[Bl][Bl][0.59]   {\ac{FD}, $\LBt{2}=0$}
	\psfrag{L2FiveFour}[Bl][Bl][0.59]   {\ac{FD}, $\LBt{2}=5 \cdot 10^{-4}$}
	\psfrag{L2OneThree}[Bl][Bl][0.59]   {\ac{FD}, $\LBt{2}=1 \cdot 10^{-3}$}
	\psfrag{L2TwoThree}[Bl][Bl][0.59]   {\ac{FD}, $\LBt{2}=2 \cdot 10^{-3}$}
	\psfrag{L2FiveThree}[Bl][Bl][0.59]   {\ac{FD}, $\LBt{2}=5 \cdot 10^{-3}$}
	\psfrag{L2OneTwoOneTwo}[Bl][Bl][0.59]   {\ac{FD}, $\LBt{2}=1 \cdot 10^{-2}$}
	\psfrag{HD}[Bl][Bl][0.59]   {\ac{HD}}
	%
	% \psfrag{10}[Bl][Bl][0.59]   {$-10$}
	%
	\psfrag{SelfIC}[tc][bc][0.7] {$\SICdB{1}$ [dB]}
	\psfrag{SAE1}[bc][tc][0.7] {$\SE_1$ [bits/sec/Hz/m$^2$]}
	 \includegraphics[width=1.00\columnwidth]{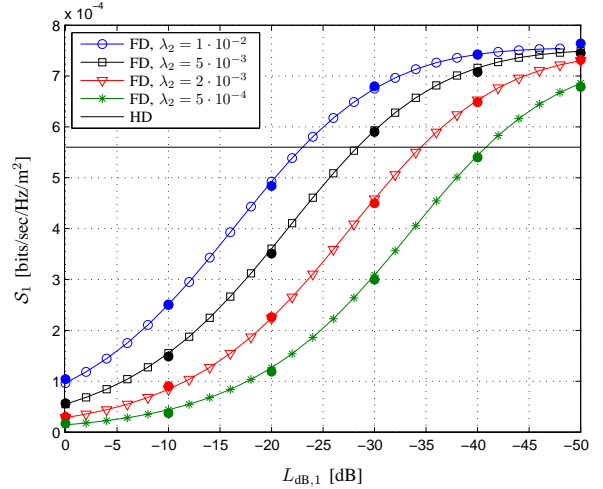}
    }
    \end{center}
    \caption{
    		\ac{HDHN} throughput of network $1$ $\SE_1$ in bits/sec/Hz/m$^2$
		as a function of the self-\ac{IC} capability $\SICdB{1}$ in dB
		for %different duplex modes 
		\ac{FD} mode ($\Pfd{1}$ = 1) and \ac{HD} mode ($\Pfd{1}$ = 0) in network $1$ and 
		different \ac{AP} spatial densities of network $2$, $\LB{2}{}$, in nodes/m$^2$
		when $\Pbs{1} = 30$ W.
		Simulation results are marked by filled circles.
%		different values of \ac{AP} spatial density ratios $\LRatio{ij} = \LB{i}{} /\LB{j}{}$ for a given $\LB{j}{} = 10^{-3}$.  		
		 }
		 \vspace{0.7mm}
   \label{fig:SE_SIC}
\end{figure}
%\/\/\/\/\/\/\/\/\/\/\/\/\/\/\/\/\/\/\/\/\/\/\/\/\/\/\/\/\/\/\/\/\/\/\/\/\/\/\/\/\/\/\/\/\/\/\/\/\/\/\/\/\/\/\/\/\/\/\/\/\/\/\/\/\/\/\/\/\/\/\/\/\/\/\/\/\/\/\/\/\/\/\/\/\/\/\/

%\/\/\/\/\/\/\/\/\/\/\/\/\/\/\/\/\/\/\/\/\/\/\/\/\/\/\/\/\/\/\/\/\/\/\/\/\/\/\/\/\/\/\/\/\/\/\/\/\/\/\/\/\/\/\/\/\/\/\/\/\/\/\/\/\/\/\/\/\/\/\/\/\/\/\/\/\/\/\/\/\/\/\/\/\/\/\/
\begin{figure}[t!]
    \begin{center}   
    { 
	\psfrag{L2Zero}[Bl][Bl][0.59]   {$\LBt{2}=0$}
	\psfrag{L2FiveFour}[Bl][Bl][0.59]   {$\LBt{2}=5 \cdot 10^{-4}$}
	\psfrag{L2OneThree}[Bl][Bl][0.59]   {$\LBt{2}=1 \cdot 10^{-3}$}
	\psfrag{L2TwoThree}[Bl][Bl][0.59]   {$\LBt{2}=2 \cdot 10^{-3}$}
	\psfrag{L2FiveThree}[Bl][Bl][0.59]   {$\LBt{2}=5 \cdot 10^{-3}$}
	\psfrag{L2OneTwoOneTwo}[Bl][Bl][0.59]   {$\LBt{2}=1 \cdot 10^{-2}$}
	\psfrag{HD}[Bl][Bl][0.59]   {\ac{HD}}
	\psfrag{FD}[Bl][Bl][0.59]   {\ac{FD}}
	\psfrag{10}[Bc][Bc][0.59]   {$-10$}
	\psfrag{20}[Bc][Bc][0.59]   {$-20$}
	\psfrag{30}[Bc][Bc][0.59]   {$-30$}
	\psfrag{40}[Bc][Bc][0.59]   {$-40$}
	\psfrag{50}[Bc][Bc][0.59]   {$-50$}
	\psfrag{5}[Bc][Bc][0.59]   {$-5$}
	\psfrag{15}[Bc][Bc][0.59]   {$-15$}
	\psfrag{25}[Bc][Bc][0.59]   {$-25$}
	\psfrag{35}[Bc][Bc][0.59]   {$-35$}
	\psfrag{45}[Bc][Bc][0.59]   {$-45$}
	\psfrag{55}[Bc][Bc][0.59]   {$-55$}
	\psfrag{SelfIC}[tc][bc][0.7] {$\SICdB{1}$ [dB]}
	\psfrag{SAE1}[bc][tc][0.7] {$\SE_1$ [bits/sec/Hz/m$^2$]}
	 \includegraphics[width=1.00\columnwidth]{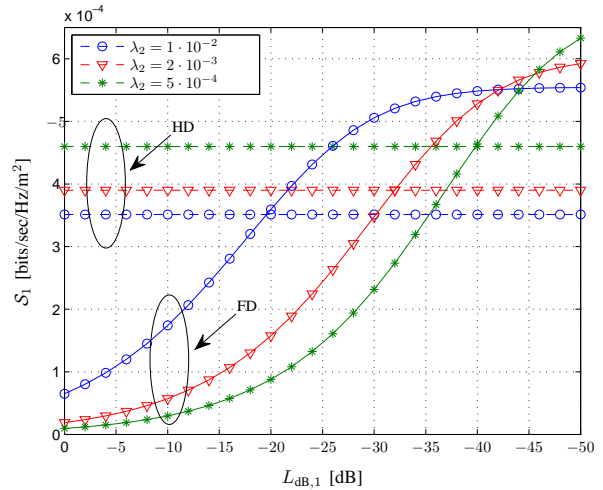}
    }
    \end{center}
    \caption{
	\ac{HDHN} throughput of network $1$ $\SE_1$ in bits/sec/Hz/m$^2$
		as a function of the self-\ac{IC} capability $\SICdB{1}$ in dB
		for %different duplex modes 
		\ac{FD} mode ($\Pfd{1}$ = 1) and \ac{HD} mode ($\Pfd{1}$ = 0) in network $1$ and 
		different \ac{AP} spatial densities of network $2$, $\LB{2}{}$, in nodes/m$^2$
		when $\Pbs{1} = 9$ W.
		% Simulation results are marked by filled circles.
%		different values of \ac{AP} spatial density ratios $\LRatio{ij} = \LB{i}{} /\LB{j}{}$ for a given $\LB{j}{} = 10^{-3}$.  		
		 }
		  \vspace{0.7mm}
   \label{fig:SE_SIC2}
\end{figure}
%\/\/\/\/\/\/\/\/\/\/\/\/\/\/\/\/\/\/\/\/\/\/\/\/\/\/\/\/\/\/\/\/\/\/\/\/\/\/\/\/\/\/\/\/\/\/\/\/\/\/\/\/\/\/\/\/\/\/\/\/\/\/\/\/\/\/\/\/\/\/\/\/\/\/\/\/\/\/\/\/\/\/\/\/\/\/\/

\begin{figure}[t!]
    \begin{center}   
    { 
	\psfrag{Pdf1}[bc][tc][0.7] {$\Pfd{1}$}
	\psfrag{Pdf2}[bc][tc][0.7] {$\Pfd{2}$}
	\psfrag{Pdf10}[cc][bc][0.6] {$\Pfd{3}=1.0$}
	\psfrag{Pdf25}[cc][bc][0.6] {$\Pfd{3}=0.25$}
	\psfrag{Pdf50}[cc][bc][0.6] {$\Pfd{3}=0.50$}
	\psfrag{Pdf75}[cc][bc][0.6] {$\Pfd{3}=0.75$}
	\psfrag{Pdf00}[cc][bc][0.6] {$\Pfd{3}=0$}
	% \psfrag{SelfICTwoZero}[Bl][Bl][0.7] {$\FnC{\FD}{\Prx} \!=\! -20$dB}
	% \psfrag{SelfICPerfect}[Bl][Bl][0.7] {$\FnC{\FD}{\Prx} \!=\! 0$}
	%
	 \includegraphics[width=1.00\columnwidth]{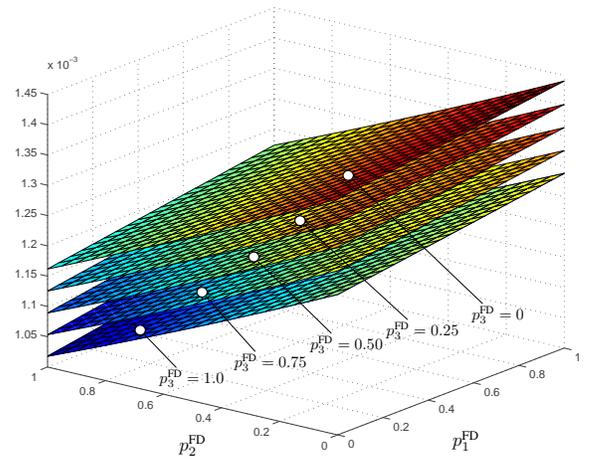}
    }
    \end{center}
    \caption{
    		\ac{HDHN} throughput $\SE$ in bits/sec/Hz/m$^2$
		of three tier networks 
		as a function of $\Pfd{1}$, $\Pfd{2}$, and $\Pfd{3}$.
		For network 3, $\LB{3}{}=5 \cdot 10^{-4}$ nodes/m$^2$, $\Pwr_{\BS,3} = 15$ W, $\Pwr_{\US,3}= 3$ W, and 
		$\SICdB{3} = - 20$ dB are used. 
%		for difference values of 	
		 }
   \label{fig:Contour_Pfd1_Pfd2_Pfd3}
\end{figure}

In this section, we evaluate the throughput of two tier \acp{HDHN} consisted of network $1$ and network $2$ 
(except for Fig.~\ref{fig:Contour_Pfd1_Pfd2_Pfd3} that considers three tier network), and 
present the effect of network parameters on the \ac{HDHN} throughput.
Specifically, 
we first show the \ac{HDHN} throughput of network $1$ % according to network parameters 
in the presence of interference from network $2$ as well as network $1$ 
to explore 
the environment that \ac{FD} achieves better throughput compared to \ac{HD}. 
We then show
how to determine the portions of \ac{FD}-mode \acp{AP} in two (or three) networks 
to maximize the \ac{HDHN} throughput.
Note that we use the self-\ac{IC} capability of nodes in the $k$th-tier network 
as $\FnCk{\FD}{\Prx}{k} = \Prx \cdot 10^{\SICdB{k}/10}$, 
where $\SICdB{k}$ [dB] is the ratio of the residual self-interference after \ac{IC} to the transmission power at the receiver. 
Unless otherwise specified, the values of network parameters presented in Table~\ref{table:parameter} are used.
% , especially, the duplex modes.

% It can be seen that 
% This can be attributed to the fact that 
% By comparing A and B, it is evident that 

\blue{Figures~\ref{fig:SE_SIC} and \ref{fig:SE_SIC2} display the \ac{HDHN} throughput of network $1$ $\SE_1$
as a function of the self-\ac{IC} capability $\SICdB{1}$ 
for different duplex modes in network $1$ and 
different values of \ac{AP} spatial density of network $2$ $\LB{2}{}$. 
Here, $\Pbs{1} = 30$ W is used for Fig.~\ref{fig:SE_SIC} while $\Pbs{1}=9$ W is used for Fig.~\ref{fig:SE_SIC2}. 
%different values of \ac{AP} spatial density ratios $\LRatio{ij}$
%when $\LB{i}{}$ is varied as $\LB{i}{} =\LRatio{ij}\LB{j}{}$ for a given $\LB{j}{} = 10^{-3}$. 
% values of $\LB{2}{}$ and $\Pfd{1}$. 
%
Simulation results are marked by filled circles in Fig.~\ref{fig:SE_SIC} and they show a good agreement with the analysis.
%
% From Fig.~\ref{fig:SE_SIC}, it can be seen 
Note that in Fig.~\ref{fig:SE_SIC}, 
$\SE_1$ in \ac{HD} mode is not changed according to $\LB{2}{}$ 
% as $\FnM{ik}{\HD}{\PL{}, \Pwr_{\BS,k}}$ in \eqref{eq:STPsHD} is equal for all $i,\,j$
% in this network setting and 
as $\SE_k$ in \ac{HD} mode is given by
\begin{align}\label{eq:SEk_B}	
	\SE_k % ^{\HD}
	=
		\frac{
			\LBt{k} \tRate{\BS} \Bratio{}^{2/\PL{}}
		}{
			2 \Ws{}{} \FnM{}{\HD}{\PL{}, \Pwr_{\BS}}
		}
\end{align}
which is not affected by $\LBt{i}$, $\forall i \neq k$.}\footnote{\blue{Specifically, 
	for $\PL{k}=\PL{}$, $\Bratio{ik}=\Bratio{}$, and $\Pbs{k}=\Pwr_{\BS}$, $\forall k, i$,
	we have 
	$\FnMA{ik}{\HD}{\PL{}, \Pwr_{\BS}} = \FnMA{}{\HD}{\PL{}, \Pwr_{\BS}}$ and 
	$\FnM{ik}{\HD}{\PL{}, \Pwr_{\BS}} = \FnM{}{\HD}{\PL{}, \Pwr_{\BS}}$, $\forall k, i$. 
	Hence, from \eqref{eq:STPsHD} and \eqref{eq:SERay}, 
	$\SE_k$ in \ac{HD} mode is represented by 
	\begin{align}\nonumber	
		\SE_k
		& =
		\frac{ \tRate{\BS}}{\Ws{}{}}
		\sum_{k=1}^2 \LBt{k} 
		\frac{
			 \Bratio{}^{2/\PL{}} \left( \sum_{k=1}^2 \LBt{k} \right)
		}{
			2 \FnM{}{\HD}{\PL{}, \Pwr_{\BS}} \left( \sum_{k=1}^2 \LBt{k} \right)
		}
%		\nonumber\\
%		& 
		=
		\frac{
			\LBt{k} \tRate{\BS} \Bratio{}^{2/\PL{}}
		}{
			2 \Ws{}{} \FnM{}{\HD}{\PL{}, \Pwr_{\BS}}
		}\,.
	\end{align}	
}} \blue{However, in Fig.~\ref{fig:SE_SIC2}, 
$\SE_1$ in \ac{HD} mode is altered by $\LB{2}{}$ 
as $\Pwr_{\BS,1} \neq \Pwr_{\BS,2}$.
% as $\FnM{ik}{\HD}{\PL{}, \Pwr_{\BS,k}}$ is not equal for all $k$ when $\Pwr_{\BS,1} \neq \Pwr_{\BS,2}$.   
} 

\begin{figure}[t!]
    \begin{center}   
    { 
	\psfrag{L2toL1Four}[Bl][Bl][0.59]   {$\LRatio{21} = 4.0$}
	\psfrag{L2toL1Two}[Bl][Bl][0.59]   {$\LRatio{21} = 2.0$}
	\psfrag{L2toL1One}[Bl][Bl][0.59]   {$\LRatio{21} = 1.0$}
	\psfrag{L2toL1ZeroFive}[Bl][Bl][0.59]   {$\LRatio{21} = 0.5$}
	\psfrag{L2toL1ZeroOne}[Bl][Bl][0.59]   {$\LRatio{21}= 0.1$}
	\psfrag{L1toL2Four}[Bl][Bl][0.59]   {$\LRatio{12} = 4.0$}
	\psfrag{L1toL2Two}[Bl][Bl][0.59]   {$\LRatio{12} = 2.0$}
	\psfrag{L1toL2ZeroFive}[Bl][Bl][0.59]   {$\LRatio{12} = 0.5$}
	\psfrag{L1toL2ZeroOne}[Bl][Bl][0.59]   {$\LRatio{12} = 0.1$}
%		\psfrag{L2toL1Four}[Bl][Bl][0.59]   {$\LB{1}{} = 10^{-3}$, $\LB{2}{}/\LB{1}{} = 4.0$}
%	\psfrag{L2toL1Two}[Bl][Bl][0.59]   {$\LB{1}{} = 10^{-3}$, $\LB{2}{}/\LB{1}{} = 2.0$}
%	\psfrag{L2toL1One}[Bl][Bl][0.59]   {$\LB{1}{} = 10^{-3}$, $\LB{2}{}/\LB{1}{} = 1.0$}
%	\psfrag{L2toL1ZeroFive}[Bl][Bl][0.59]   {$\LB{1}{} = 10^{-3}$, $\LB{2}{}/\LB{1}{} = 0.5$}
%	\psfrag{L2toL1ZeroOneOneOneOne}[Bl][Bl][0.59]   {$\LB{1}{} = 10^{-3}$, $\LB{2}{}/\LB{1}{} = 0.1$}
%	\psfrag{L1toL2Four}[Bl][Bl][0.59]   {$\LB{2}{} = 10^{-3}$, $\LB{1}{}/\LB{2}{} = 4.0$}
%	\psfrag{L1toL2Two}[Bl][Bl][0.59]   {$\LB{2}{} = 10^{-3}$, $\LB{1}{}/\LB{2}{} = 2.0$}
%	\psfrag{L1toL2ZeroFive}[Bl][Bl][0.59]   {$\LB{2}{} = 10^{-3}$, $\LB{1}{}/\LB{2}{} = 0.5$}
%	\psfrag{L1toL2ZeroOne}[Bl][Bl][0.59]   {$\LB{2}{} = 10^{-3}$, $\LB{1}{}/\LB{2}{} = 0.1$}
	%
	\psfrag{Pfd1}[tc][bc][0.7] {$\Pfd{1}$}
	\psfrag{SAE1}[bc][tc][0.7] {${\SE}_{1}/ \SE_1^{\HD}$} % {${\SE}_{1}^{\text{n}}$}
	%
	% \psfrag{SelfICTwoZero}[Bl][Bl][0.7] {$\FnC{\FD}{\Prx} \!=\! -20$dB}
	% \psfrag{SelfICPerfect}[Bl][Bl][0.7] {$\FnC{\FD}{\Prx} \!=\! 0$}
	%
	 \includegraphics[width=1.00\columnwidth]{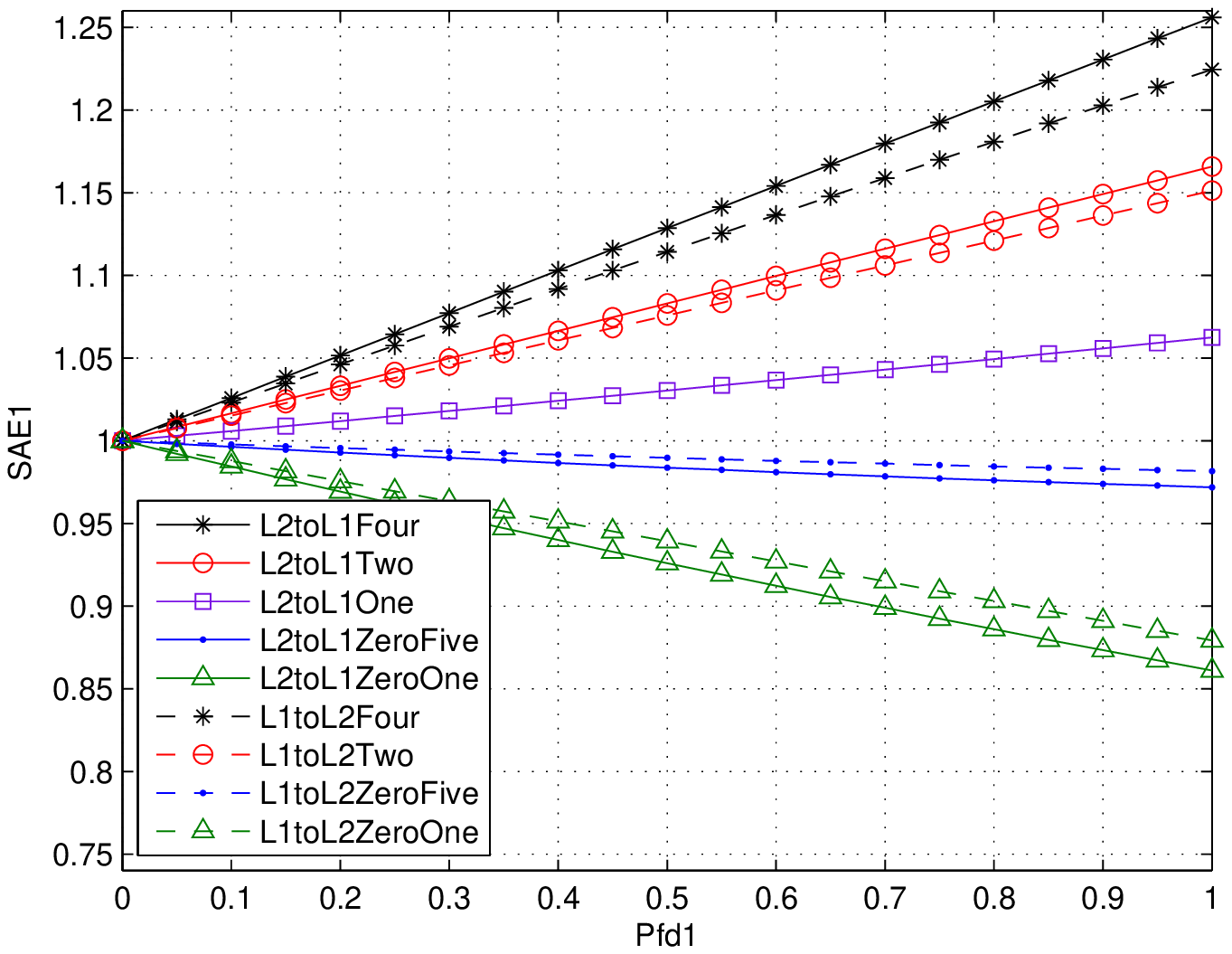}
    }
    \end{center}
    \caption{
    		Ratio of $\SE_1$ to the achievable $\SE_1$ in \ac{HD} mode, ${\SE}_{1}/ \SE_1^{\HD}$, %$\NSE_1$,
		as a function of $\Pfd{1}$	
		for different duplex modes
		and different values of $\LRatio{ij}$, 
		where $\LRatio{ij} = \LB{i}{}/\LB{j}{}$ for a given $\LB{j}{} = 10^{-3}$ nodes/m$^2$. 		
		 }
   \label{fig:SE1_Pfd1_L2L1}
\end{figure}
%\/\/\/\/\/\/\/\/\/\/\/\/\/\/\/\/\/\/\/\/\/\/\/\/\/\/\/\/\/\/\/\/\/\/\/\/\/\/\/\/\/\/\/\/\/\/\/\/\/\/\/\/\/\/\/\/\/\/\/\/\/\/\/\/\/\/\/\/\/\/\/\/\/\/\/\/\/\/\/\/\/\/\/\/\/\/\/

\blue{From Fig.~\ref{fig:SE_SIC}, it can be seen that for large $\LB{2}{}$ and low $\SICdB{1}$, 
$\SE_1$ in \ac{FD} mode is higher than that of \ac{HD} mode. 
% $\SE_1$ in \ac{FD} mode increases as $\LB{2}{}$ increases or $\SICdB{1}$ decreases. 
The effect of low $\SICdB{1}$ on $\SE_1$ is obvious as it means we have smaller residual self-interference. 
On the other hand, large $\LB{2}{}$ affects the network throughput in two aspects: 
1) increasing the network interference (negative effect); and 
2) making an user associate to closer \ac{AP} with higher probability (positive effect). 
As $\LB{2}{}$ increases, we have large network interference, 
which makes the effect of self-interference on $\SE_1$ less in \ac{FD} mode
and the successful transmission probabilities in \ac{FD} and \ac{HD} modes relatively similar. 
Hence, for large $\LB{2}{}$ or low $\SICdB{1}$, operating \acp{AP} in \ac{FD} achieves
higher $\SE_1$ compared to that in $\HD$ due to additionally communicating users in \ac{FD} mode. 
} 

\blue{From Fig.~\ref{fig:SE_SIC}, we can also see that for a fixed $\SICdB{1}$, 
$\SE_1$ in \ac{FD} mode increases with $\LB{2}{}$. % increases. 
This is due to the fact that when the self-interference is large, as $\LB{2}{}$ increases, 
the increased network interference affects less than the shorter distance to associated \ac{AP}. 
However, this results becomes different when the self-interference is small as shown in Fig.~\ref{fig:SE_SIC2}. 
In Fig.~\ref{fig:SE_SIC2}, we can see that $\SE_1$ in \ac{FD} mode decreases as $\LB{2}{}$ increases
when $\SICdB{1} < - 45$. 
This can be attributed to the fact that for small self-interference,
$\SE_1$ is changed more by the increased network interference than the shorter communication link distance. 
Hence, in this case, having smaller $\LB{2}{}$ can enhance $\SE_1$. 
This result is also applied for the \ac{HD} mode case. From Fig.~\ref{fig:SE_SIC2}, 
we can see that $\SE_1$ in HD mode decreases as $\LBt{2}$ increases 
due to the large effect of the increased network interference. 
%However, in Fig.~\ref{fig:SE_SIC2}, $\SE_1$ in HD mode decreases as $\LBt{2}$ increases
% as the increased network interference 
% the effect of increased interference is less than the shorter communication links 
}

%From Fig.~\ref{fig:SE_SIC}, it can be also seen that 
%$\SE_1$ in \ac{FD} mode increases as $\LB{2}{}$ increases or $\SICdB{1}$ decreases. 
%%
%This can be attributed to the fact that, for larger $\LB{2}{}$ or lower $\SICdB{1}$,  
%the self-interference becomes less significant compared to the network interference. 
%% while the communication link distance becomes shorter. 
%%
%Therefore, for large $\LB{2}{}$ or low $\SICdB{1}$,
%%having more communicating nodes by 
%operating all \acp{AP} in \ac{FD} achieves
%higher $\SE_1$ compared to that in $\HD$. 
%% operating \acp{AP} in \ac{FD} mode can achieve higher $\SE_1$ compared to $\HD$ mode. 

%***************************
Figure~\ref{fig:SE1_Pfd1_L2L1} shows the ratio of $\SE_1$ to the achievable $\SE_1$ when $\Pfd{1}=0$ 
(i.e., when all \acp{AP} are operating in \ac{HD} mode),
$\NSE_1$, 
as a function of the \ac{FD}-mode \ac{AP} portion $\Pfd{1}$
for different values of \ac{AP} spatial density ratios $\LRatio{12}$ and $\LRatio{21}$
when $\LB{i}{}$ is varied as $\LB{i}{} =\LRatio{ij}\LB{j}{}$ for a given $\LB{j}{} = 10^{-3}$. 
% for different values of $\LRatio{12}$ and $\LRatio{21}$. 
% to show how $\SE_1$ is changed as $\Pfd{1}$ increases. 
% In this figure, $\NSE_1$ is shown 
%for different values of \ac{AP} spatial density ratios $\LRatio{12}$ and $\LRatio{21}$
%when $\LB{i}{}$ is varied as $\LB{i}{} =\LRatio{ij}\LB{j}{}$ for a given $\LB{j}{} = 10^{-3}$. 
%
%
This figure shows how much $\SE_1$ increases or decreases in \ac{FD} compared to achievable $\SE_1$ in \ac{HD}.
Furthermore, larger $\LRatio{12}$ or $\LRatio{21}$ means larger $\LT$ where $\LT = \LB{1}{}+\LB{2}{}$, 
and $\LT=\left(1+\LRatio{ij}\right) \cdot 10^{-3}$ in this figure. 
% 
% From Fig.~\ref{fig:SE1_Pfd1_L2L1}, %according to $\LRatio{12}$ and $\LRatio{21}$. 
From Fig.~\ref{fig:SE1_Pfd1_L2L1}, it can be seen that $\SE_1$ increases with $\Pfd{1}$ for high $\LT$ while it decreases for low $\LT$. 
This can be attributed to the fact that
the \ac{FD} mode achieves higher throughput than the \ac{HD} mode for large $\LT$ as also shown in Fig.~\ref{fig:SE_SIC}. 
%
%It can be also seen that
%$\NSE$ is increased more by $\LRatio{21}$ than by $\LRatio{12}$ for the same $\LT$.
%This is due to the fact that 
%% increasing 
%% network interference increases 
%$\LB{1}{}$ increases network interference more than $\LB{2}{}$
%since the interference from \ac{FD}-mode users also increases with $\LB{1}{}$, but not with $\LB{2}{}$. 
%from \ac{FD}-mode users 
%while increasing $\LB{2}{}$ does not. 
%
% it can be seen that 
% $\NSE_1$ is either increasing or decreasing over all range of $\Pfd{1}$.
%
In Fig.~\ref{fig:SE1_Pfd1_L2L1}, 
$\NSE_1$ either increases or decreases with $\Pfd{1}$ over all range of $\Pfd{1}$.
This shows, in terms of the throughput of a tier network,
operating all \acp{AP} either in \ac{FD} mode or \ac{HD} mode achieves the maximum throughput compared to
the mixture of two mode \acp{AP}. % This also can be seen in Figs.~\ref{}, which shows 
For example, when $\LRatio{ij}$ is greater than $1$ in Fig.~\ref{fig:SE1_Pfd1_L2L1}, 
deploying \ac{FD}-mode \acp{AP} in all cells of network $1$
achieves the maximum $\SE_1$. 

%\/\/\/\/\/\/\/\/\/\/\/\/\/\/\/\/\/\/\/\/\/\/\/\/\/\/\/\/\/\/\/\/\/\/\/\/\/\/\/\/\/\/\/\/\/\/\/\/\/\/\/\/\/\/\/\/\/\/\/\/\/\/\/\/\/\/\/\/\/\/\/\/\/\/\/\/\/\/\/\/\/\/\/\/\/\/\/
%***** x-axis:  [tc][bc][0.7] y-axis: [bc][tc][0.7], legend: [Bl][Bl][0.59]
\begin{figure}[t!]
    \begin{center}   
    { 
	%\psfrag{FDPusPbsZeroOne}[Bl][Bl][0.59]   {\ac{FD}, $\Pus{1} = 0.1$}
	%\psfrag{FDPusPbsZeroFive}[Bl][Bl][0.59]   {\ac{FD}, $\Pus{1} = 0.5$}
	%\psfrag{FDPusPbsOne}[Bl][Bl][0.59]   {\ac{FD}, $\Pus{1} = 1.0$}
	\psfrag{FDPusPbsZeroOne}[Bl][Bl][0.59]   {\ac{FD}, $\Pus{1} = 3$ }
	\psfrag{FDPusPbsZeroFive}[Bl][Bl][0.59]   {\ac{FD}, $\Pus{1} = 15$ }
	\psfrag{FDPusPbsOne}[Bl][Bl][0.59]   {\ac{FD}, $\Pus{1} = 30$}
	\psfrag{HD}[Bl][Bl][0.59]   {\ac{HD}}
	\psfrag{LOnetoLTwo}[tc][bc][0.7] {$\LRatio{12}$}
	\psfrag{SAE1}[bc][tc][0.7] {$\SE_1$ [bits/sec/Hz/m$^2$]}
	\psfrag{SelfICThreeZero}[Bl][Bl][0.7] {$\SICdB{1} \!=\! -30$}
	\psfrag{SelfICPerfect}[Bl][Bl][0.7] {$\SICdB{1} \!=\! -\infty$}
	\psfrag{Pusincrease}[Bc][Bc][0.7] {$\Pus{1}$ increases}
	\psfrag{Pusincrease2}[Bl][Bl][0.7] {$\Pus{1}$ increases}
	 \includegraphics[width=1.00\columnwidth]{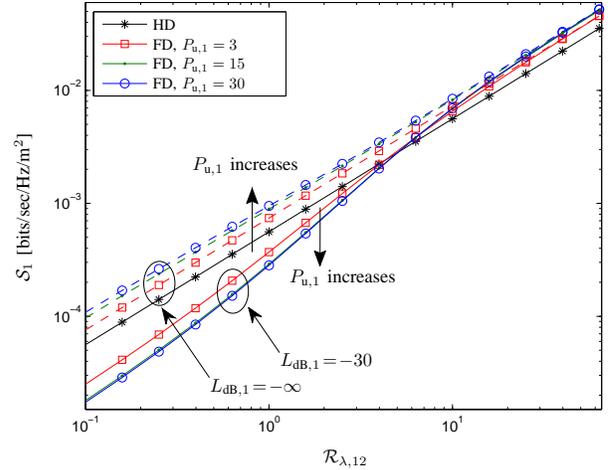}
    }
    \end{center}
    \caption{
    		\ac{HDHN} throughput of network $1$ $\SE_1$ in bits/sec/Hz/m$^2$
		as a function of $\LRatio{12}$ with $\LB{2}{}=10^{-3}$ nodes/m$^2$
		for different values of $\Pus{1}$ in W and $\SICdB{1}$ in dB 
		and different duplex modes. 		
		 }
   \label{fig:SE1_L1L2_PusPbs_SIC}
\end{figure}
%\/\/\/\/\/\/\/\/\/\/\/\/\/\/\/\/\/\/\/\/\/\/\/\/\/\/\/\/\/\/\/\/\/\/\/\/\/\/\/\/\/\/\/\/\/\/\/\/\/\/\/\/\/\/\/\/\/\/\/\/\/\/\/\/\/\/\/\/\/\/\/\/\/\/\/\/\/\/\/\/\/\/\/\/\/\/\/

Figure~\ref{fig:SE1_L1L2_PusPbs_SIC} displays $\SE_1$ 
as a function of $\LRatio{12}$ with $\LB{2}{}=10^{-3}$
for different values of $\Pus{1}$ and $\SICdB{1}$ and different duplex modes. 
%
% From Fig.~\ref{fig:SE1_L1L2_PusPbs_SIC}, 
Note that $\SE_1$ increases with $\LB{1}{}$ as shown in \eqref{eq:SEk_B}.
% since more communicating nodes exist in the network. 
%
From Fig.~\ref{fig:SE1_L1L2_PusPbs_SIC}, 
it can be seen that for $\LRatio{12}<4$,
the \ac{HD} mode achieves higher $\SE_1$ than the \ac{FD} mode when $\SICdB{1} = -30$,
but it becomes opposite for the perfect self-\ac{IC}, i.e., $\SICdB{1} = -\infty$. 
%However, \ac{FD} mode obtains higher $\SE_1$ when the self-\ac{IC} is perfect, i.e., $\SICdB{1} = -\infty$, 
This is also verified in Corollary~\ref{cor:PfdOIC}, which shows the optimal portion of \ac{FD}-mode \acp{AP}
is $\PfdO{k}=1$, $\forall k$, for $\FnCk{\FD}{\Prx}{k}=0$. 
%
%This is %due to the effect of self-\ac{IC} capability on the performance of \ac{FD} mode,
%also verified in Corollary~\ref{cor:PfdOIC}, 
%which shows the \ac{FD} mode always achieves larger \ac{HDHN} throughput than the \ac{HD} mode 
%when the self-\ac{IC} is perfect. 
%
From Fig.~\ref{fig:SE1_L1L2_PusPbs_SIC},
it can be also seen that, for $\LRatio{12} < 4$,  
$\SE_1$ in \ac{FD} mode increases as $\Pus{1}$ increases for $\SICdB{1} = -\infty$
while it decreases for $\SICdB{1} = - 30$. 
This is due to the fact that, for high self-\ac{IC} capability,
the network throughput in \ac{FD} mode increases with $\Pus{1}$
since higher $\Pus{1}$ provides more reliable communication between a user and its associated \ac{AP}. % in \ac{FD} mode.  
% the communication link from a user to an \ac{AP} becomes more reliable
% with small increment of self-interference. 
% 
On the other hand, for low self-\ac{IC} capability,
the self-interference mainly determines the network throughput,
so lower $\Pus{1}$ achieves higher $\SE_1$. 
\blue{From Fig.~\ref{fig:SE1_L1L2_PusPbs_SIC},
we can also see that the 
$\SE_1$ in \ac{FD} mode with $\SICdB{1} = - 30$ converges to 
that with $\SICdB{1} = -\infty$ as $\LB{1}{}$ increases. 
This is due to the fact that as we have large network interference (i.e., large $\LB{1}{}$),
the network interference mainly determines the network throughput while the effect of residual self-interference becomes marginal. 
Due to the relatively weak effect of self-interference for large $\LB{1}{}$, when $\LRatio{12} > 4$, 
$\SE_1$ in \ac{FD} mode with $\SICdB{1} = - 30$
becomes larger than $\SE_1$ in \ac{HD} mode as we have additional communicating users in \ac{FD} mode. 
}

%\/\/\/\/\/\/\/\/\/\/\/\/\/\/\/\/\/\/\/\/\/\/\/\/\/\/\/\/\/\/\/\/\/\/\/\/\/\/\/\/\/\/\/\/\/\/\/\/\/\/\/\/\/\/\/\/\/\/\/\/\/\/\/\/\/\/\/\/\/\/\/\/\/\/\/\/\/\/\/\/\/\/\/\/\/\/\/
%***** x-axis:  [tc][bc][0.7] y-axis: [bc][tc][0.7], legend: [Bl][Bl][0.59]
\begin{figure}[t!]
    \begin{center}   
    { 
	\psfrag{HDHDHDHD}[Bl][Bl][0.59]   {(\ac{HD}, \ac{HD})}
	\psfrag{HDFD}[Bl][Bl][0.59]   {(\ac{HD}, \ac{FD})}
	\psfrag{FDHD}[Bl][Bl][0.59]   {(\ac{FD}, \ac{HD})}
	\psfrag{FDFD}[Bl][Bl][0.59]   {(\ac{FD}, \ac{FD})}
	\psfrag{OHDHD}[Br][Bl][0.59]   {$\mathcal{M}^{\text{b}}=$(\ac{HD}, \ac{HD})}
	\psfrag{OFDHD}[Bc][Bc][0.59]   {$\mathcal{M}^{\text{b}}=$(\ac{FD}, \ac{HD})}
	\psfrag{OFDFD}[Bc][Bc][0.59]   {$\mathcal{M}^{\text{b}}=$(\ac{FD}, \ac{FD})}
	\psfrag{LTwotoLOne}[tc][bc][0.7] {$\LRatio{21}$}
	\psfrag{CellSAE}[bc][tc][0.7] {$\CSE$ [bits/sec/Hz/cell]}
	%
	% \psfrag{SelfICTwoZero}[Bl][Bl][0.7] {$\FnC{\FD}{\Prx} \!=\! -20$dB}
	% \psfrag{SelfICPerfect}[Bl][Bl][0.7] {$\FnC{\FD}{\Prx} \!=\! 0$}
	%
	 \includegraphics[width=1.00\columnwidth]{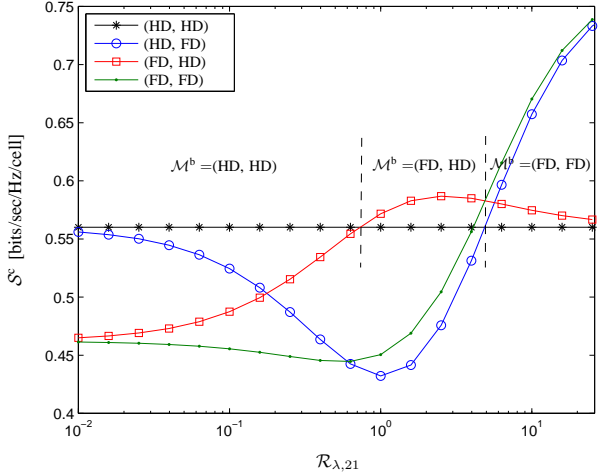}
    }
    \end{center}
    \caption{
    		Cell \ac{HDHN} throughput $\CSE$ in bits/sec/Hz/cell 
		as a function of $\LRatio{21}$ with $\LB{1}{}=10^{-3}$ nodes/m$^2$
		for different sets $(m_1, m_2)$ of duplex modes   
		in network $1$ $m_1$ and network $2$ $m_2$ . 		
		 }
   \label{fig:CellSE_L2L1_HDFD}
\end{figure}
%\/\/\/\/\/\/\/\/\/\/\/\/\/\/\/\/\/\/\/\/\/\/\/\/\/\/\/\/\/\/\/\/\/\/\/\/\/\/\/\/\/\/\/\/\/\/\/\/\/\/\/\/\/\/\/\/\/\/\/\/\/\/\/\/\/\/\/\/\/\/\/\/\/\/\/\/\/\/\/\/\/\/\/\/\/\/\/

Now, we present the \ac{HDHN} throughput for two networks. 
Figure~\ref{fig:CellSE_L2L1_HDFD} shows the cell \ac{HDHN} throughput $\CSE$ 
as a function of $\LRatio{21}$ with $\LB{1}{}=10^{-3}$
for different duplex mode set $(m_1, m_2)$, where $m_i$ is the duplex mode of network $i$, $\forall i \in \{1,2\}$. 
Here, $\mathcal{M}^{\text{b}}$ is the best duplex mode set $(m_1, m_2)$ that achieves the highest $\CSE$. 
\blue{Note that due to the network parameters used for this figure, from \eqref{eq:CSE} and \eqref{eq:SEk_B}, 
$\CSE$ in (\ac{HD}, \ac{HD}) is presented by}
\begin{align}\nonumber	
	\CSE
	=
		\frac{
			\tRate{\BS} \Bratio{}^{2/\PL{}}
		}{
			2 \Ws{}{} \FnM{}{\HD}{\PL{}, \Pwr_{\BS}}
		}
\end{align}
which is not affected by any $\LB{i}{}$, $\forall i$ (nor by $\LRatio{ij}$).   
Note also that, for a given $\LRatio{ij}$, the best duplex mode set in terms of $\CSE$
is equal to that in terms of $\SE$ as $\SE$ is {the scaling of $\CSE$ with $\LT$}. 
%the $\mathcal{R}_{\lambda2}$ ranges that make a specific $(m_1, m_2)$ achieves highest $\CSE$
%are equal to those that achieves highest $\SE$ since $\SE$ is {the scaling of $\CSE$ with $\LT$}. 
% 
From Fig.~\ref{fig:CellSE_L2L1_HDFD}, 
% As also shown in Figs~\ref{fig:CellSE_L2L1_HDFD}, 
it can be seen that the best duplex mode set is (\ac{FD}, \ac{FD}) for large $\LRatio{21}$
because   
% This is due to the fact that 
the \ac{FD} mode achieves better throughput than the \ac{HD} mode for large $\LT$.
% as shown in Figs.~\ref{fig:SE_SIC} and \ref{fig:SE1_Pfd1_L2L1}.
%
It can be also seen that the $\LRatio{21}$ value that changes the best duplex mode from \ac{HD} to \ac{FD} 
in the network $1$ is generally smaller than that in the network $2$. 
This can be attributed to the fact that 
the network $1$ has better self-\ac{IC} capability and lower $\Pus{1}$, 
so the self-interference in the network $1$ is smaller than that in the network $2$.
Hence, the \ac{FD} mode is preferred to \ac{HD} mode even for small $\LB{2}{}$ in the network $1$. 
From this figure, it can be seen that
the hybrid-duplex mode set can enhance the throughput of heterogeneous network
for the $\LRatio{21}$ range of $\mathcal{M}^{\text{b}}=$(\ac{FD}, \ac{HD}).
% e.g., $\mathcal{M}^{\text{b}}=$(\ac{FD}, \ac{HD}).
% ,when networks with different network parameters and the self-\ac{IC} capabilities coexist. 
This is also verified in the following figure. 
% Columns 1 through 8
%
%    0.0100    0.0158    0.0251    0.0398    0.0631    0.1000    0.1585    0.2512
%
%  Columns 9 through 16
%
%    0.3981    0.6310    1.0000    1.5849    2.5119    3.9811    6.3096   10.0000
%
%  Columns 17 through 21
%
%   15.8489   25.1189   39.8107   63.0957  100.0000

%\/\/\/\/\/\/\/\/\/\/\/\/\/\/\/\/\/\/\/\/\/\/\/\/\/\/\/\/\/\/\/\/\/\/\/\/\/\/\/\/\/\/\/\/\/\/\/\/\/\/\/\/\/\/\/\/\/\/\/\/\/\/\/\/\/\/\/\/\/\/\/\/\/\/\/\/\/\/\/\/\/\/\/\/\/\/\/
%%%  x-axis:  [tc][bc][0.7] y-axis: [bc][tc][0.7], legend: [Bl][Bl][0.59]
\begin{figure}[t!]
    \begin{center}   
    { 
	\psfrag{Pdf1}[tc][bc][0.7] {$\Pfd{1}$}
	\psfrag{Pdf2}[bc][tc][0.7] {$\Pfd{2}$}
	\psfrag{0.0009}[cc][bc][0.6] {$0.9\cdot10^{-3}$}
	\psfrag{0.00095}[cc][cc][0.6] {$0.95\cdot10^{-3}$}
	\psfrag{0.001}[cc][cc][0.6] {$1.0\cdot10^{-3}$}
	\psfrag{0.00105}[cc][cc][0.6] {$1.05\cdot10^{-3}$}
	\psfrag{0.0011}[cc][cc][0.6] {$1.1\cdot10^{-3}$}
	% \psfrag{SelfICTwoZero}[Bl][Bl][0.7] {$\FnC{\FD}{\Prx} \!=\! -20$dB}
	% \psfrag{SelfICPerfect}[Bl][Bl][0.7] {$\FnC{\FD}{\Prx} \!=\! 0$}
	%
	 \includegraphics[width=1.00\columnwidth]{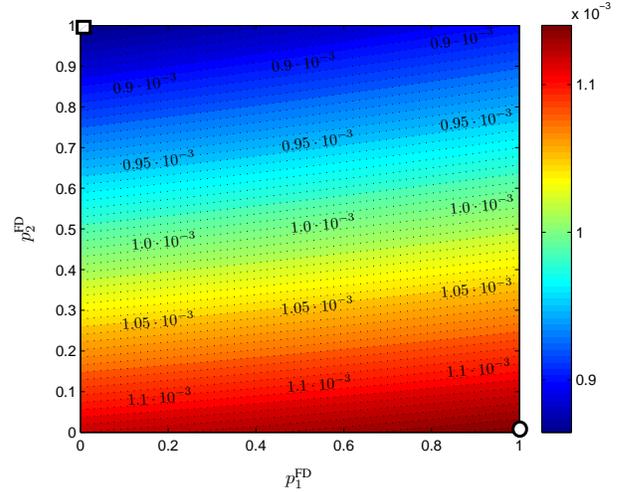}
    }
    \end{center}
    \caption{
    		\ac{HDHN} throughput $\SE$ in bits/sec/Hz/m$^2$
		as a function of $\Pfd{1}$ and $\Pfd{2}$
		(the square and the circle are the points achieving the minimum and the maximum $\SE$, respectively).  	
		 }
   \label{fig:Contour_Pfd1_Pfd2}
\end{figure}
%\/\/\/\/\/\/\/\/\/\/\/\/\/\/\/\/\/\/\/\/\/\/\/\/\/\/\/\/\/\/\/\/\/\/\/\/\/\/\/\/\/\/\/\/\/\/\/\/\/\/\/\/\/\/\/\/\/\/\/\/\/\/\/\/\/\/\/\/\/\/\/\/\/\/\/\/\/\/\/\/\/\/\/\/\/\/\/

%\/\/\/\/\/\/\/\/\/\/\/\/\/\/\/\/\/\/\/\/\/\/\/\/\/\/\/\/\/\/\/\/\/\/\/\/\/\/\/\/\/\/\/\/\/\/\/\/\/\/\/\/\/\/\/\/\/\/\/\/\/\/\/\/\/\/\/\/\/\/\/\/\/\/\/\/\/\/\/\/\/\/\/\/\/\/\/
%%%  x-axis:  [tc][bc][0.7] y-axis: [bc][tc][0.7], legend: [Bl][Bl][0.59]
%\/\/\/\/\/\/\/\/\/\/\/\/\/\/\/\/\/\/\/\/\/\/\/\/\/\/\/\/\/\/\/\/\/\/\/\/\/\/\/\/\/\/\/\/\/\/\/\/\/\/\/\/\/\/\/\/\/\/\/\/\/\/\/\/\/\/\/\/\/\/\/\/\/\/\/\/\/\/\/\/\/\/\/\/\/\/\/
%    Pbs3 = 0.5;
%    Pus3 = 0.1;
%        SelfCancel_dB3 = 20;

Figure~\ref{fig:Contour_Pfd1_Pfd2} displays the contour of $\SE$ 
as a function of $\Pfd{1}$ and $\Pfd{2}$. 
It can be seen that $\SE$ is maximized when $\Pfd{1}=1$ and $\Pfd{2}=0$ (the point marked by a circle in the figure), i.e., (\ac{FD}, \ac{HD}), 
which is the same result for $\LRatio{21}=1$ in Fig.~\ref{fig:CellSE_L2L1_HDFD}. 
Hence, by operating \acp{AP} of network $1$ in \ac{FD} mode and \acp{AP} of network $2$ in \ac{HD} mode,
we achieve the maximum \ac{HDHN} throughput. 
From Fig.~\ref{fig:Contour_Pfd1_Pfd2},
it can be also seen that $\SE$ keeps increasing with $\Pfd{1}$ and decreasing with $\Pfd{2}$.
This also verifies that, within a tier network, 
operating all \acp{AP} either in \ac{FD} or \ac{HD} achieves higher $\SE$
compared to having \acp{AP} in both modes.
%but the hybrid-duplex networks, i.e., the mixture of networks in different duplex modes, can enhance $\SE$. 
% (\ac{FD}, \ac{HD}) 
%
%
This can be also verified in Fig.~\ref{fig:Contour_Pfd1_Pfd2_Pfd3},
which displays the \ac{HDHN} throughput $\SE$
for three tier networks 
as a function of $\Pfd{1}$ and $\Pfd{2}$ for different values of $\Pfd{3}$. 
From this figure, we can see that the maximum $\SE$ can be achieved 
by operating all \acp{AP} in network 1, 2, and network 2 
in \ac{HD} ($\Pfd{1}$ = 0), \ac{FD} ($\Pfd{2}$ = 1), and \ac{HD}  ($\Pfd{3}$ = 0) modes, respectively. 

%\/\/\/\/\/\/\/\/\/\/\/\/\/\/\/\/\/\/\/\/\/\/\/\/\/\/\/\/\/\/\/\/\/\/\/\/\/\/\/\/\/\/\/\/\/\/\/\/\/\/\/\/\/\/\/\/\/\/\/\/\/\/\/\/\/\/\/\/\/\/\/\/\/\/\/\/\/\/\/\/\/\/\/\/\/\/\/
%***** x-axis:  [tc][bc][0.7] y-axis: [bc][tc][0.7], legend: [Bl][Bl][0.59]
\begin{figure}[t!]
    \begin{center}   
    { 
	\psfrag{HDHDHDHDHDHDHDHD}[Bl][Bl][0.59]   {(\ac{HD}, \ac{HD}), $\LT = 2 \cdot 10^{-3}$}
	\psfrag{HDFD}[Bl][Bl][0.59]   {(\ac{HD}, \ac{FD}),   $\,\LT = 2 \cdot 10^{-3}$}
	\psfrag{FDHD}[Bl][Bl][0.59]   {(\ac{FD}, \ac{HD}),   $\,\LT = 2 \cdot 10^{-3}$}
	\psfrag{FDFD}[Bl][Bl][0.59]   {(\ac{FD}, \ac{FD}), $\,\,\LT = 2 \cdot 10^{-3}$}
	\psfrag{HDHD2}[Bl][Bl][0.59]   {(\ac{HD}, \ac{HD}), $\LT = 1 \cdot 10^{-2}$}
	\psfrag{HDFD2}[Bl][Bl][0.59]   {(\ac{HD}, \ac{FD}), $\,\LT = 1 \cdot 10^{-2}$}
	\psfrag{FDHD2}[Bl][Bl][0.59]   {(\ac{FD}, \ac{HD}), $\,\LT = 1 \cdot 10^{-2}$}
	\psfrag{FDFD2}[Bl][Bl][0.59]   {(\ac{FD}, \ac{FD}), $\,\,\LT = 1 \cdot 10^{-2}$}
	\psfrag{LOnetoLTwo}[tc][bc][0.7] {$\LB{1}{}/\LB{2}{}$}
	\psfrag{CellSAE}[bc][tc][0.7] {$\CSE$ [bits/sec/Hz/cell]}
	%
	% \psfrag{SelfICTwoZero}[Bl][Bl][0.7] {$\FnC{\FD}{\Prx} \!=\! -20$dB}
	% \psfrag{SelfICPerfect}[Bl][Bl][0.7] {$\FnC{\FD}{\Prx} \!=\! 0$}
	%
	 \includegraphics[width=1.00\columnwidth]{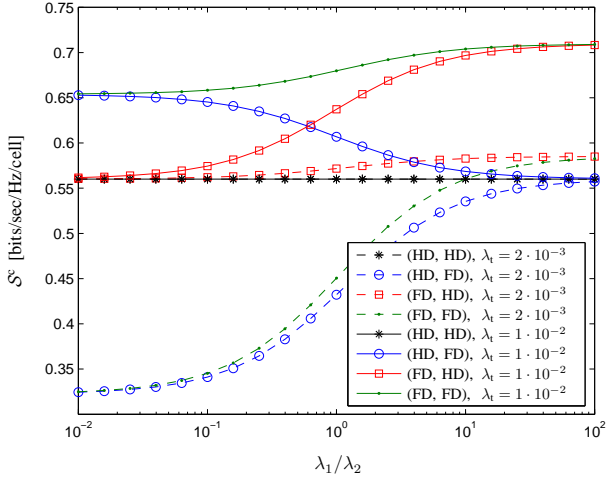}
    }
    \end{center}
    \caption{
    		Cell \ac{HDHN} throughput $\CSE$ in bits/sec/Hz/cell 
		as a function of  the ratio $\LB{1}{}/\LB{2}{}$ 
		for different values of $\LT = \LB{1}{} + \LB{2}{}$ in nodes/m$^2$ and 
		different duplex mode set $(m_1, m_2)$. 			
		 }
   \label{fig:CellSE_LRatio_LT}
\end{figure}
%\/\/\/\/\/\/\/\/\/\/\/\/\/\/\/\/\/\/\/\/\/\/\/\/\/\/\/\/\/\/\/\/\/\/\/\/\/\/\/\/\/\/\/\/\/\/\/\/\/\/\/\/\/\/\/\/\/\/\/\/\/\/\/\/\/\/\/\/\/\/\/\/\/\/\/\/\/\/\/\/\/\/\/\/\/\/\/

Figure~\ref{fig:CellSE_LRatio_LT} shows the cell \ac{HDHN} throughput $\CSE$ 
for two tier network 
as a function of  the ratio $\LB{1}{}/\LB{2}{}$ 
for different values of $\LT = \LB{1}{} + \LB{2}{}$ and different 
duplex mode set $(m_1, m_2)$. 	
From Fig.~\ref{fig:CellSE_LRatio_LT}, it can be seen that %which is the same environment of Fig.~\ref{} with $\LRatio{21}=1$,
the best duplex mode sets for $\LT = 2 \cdot 10^{-3}$ and $\LT = 1 \cdot 10^{-2}$
are (\ac{FD}, \ac{HD}) and (\ac{FD}, \ac{FD}), respectively, for all the range of $\LB{1}{}/\LB{2}{}$. 
Hence, we can see that the best duplex mode is determined more by the total density of \acp{AP} in \acp{HDHN} $\LT$ 
than the \ac{AP} density ratio, $\LB{1}{}/\LB{2}{}$.

%---------------------------------------------------------------------------%
%                            Sec: Conclusion                                    %
%---------------------------------------------------------------------------%

\section{Conclusion}\label{sec:conclusion}

This paper establishes a foundation for \acp{HDHN}
accounting for the spatial \ac{AP} distribution, the self-\ac{IC} capability, 
and the network interference.
After newly characterizing the network interference generated by \ac{FD}-mode cells, 
we define and derive the \ac{HDHN} throughput. % as a new network throughput measurement.
By quantifying the \ac{HDHN} throughput, we show the effect of network parameters and self-\ac{IC} capabilities on the \ac{HDHN} throughput,
and present how to optimally determine the duplex mode to maximize the \ac{HDHN} throughput. 
Specifically, our results demonstrate that  
the \ac{FD} mode achieves higher network throughput than the \ac{HD} mode 
for high self-\ac{IC} capability and large \ac{AP} density of \acp{HDHN}. % , and  low 
%
% nn the throughput aspect of a network in the presence of network interference,
In order to maximize the throughput of a tier network, 
operating all \acp{AP} in either \ac{HD} or \ac{FD} is better than having two mode \acp{AP}.
On the other hand, in terms of the total throughput of heterogeneous networks,
making different tier networks operate in different duplex modes can enhance the throughput. 
The outcomes of our work provide insights on the efficient design of \acp{HDHN}, 
and opens several issues for future research on \acp{HDHN}
including the transmission power control for \ac{FD}- and \ac{HD}-mode nodes,  
the throughput of \ac{MIMO} \ac{FD} system in the presence of network interference, 
the effect of network interference cancellation on the \ac{HDHN} throughput, 
%the efficiency of \ac{FD}-mode nodes equipped with multiple antennas
and the communication secrecy of \acp{HDHN}. 
%n the overall efficiency of multi- antenna equipped nodes.
%\mysubnote{ADD future works}. 

%This paper opens several issues for future research on spectrum sharing, including the impact of spectrum sensing performance on the overlay method and the effect of spatial interference cancellation on the overall efficiency of multi- antenna equipped nodes. Also, the extension of this work should make it possible to analyze the area spectral efficiency of overlaid networks that may consist of combinations of macrocells, picocells, femtocells, and perhaps other architec- tural elements.
%
%---------------------------------------------------------------------------%
%                                Appendix                                            %
%---------------------------------------------------------------------------%

\begin{appendix}

\subsection{Proof of Lemma~\ref{lem:Laplace}}\label{app:lem1}
The Laplace transform $\Lap{\Int{i}{\FD}}{s}$ is given by
\begin{align}	
& \Lap{\Int{i}{\FD}}{s}  
\nonumber \\ 
	&  \mathop = \limits^{\mathrm{(a)}} 
		\exp
		\left\{
			- 2 \pi \LB{i}{\FD}
			\int_{\hat{\rDis}}^\infty
				x\,  \EXs{\RV{G}_i}
				{
					1 - e^{ - s \RV{G}_i x^{-\PL{i}}}
				}
			dx
		\right\}
	\nonumber \\
	& = 
	%& \mathop = \limits^{\mathrm{(b)}} 
	\exp
	\left\{
			- 2 \pi \LB{i}{\FD} \,
			\EXs{\RV{G}_i}
			{
						\int_{\hat{\rDis}}^\infty
				x\, 
				\left(
					1 - e^{ - s \RV{G}_i x^{-\PL{i}}}
				\right)
			dx
			}
	\right\}
	\nonumber \\
	&  \mathop = \limits^{\mathrm{(b)}} 
	%& \mathop = \limits^{\mathrm{(b)}} 
	\exp
	\left\{
			- \frac{2}{\PL{i}} \pi \LB{i}{\FD} \,
			\EXs{\RV{G}_i}
			{
						\int_{\hat{\rDis}^{\PL{i}}}^\infty
				y^{2/\PL{i}-1} 
				\left(
					1 - e^{ - s \RV{G}_i /y}
				\right)
			dy
			}
	\right\}
	\label{eq:Pro40}
\end{align}
where (a) is from the Campbell's theorem \cite{Kin:B93},\footnote{\blue{For a 
homogeneous \ac{PPP} $\PPP_{\lo} \in \R^d$ 
with density $\lambda_{\lo}$, the Campbell's theorem shows the following property holds\cite{Kin:B93} 
\begin{align}\nonumber	
	\EXs{\PPP_\lo}{
		\sum_{\y \in \PPP_\lo \bigcup \mathcal{A}}
		%\IndF{\mathcal{A}}{\y}
		g(\y)
	}
%	&\hspace{-2.5cm} =
%		\EXs{\Pgn}{\EXs{\Pnode|\Pgn} {
%			\sum_{\Ndy \in \Pnode}\IndF{\mathcal{A}}{\Ndy}g\left(\Ndy, \Pgn\right)|\Pgn
%		}}
%	\nonumber\\
%	&   = %\mathop = \limits^{\mathrm{(a)}} 
%		\int_{\mathcal{A}} \int_{\mathbbmss{M}_{l-1}} g(y, \pi) f\left( d\pi \right) \mathsf{\Lambda} \left({dy}\right)
%		\nonumber \\
	=
		{\lambda}_{\lo} \int_{\mathcal{A}} g(\ly)\,{d\ly}
\end{align}
where $\mathcal{A}$ is a bounded space and $g(\ly)$ is a bounded measurable function for $\ly \in \R^d$. }
}
and (b) is obtained by replacing $y$ for $x^{\PL{i}}$. 	
In \eqref{eq:Pro40}, by replacing $z$ for $1/y$, the integral inside of the expectation is represented as
\begin{align}
& \int_{\hat{\rDis}^{\PL{i}}}^\infty
				y^{2/\PL{i}-1} 
				\left(
					1 - e^{ - s \RV{G}_i /y}
				\right)
			dy
\nonumber \\
&
	= 	
	\int_{0}^{\frac{1}{\hat{\rDis}^{\PL{i}}}}
				z^{-2/\PL{i}-1} 
				\left(
					1 - e^{ - s \RV{G}_i z}
				\right)
	dz
\nonumber \\
&
= 
	\int_{0}^\infty
		z^{-\frac{2}{\PL{i}}-1} \left( 1 - e^{- s \RV{G}_i z}\right)
	dz
	- 
	\int_{\frac{1}{\hat{\rDis}^{\PL{i}}}}^\infty
		z^{-\frac{2}{\PL{i}}-1}
	dz
\nonumber\\
& \quad
	+ 
	\int_{\frac{1}{\hat{\rDis}^{\PL{i}}}}^\infty
		z^{-\frac{2}{\PL{i}}-1}  e^{- s \RV{G}_i z} % \left( 1 - e^{- s \RV{G}_i z}\right)
	dz\,. 
\label{eq:Pro39}
\end{align}
In \eqref{eq:Pro39}, from \cite[eq. (3.478)]{GraRyz:B07} and \cite[eq. (3.381)]{GraRyz:B07}, 
the first and the third integrals are respectively given by  
\begin{align}	
&	 \int_{0}^\infty \!\!\!
		z^{-\frac{2}{\PL{i}}-1} \! \left( 1 - e^{- s \RV{G}_i z}\right)\!
	dz
	=
		-\left( s \RV{G}_i\right)^{\frac{2}{\PL{i}}} \GF{-\frac{2}{\PL{i}}},\, \forall \PL{i}>2
	\nonumber \\
& 
	\int_{\frac{1}{\hat{\rDis}^{\PL{i}}}}^\infty
		z^{-\frac{2}{\PL{i}}-1}  e^{- s \RV{G}_i z} % \left( 1 - e^{- s \RV{G}_i z}\right)
	dz
	=
		\left( s \RV{G}_i\right)^{\frac{2}{\PL{i}}} 
		\GF{
			-\frac{2}{\PL{i}}, 
			\frac{ s \RV{G}_i}{\hat{\rDis}^{\PL{i}} }
		}\,. 
\label{eq:Pro33}	 
\end{align}
Therefore, by using  \eqref{eq:Pro39} in \eqref{eq:Pro40} 
after representing it using 
$
	\GF{-\frac{2}{\PL{i}}}
	= 
		\frac{-\PL{i}}{2}
		\GF{1-\frac{2}{\PL{i}}}
$, 
$
	\int_{\frac{1}{\hat{\rDis}^{\PL{i}}}}^\infty
		z^{-\frac{2}{\PL{i}}-1}
	dz 
	= \frac{
		\PL{i} \rDis^2
	}{
		2
	}
$, and \eqref{eq:Pro33}, we can obtain $\Lap{\Int{i}{\FD}}{s}$ as 
\begin{align}	
& \Lap{\Int{i}{\FD}}{s}  
\nonumber\\
	% & \mathop = \limits^{\mathrm{(c)}} 
	& = 
		\exp
		\left\{
			- \pi \LB{i}{\FD} 
			% \frac{2}{\PL{i}}
			\left[
			-	%\frac{\PL{i}}{2} 
				\hat{\rDis}^{2}
			+ 	%\frac{\PL{i}}{2}
				s^{2/\PL{i}} \EXs{\RV{G}_i}{\RV{G}_i^{2/\PL{i}}}
				\GF{1-\frac{2}{\PL{i}}}
		\right.\right.
		\nonumber \\
		& \quad\quad\quad %\quad\quad\quad \quad\quad\quad \quad\quad\quad \quad% \quad\quad
		\left.\left.	
			+ 	\frac{2}{\PL{i}}
				s^{2/\PL{i}} \EXs{\RV{G}_i}{\RV{G}_i^{2/\PL{i}} \GF{-\frac{2}{\PL{i}}, \frac{s\RV{G}_i}{\hat{\rDis}^{\PL{i}}}}}
			\right]
		\right\}
	\label{eq:Pro4}
\end{align}
for $\PL{i}>2$. 
% and (b) is obtained by changing the order of integrals. 
%
%In \eqref{eq:Pro4}, from \eqref{eq:HypoMoment}, we have
%%
%\begin{align}	
%	\EXs{\RV{G}_i}{\RV{G}_i^{2/\PL{i}}}
%	& =
%		\frac{2}{\PL{i}}\GF{\frac{2}{\PL{i}}}
%		\frac
%		{
%			\Pus{i}^{\frac{2}{\PL{i}} + 1}  - \Pbs{i}^{\frac{2}{\PL{i}} + 1}
%		}
%		{
%			\Pus{i} - \Pbs{i}
%		}
%		%\left( \Pus{i}^{\frac{2}{\PL{i}} + 1}  - \Pbs{i}^{\frac{2}{\PL{i}} + 1} \right)
%	\label{eq:HypoMoment2}
%\end{align}

From \eqref{eq:HypoPDF}, for $\Pbs{i} \neq \Pus{i}, \, \forall i \in \SetI$, we have  
\begin{align}	
	& \EXs{\RV{G}_i}{\RV{G}_i^{2/\PL{i}} \GF{-\frac{2}{\PL{i}}, \frac{s\RV{G}_i}{\hat{\rDis}^{\PL{i}}}}}
\nonumber\\
	&
	% \quad \quad \quad \quad\quad \quad\quad 
	= 	
		\frac{1}{\Pus{i} \!-\! \Pbs{i}}\!\!
		\int_{0}^{\infty} \!\!\!
			g^{2/\PL{i}}
			\GF{
				\frac{-2}{\PL{i}}, 
				\frac{g s 
				}{
				\hat{\rDis}^{\PL{i}}}
			} \!\!
		 % \frac{g \TSIR{k}{m} }{\Ptx \Bratio{ik}}}
		 \left(e^{- \frac{g}{\Pus{i}}} - e^{- \frac{g}{\Pbs{i}}}\right)
%			\left(e^{- g/\Pus{i}} - e^{- g/\Pbs{i}}\right)
		dg
	\nonumber \\
	& 
	% \quad \quad \quad \quad\quad \quad\quad
	\mathop = \limits^{\mathrm{(a)}} 
		\frac{1}{\Pus{i} - \Pbs{i}}
		\left\{
			\FnI{1}{
				\frac{2}{\PL{i}}+1,
				\frac{1}{\Pus{i}},
				\frac{-2}{\PL{i}},
				\frac{s}{
					\hat{\rDis}^{\PL{i}}
				} 
				%\frac{\TSIR{k}{m}}{\Ptx \Bratio{ik}}
				 }
		\right. 
	\nonumber	\\
		& \left.
		\quad \quad \quad \quad \quad \quad \quad\quad %\quad %\quad \quad 
			- 
			\FnI{1}{
				\frac{2}{\PL{i}}+1,
				\frac{1}{\Pbs{i}},
				\frac{-2}{\PL{i}},
				\frac{s}{\hat{\rDis}^{\PL{i}}}
				% \frac{\TSIR{k}{m}}{\Ptx \Bratio{ik}}
				}
		\right\}
	\label{eq:ProG2} 
\end{align}
where $\FnI{1}{x, y, z, \nu}$ is defined in \eqref{eq:Integral1} and 
(a) is obtained by \cite[eq. 6.455]{GraRyz:B07}. 
For $\Pbs{i} = \Pus{i}$, we have
\begin{align}	
	& \EXs{\RV{G}_i}{\RV{G}_i^{2/\PL{i}} \GF{-\frac{2}{\PL{i}}, \frac{s\RV{G}_i}{\hat{\rDis}^{\PL{i}}}}}
\nonumber \\
	& \quad \quad \quad  \quad \quad 
	=
		\frac{1}{\Pbs{i}^2}
		\int_{0}^\infty
			g^{2/\PL{i}+1} \GF{\frac{-2}{\PL{i}},
			\frac{s g}{ \hat{\rDis}^{\PL{i}}} } e^{-g/\Pbs{i}} dg
	\nonumber \\
	& \quad \quad \quad  \quad \quad 
	= 
		\frac{1}{\Pbs{i}^2}
			\FnI{1}{
				\frac{2}{\PL{i}}+2,
				\frac{1}{\Pbs{i}},
				\frac{-2}{\PL{i}},
				\frac{s}{\hat{\rDis}^{\PL{i}}}
				% \frac{\TSIR{k}{m}}{\Ptx \Bratio{ik}}
			}\,. 
	\label{eq:ProG3}
\end{align}
Using \eqref{eq:HypoPDF},
we can readily obtain $\EXs{\RV{G}_i}{\RV{G}_i^{2/\PL{i}}}$ in \eqref{eq:Pro4}. 
Finally, by substituting \eqref{eq:HypoPDF},  \eqref{eq:ProG2}, and \eqref{eq:ProG3} into \eqref{eq:Pro4}, we obtain \eqref{eq:LapDiff}. % and \eqref{eq:LapEqual}. 

%==============================================================

\subsection{Proof of Theorem~\ref{trm:STP}}\label{app:trm1}
The \ac{CCDF} of \ac{SIR} in \eqref{eq:SIR} is presented for Rayleigh fading channels as
\begin{align}	
	& \PX{\SIR{k}{m} \ge \TSIR{k}{m}}
	\nonumber\\
	&	
	=
		\EXs{}
		{
			\exp\left\{ 
				- \frac{\rDis_{}^{\PL{k}}  \TSIR{k}{m}}{ \Ptx}
				\left(
					\FnCk{m}{\Prx}{k} + \sum_{i \in \SetI} \left( \Int{i}{\HD}+ \Int{i}{\FD} \right) 
				\right)
			\right\} 
		}
	\nonumber \\
	& = 
		\EXs{\rDis_{}}
		{ 
			e^{- \frac{\rDis_{}^{\PL{k}} \TSIR{k}{m} \FnCk{m}{\Prx}{k}  }{ \Ptx}}
			\prod_{i \in \SetI}
				\Lap{\Int{i}{\HD}}{\frac{\rDis_{}^{\PL{k}} \TSIR{k}{m} }{ \Ptx}}
				\Lap{\Int{i}{\FD}}{\frac{\rDis_{}^{\PL{k}} \TSIR{k}{m} }{ \Ptx}}
				%
%				\EXs{\Int{i}{\HD}}
%				{ e^{- \frac{\rDis_{}^{\PL{k}}  }{ \Ptx}\Int{i}{\HD}} }
%				\EXs{\Int{i}{\FD}}
%				{
%				e^{- \frac{\rDis_{}^{\PL{k}}  }{ \Ptx}\Int{i}{\FD}} 
%				}
		}
		\label{eq:Pro1}
\end{align}
where 
$\rDis_{}$ is the distance from a typical user to its associated \ac{AP} and 
$\Lap{\RV{Z}}{s} = \EXs{\RV{Z}}{e^{-s\RV{Z}}}$ is the Laplace transform of $\RV{Z}$. 
Using the \ac{PDF} of $\rDis_{}$ in \eqref{eq:Pa}, we can represent \eqref{eq:Pro1} as 
\begin{align}\label{eq:STP2}
	&  \PX{\SIR{k}{m} > \TSIR{k}{m}}
	\! =\!
		\frac{2 \pi \LB{k}{m}}{\Pa{k}{m}} \!
		\int_{0}^\infty \!\!\!\!
			r 
			e^{- \frac{\FnCk{m}{\Prx}{k} \TSIR{k}{m} }{ \Ptx}r^{\PL{k}} 
			- \pi \sum_{i \in \SetI} \LB{i}{} \Bratio{ik}^{\frac{2}{\PL{i}}} r^{ \frac{2\PL{k}}{\PL{i}}}
			}
	\nonumber	\\
	& \quad \quad \quad \quad 
	\times
			\prod_{i \in \SetI}
				\Lap{\Int{i}{\HD}}{\frac{r^{\PL{k}} \TSIR{k}{m} }{ \Ptx}}
				\Lap{\Int{i}{\FD}}{\frac{r^{\PL{k}} \TSIR{k}{m} }{ \Ptx}}
				%
%				\EXs{\Int{i}{\HD}}
%				{ e^{- \frac{r^{\PL{k}}  }{ \Ptx}\Int{i}{\HD}} }
%				\EXs{\Int{i}{\FD}}
%				{
%				e^{- \frac{r^{\PL{k}}  }{ \Ptx}\Int{i}{\FD}} 
%				}
%		e^{ 
%			- \pi \sum_{i \in \SetI} \LB{i}{} \Bratio{ik}^{2/\PL{i}} x^{2 \PL{k}/\PL{i}}
%		}
		dr \,.
%	f_{\rDis_k^{m}}(x)
%	=
%		\frac{2 \pi \LB{k}{m} x}{\Pa{k}{m}}
%		\exp
%		\left\{ 
%			- \pi \sum_{i \in \SetI} \LB{i}{} \Bratio{ik}^{2/\PL{i}} x^{2 \PL{k}/\PL{i}}
%		\right\}
\end{align}
In \eqref{eq:STP2}, 
$\Lap{\Int{i}{\HD}}{s}, \forall s > 0$ can be represented by 
\begin{align}	
	&\Lap{\Int{i}{\HD}}{s}
	  \mathop = \limits^{\mathrm{(a)}} 
		\exp
		\left\{
			- 2 \pi \LB{i}{\HD}
			\int_{\hat{\rDis}_{}}^\infty
				x\,  \EXs{\ChG{}}
				{
					1 - e^{ - s \Pbs{i} \ChG{} x^{\PL{i}}}
				}
			dx
		\right\}
	\nonumber \\
	&  \quad \quad\quad
	= % \hspace{-0cm} \mathop = \limits^{\mathrm{(b)}} 
		\exp
		\left\{
			- 2 \pi \LB{i}{\HD}
			\int_{\hat{\rDis}_{}}^\infty
				\frac{x}
				{
					1 + (s^{-1} \Pbs{i}^{-1} x^{\PL{i}})	
				}
			dx
		\right\}
	\label{eq:Pro22}
\end{align}
where $\hat{\rDis}_{}$ is the distance to nearest unassociated \ac{AP}, given by\cite{SinDhiAnd:13}
\begin{align}	
	\hat{\rDis}_{} = \Bratio{ik}^{1/\PL{i}} \rDis_{}^{\PL{k}/\PL{i}}\,.
\end{align}
In \eqref{eq:Pro22}, (a) is from the Campbell's theorem \cite{Kin:B93}
and the integral over $x$ is performed from $\hat{\rDis}_{}$ 
since, due to the association rule, 
$\Bias{k} \rDis_{k}^{-\PL{k}}$ is the minimum value among 
$\Bias{i} \rDis_{i}^{-\PL{i}}, \, \forall i \in \SetI$,  
where $\rDis_{i}$ is the distance between a typical user to the nearest \ac{AP} in the $i$th-tier network. 
%due to the association rule. 
%
By replacing $s = {\rr^{\PL{k}} \TSIR{k}{m} }{ \Ptx}^{-1}$ and $\hat{\rDis}_{} = \Bratio{ik}^{1/\PL{i}} \rr^{\PL{k}/\PL{i}}$ in \eqref{eq:Pro22}, we have 
\begin{align}
& \Lap{\Int{i}{\HD}}{\frac{\rr^{\PL{k}} \TSIR{k}{m} }{ \Ptx}}
	% \hspace{-0cm} \mathop = \limits^{\mathrm{(b)}} 
\nonumber \\
& \!= \!
		\exp \!
		\left\{
			- 2 \pi \LB{i}{\HD} \!\!
%		\right.
%	\nonumber \\
%		& \left. % \quad \quad 
%		%\times 
%		\right.
			\int_{ \Bratio{ik}^{\frac{1}{\PL{i}}} \rr^{\frac{\PL{k}}{\PL{i}}}}^\infty
				\frac{x}
				{
					1 + \Ptx (\rr^{\PL{k}} \TSIR{k}{m} \Pbs{i})^{-1} x^{\PL{i}}
				}
			dx
		\right\}		
	\nonumber \\
& \!= \! % \hspace{-0cm} \mathop = \limits^{\mathrm{(c)}} 
		\exp
		\left\{
			- 2 \pi \LB{i}{} (1-\Pfd{i})%\LB{i}{\HD}
			\rr^{2\PL{k}/\PL{i}}
			\FnI{0}{\rr^{\PL{k}}, \frac{\Ptx}{ \TSIR{k}{m} \Pbs{i}}, \Bratio{ik}, \PL{i} }
		\right\}
	\nonumber % \label{eq:Pro3}		
\end{align}
where $\FnI{0}{x, y, z, \nu}$ is defined in \eqref{eq:Integral0} using \cite[eq. (3.194)]{GraRyz:B07}. 
\blue{From \eqref{eq:Integral0}, we can see that the parameter $x$ does not affect $\FnI{0}{x, y, z, \nu}$. 
%i.e. $\FnI{0}{x, y, z, \nu} = \FnI{0}{c, y, z, \nu}$ for $c>0$.
Hence, we have $\FnI{0}{x, y, z, \nu} = \FnI{0}{1, y, z, \nu}$ and}
%Since $\FnI{0}{x, y, z, \nu} = \FnI{0}{1, y, z, \nu}$, we have
%
%
\begin{align}	
	& \prod_{i \in \SetI} 
	\Lap{\Int{i}{\HD}}{\frac{\rr^{\PL{k}} \TSIR{k}{m}  }{ \Ptx}}
	 =
		\exp
		\Bigg\{
			- 2 \pi 
			\sum_{i \in \SetI}
			\LB{i}{} (1-\Pfd{i})
			%\LB{i}{\HD}
			\rr^{2\PL{k}/\PL{i}}
	\nonumber \\
	& \quad\quad\quad\quad\quad\quad\quad\quad\quad\quad
	\times
			\FnI{0}{1,\frac{ \Ptx}{\TSIR{k}{m} \Pbs{i}}, \Bratio{ik}, \PL{i} }
		\Bigg\}\,.
	\label{eq:LapHD}
\end{align}
%
%

%In \eqref{eq:STP2}, $\Lap{\Int{i}{\FD}}{s}$ is given by
%%
%%
%\begin{align}	
%	\Lap{\Int{i}{\FD}}{s}
%	&  \mathop = \limits^{\mathrm{(a)}} 
%		\exp
%		\left\{
%			- 2 \pi \LB{i}{\FD}
%			\int_{\hat{\rDis}_{}}^\infty
%				x\,  \EXs{\RV{G}_i}
%				{
%					1 - e^{ - s \Pbs{i} \RV{G}_i x^{\PL{i}}}
%				}
%			dx
%		\right\}
%	\nonumber \\
%	& \mathop = \limits^{\mathrm{(b)}} 
%	\EXs{\RV{G}_i}
%	{
%		\exp
%		\left\{
%			- 2 \pi \LB{i}{\FD} 
%						\int_{\hat{\rDis}_{}}^\infty
%				x\, 
%				\left(
%					1 - e^{ - s \Pbs{i} \RV{G}_i x^{\PL{i}}}
%				\right)
%			dx
%		\right\}
%	}
%	\nonumber \\
%	& = %\mathop = \limits^{\mathrm{(c)}} 
%		\exp
%		\left\{
%			- \pi \LB{i}{\FD} 
%			% \frac{2}{\PL{i}}
%			\left[
%			-	%\frac{\PL{i}}{2} 
%				\hat{\rDis}_{}^{2}
%			+ 	%\frac{\PL{i}}{2}
%				s^{2/\PL{i}} \EXs{\RV{G}_i}{\RV{G}_i^{2/\PL{i}}}
%				\GF{1-\frac{2}{\PL{i}}}
%			+ 	\frac{2}{\PL{i}}
%			s^{2/\PL{i}} \EXs{\RV{G}_i}{\RV{G}_i^{2/\PL{i}} \GF{-\frac{2}{\PL{i}}, \frac{s\RV{G}_i}{\hat{\rDis}_{}^{\PL{i}}}}}
%			\right]
%		\right\}
%	\label{eq:Pro4}
%\end{align}
%%
%%
%where (a) is from the Campbell's theorem \cite{Kin:B93} and 
%(b) is obtained by changing the order of integrals. 

In Lemma~\ref{lem:Laplace}, by replacing $s = {\rr^{\PL{k}}  }/{ \Ptx}$ and $\hat{\rDis} = \Bratio{ik}^{1/\PL{i}} \rr^{\PL{k}/\PL{i}}$,
% in \eqref{eq:Pro4}, 
we have\footnote{\redR{The \eqref{eq:LapFD} is obtained when the biased distance to the associated \ac{AP} from a typical user $\Bias{k} \rDis_{k}^{-\PL{k}}$ is smaller than that to any \acp{AP}. In \ac{FD}-mode cells, 
the biased distance from the typical user to a user, who associates to another \ac{AP}, 
can be smaller than that to the associated \ac{AP}. 
However, we ignore this case since a user generally transmits with smaller power than an \ac{AP} and for analytical tractability. 
}
%Since an \ac{AP} generally transmits 
%	with stronger power than a user, 
%	we ignore the interference, generated from a user 
%	that may locates closer
%	than the nearest unassociated \ac{AP} from the typical user. 
% 	from the typical user. 
}
\begin{align}	
	&\prod_{i \in \SetI}
	\Lap{\Int{i}{\FD}}{\frac{\rr^{\PL{k}} \TSIR{k}{m}  }{ \Ptx}}
	=
		\exp
		 \left\{
			- 2\pi 
			% \frac{2}{\PL{i}}
			\sum_{i \in \SetI}
			\LB{i}{} \Pfd{i}
%			 \LB{i}{\FD}
			\rr^{2\PL{k}/\PL{i}} 
		\right.
	\nonumber\\
	&\quad\quad 	
		\times
		\left. 
			\left[
			-	\frac{1}{2} 
				\Bratio{ik}^{2/\PL{i}}% \rr^{2\PL{k}/\PL{i}}
			 + 	\frac{\TSIR{k}{m}^{2/\PL{i}}  }{2 {\Ptx}^{2/\PL{i}}}
				\GF{1-\frac{2}{\PL{i}}}
				 \EXs{\RV{G}_i}{\RV{G}_i^{2/\PL{i}}}
			\right. 
		\right.
		\nonumber \\
		& \quad \quad  %\quad\quad \quad \quad% \quad
		\left. 
			\left.
			+ 	\frac{
					\TSIR{k}{m}^{2/\PL{i}}  
				}{ 
					\PL{i} {\Ptx}^{2/\PL{i}}  
				}
				%\rr^{2\PL{k}/\PL{i}} 
				\EXs{\RV{G}_i}
				{
					\RV{G}_i^{2/\PL{i}} 
					\GF{
						-\frac{2}{\PL{i}}, 
					 	\frac{\RV{G}_i \TSIR{k}{m} }{\Ptx \Bratio{ik}}
					}
				}
			\right]
		\right\}\,.
		\label{eq:LapFD}	
\end{align}
%
%
%In \eqref{eq:LapFD}, from \eqref{eq:HypoMoment}, we have
%%
%%
%\begin{align}	
%	\EXs{\RV{G}_i}{\RV{G}_i^{2/\PL{i}}}
%	& =
%		\frac{2}{\PL{i}}\GF{\frac{2}{\PL{i}}}
%		\frac
%		{
%			\Pus{i}^{\frac{2}{\PL{i}} + 1}  - \Pbs{i}^{\frac{2}{\PL{i}} + 1}
%		}
%		{
%			\Pus{i} - \Pbs{i}
%		}
%		%\left( \Pus{i}^{\frac{2}{\PL{i}} + 1}  - \Pbs{i}^{\frac{2}{\PL{i}} + 1} \right)
%	\label{eq:HypoMoment2}
%\end{align}
%and, using \eqref{eq:HypoPDF}, we can also obtain 
%\begin{align}	
%	& \EXs{\RV{G}_i}{\RV{G}_i^{2/\PL{i}} \GF{-\frac{2}{\PL{i}}, \frac{\RV{G}_i \TSIR{k}{m}}{\Ptx \Bratio{ik}}}}
%\nonumber\\
%	&\quad \quad \quad \quad\quad \quad\quad 
%	= 	
%		\frac{1}{\Pus{i} - \Pbs{i}}
%		\int_{0}^{\infty}
%			g^{2/\PL{i}}
%			\GF{-\frac{2}{\PL{i}}, 
%			\frac{g \TSIR{k}{m} }{\Ptx \Bratio{ik}}}
%			\left(e^{- g/\Pus{i}} - e^{- g/\Pbs{i}}\right)
%		dg
%	\label{eq:ProG2} \\
%	& \quad \quad \quad \quad\quad \quad\quad
%	\mathop = \limits^{\mathrm{(a)}} 
%		\frac{1}{\Pus{i} - \Pbs{i}}
%		\left\{
%			\FnI{1}{
%				\frac{2}{\PL{i}}+1,
%				\frac{1}{\Pus{i}},
%				\frac{-2}{\PL{i}},
%				 \frac{\TSIR{k}{m}}{\Ptx \Bratio{ik}}
%				 }
%			- 
%			\FnI{1}{
%				\frac{2}{\PL{i}}+1,
%				\frac{1}{\Pbs{i}},
%				\frac{-2}{\PL{i}},
%				 \frac{\TSIR{k}{m}}{\Ptx \Bratio{ik}}
%				}
%		\right\}
%	\nonumber
%\end{align}
%%
%%
%where $\FnI{1}{x, y, z, \nu}$ is defined in \eqref{eq:Integral1} and 
%(a) is obtained by \cite[eq. 6.455]{GraRyz:B07}. 
Finally, substituting \eqref{eq:Paa}, \eqref{eq:LapHD} and \eqref{eq:LapFD} into \eqref{eq:STP2} results in \eqref{eq:STP}.
%

%-----------------------------------------

\subsection{Proof of Corollary~\ref{cor:PfdOIC} }\label{app:cor5}

%When $\TSIR{k}{\HD} = \TSIR{k}{\FD},\, \forall k\in \SetI$, 
When $\tRate{\US}=\tRate{\BS}=\TRate$, 
the throughput of $k$th-tier network $\SE_k$ in \eqref{eq:SAEIC}
is represented by 
\begin{align}\label{eq:SAEIC3}	
	\SE_k 
	=
		\frac{\TRate}{2\Ws{}{}}
		%\sum_{k=1}^K 
			\LB{k}{}\!
			\left( \sum_{t \in \SetI} \LB{t}{} \Bratio{tk}^{2/\PL{}}\right)\!
			\SE_k^{\text{o}}	
\end{align}
where $\SE_k^{\text{o}}$ is given by
\begin{align}	
	\SE_k^{\text{o}}
	& = 
%		\TRate
%		% \sum_{k=1}^K
%		\LB{k}{}\!
%		\left( \sum_{t \in \SetI} \LB{t}{} \Bratio{tk}^{2/\PL{}}\right)\!
%		\Bigg\{
			\frac{
				1
			}{
				\sum_{i \in \SetI} \LB{i}{} \FnM{ik}{\FD}{\PL, \Pbs{k}}
			}
%	 \nonumber\\
%	& \quad\quad\quad\quad\quad\quad\quad\quad \quad%\quad\quad\quad\quad
%	\times  		
%		\Bigg\{
%			\frac{
%				1 - \Pfd{k}
%			}{
%				\sum_{i \in \SetI} \LB{i}{} \FnM{ik}{\HD}{\PL, \Pbs{k}}
%			}
%		\right.
%		\nonumber \\
%	& \quad\quad\quad\quad\quad\quad\quad\quad 
%		\left.
			+
			\frac{
				\Pfd{k}
			}{
				\sum_{i \in \SetI} \LB{i}{} \FnM{ik}{\FD}{\PL, \Pus{k}}
			}
%		\Bigg\} 
		\label{eq:SAEIC2}
\end{align}
From \eqref{eq:term1}, $\FnM{ik}{m}{\PL, \Ptx}$ in \eqref{eq:SAEIC2} can be represented by 
\begin{align}	
	\FnM{ik}{m}{\PL, \Ptx} 
	=
		 \FnMa{ik}{m}{\Ptx} \Pfd{i} + \FnMb{ik}{m}{\Ptx}
\end{align}
where $\FnMa{ik}{m}{\Ptx}$ and $\FnMb{ik}{m}{\Ptx}$ are given by 
\begin{align}	
	\FnMa{ik}{m}{\Ptx} & = \FnMB{ik}{m}{\PL{}, \Ptx} -  \FnMA{ik}{m}{\PL{}, \Ptx}
	\nonumber\\ 
	\FnMb{ik}{m}{\Ptx} & = \FnMA{ik}{m}{\PL{}, \Ptx} + \frac{\Bratio{ik}^{2/\PL{}}}{2}\,.
	\nonumber
\end{align}
Here, \blue{for the equal density $\lambda_\lo$ of \ac{FD}-mode and \ac{HD}-mode \ac{AP}s}, $\Lap{\Int{i}{\HD}}{s}^2  \leq \Lap{\Int{i}{\FD}}{s} \leq \Lap{\Int{i}{\HD}}{s}$.\footnote{For the equal density of \acp{AP}, 
	from \eqref{eq:HDint} and \eqref{eq:FDint},  
	$\Int{i}{\FD}$ is always not less than $\Int{i}{\HD}$, 
	which results in $\Lap{\Int{i}{\FD}}{s} \leq \Lap{\Int{i}{\HD}}{s}$. 
	In addition, 
	$\Lap{\Int{i}{\FD}}{s} = \EXs{\Int{i}{\FD}}{e^{-s \Int{k}{\FD}}} \leq \EXs{\Int{i}{\HD}}{e^{-s 2 \Int{i}{\HD}}}$,
	which gives $\Lap{\Int{i}{\HD}}{s}^2  \leq \Lap{\Int{i}{\FD}}{s}$. 	
} 
\blue{Since $\Lap{\Int{i}{\HD}}{\frac{\rr^{\PL{}} \TSIR{}{}  }{ \Ptx}} =\exp\left\{-2 \pi \lambda_\lo r^{2} \FnMA{ik}{m}{\PL{}, \Ptx} \right\}$ and 
$\Lap{\Int{i}{\FD}}{\frac{\rr^{\PL{}} \TSIR{}{}  }{ \Ptx}} =\exp\left\{-2 \pi \lambda_\lo r^{2} \FnMB{ik}{m}{\PL{}, \Ptx} \right\}$,}
we have 
\begin{align}\nonumber 	
	\FnMA{ik}{m}{\PL{}, \Ptx} \leq \FnMB{ik}{m}{\PL{}, \Ptx} \leq 2 \FnMA{ik}{m}{\PL{}, \Ptx}\,. 
\end{align}
%
%
% $\FnMA{ik}{m}{\PL{}, \Ptx} \leq \FnMB{ik}{m}{\PL{}, \Ptx} \leq 2 \FnMA{ik}{m}{\PL{}, \Ptx}$ 
Hence, $\FnMa{ik}{m}{\Ptx} >0$ and % $\FnMa{ik}{m}{\Ptx} \leq \FnMb{ik}{m}{\Ptx}$. 
\begin{align}\label{eq:relation}	
	\FnMa{ik}{m}{\Ptx} \leq \FnMb{ik}{m}{\Ptx}\,.
\end{align}
In \eqref{eq:SAEIC2}, 
$\sum_{i \in \SetI} \LB{i}{} \FnM{ik}{\FD}{\PL, \Pus{k}}$ can be presented as a function of $\Pfd{k}$ as
\begin{align}\label{eq:SumCon1}	
	\sum_{i \in \SetI} \LB{i}{} \FnM{ik}{m}{\PL, \Ptx}
	= 
		\Con{1}{\Ptx} \Pfd{i} + \Con{2}{\Ptx}
\end{align}
 where $\Con{1}{\Ptx}$ and $\Con{2}{\Ptx}$ are given by 
\begin{align}	
	\Con{1}{\Ptx} &= \LB{k}{} \FnMa{kk}{m}{\Ptx}
	\nonumber\\
	\Con{2}{\Ptx} & = \LB{k}{} \FnMb{kk}{m}{\Ptx} + \sum_{j \neq k, j\in \SetI} \LB{j}{} \FnM{jk}{m}{\PL, \Ptx}\,.
	\nonumber
\end{align}
%
%
% $
% 	\Con{1}{\Ptx} = \LB{i}{} \FnMa{ik}{m}{\Ptx}
% $ and 
% $
% 	\Con{2}{\Ptx} = \LB{i}{} \FnMb{ik}{m}{\Ptx} + \sum_{j \neq i, j\in \SetI} \LB{j}{} \FnM{jk}{m}{\PL, \Ptx}
% $. 
 Then, using \eqref{eq:SumCon1} in \eqref{eq:SAEIC2}, 
 we can obtain the first derivative of $\SE_k^{\text{o}}$ according to $\Pfd{k}$ as
\begin{align}% \label{}	
	\frac{\partial \SE_k^{\text{o}}}{\partial \Pfd{k}}
	&\!=\! 
		\frac{
			- \Con{1}{\Pbs{k}}
%			\left(
%				\Con{1}{\Pus{k}} \Pfd{k} + \Con{2}{\Pus{k}}
%			\right)^2	
		}{
%			\left(
%				\Con{1}{\Pus{k}} \Pfd{k} + \Con{2}{\Pus{k}}
%			\right)
			\left(
				\Con{1}{\Pbs{k}}  \Pfd{k} \!+\! \Con{2}{\Pbs{k}}
			\right)^2
		}
		\!+\! 
		\frac{
			\Con{2}{\Pus{k}}
%			\left(
%				\Con{1}{\Pbs{k}} \Pfd{k} + \Con{2}{\Pbs{k}}
%			\right)^2	
		}{
			\left(
				\Con{1}{\Pus{k}}  \Pfd{k} \!+\! \Con{2}{\Pus{k}}
			\right)^2
%			\left(
%				\Con{1}{\Pbs{k}} \Pfd{k} + \Con{2}{\Pbs{k}}
%			\right)^2
		}
	\label{eq:DiffSAE}
	% \nonumber
\end{align}
for all $i \in \SetI$. 
Here, $\Con{1}{\Ptx} \leq \Con{2}{\Ptx}$ since $\FnMa{ik}{m}{\Ptx} \leq \FnMb{ik}{m}{\Ptx}$ in \eqref{eq:relation}, and\footnote{This relation can be presented using \eqref{eq:SumCon1} and 
	$\FnM{ik}{m}{\PL, \Pbs{k}} \leq \FnM{ik}{m}{\PL, \Pus{k}}$, obtained from
	$\Ps{k}{\FD}{\Pbs{k}, 0, \TSIR{k}{\FD}} \geq \Ps{k}{\FD}{\Pus{k}, 0, \TSIR{k}{\FD}}$
	in \eqref{eq:STPsFD}.
}
\begin{align}\nonumber	
	\Con{1}{\Pus{k}} \Pfd{k} +\Con{2}{\Pus{k}} \geq \Con{1}{\Pbs{k}} \Pfd{k} + \Con{2}{\Pbs{k}}, \,\,  \forall k,\,\forall\Pfd{k}\,.
\end{align}
Hence, $\Con{1}{\Ptx}$ and $\Con{2}{\Ptx}$ are decreasing function according to $\Ptx$, 
and we can see that 
$
		\frac{
			\Con{2}{\Pus{k}}
%			\left(
%				\Con{1}{\Pbs{k}} \Pfd{k} + \Con{2}{\Pbs{k}}
%			\right)^2	
		}{
			\left(
				\Con{1}{\Pus{k}}  \Pfd{k} \!+\! \Con{2}{\Pus{k}}
			\right)^2
		}
$ in \eqref{eq:DiffSAE} decreases as $\Pus{k}$ decreases (i.e., as both $\Con{1}{\Pus{k}}$ and $\Con{2}{\Pus{k}}$ increases). 
Here, when $\Pus{k} = \Pbs{k}$, from \eqref{eq:DiffSAE}, we have 
\begin{align}	
	\frac{\partial \SE_k^{\text{o}}}{\partial \Pfd{k}}
	&\!=\! 
		\frac{
			\Con{2}{\Pbs{k}} - \Con{1}{\Pbs{k}}
%			\left(
%				\Con{1}{\Pus{k}} \Pfd{k} + \Con{2}{\Pus{k}}
%			\right)^2	
		}{
%			\left(
%				\Con{1}{\Pus{k}} \Pfd{k} + \Con{2}{\Pus{k}}
%			\right)
			\left(
				\Con{1}{\Pbs{k}}  \Pfd{k} \!+\! \Con{2}{\Pbs{k}}
			\right)^2
		}
	\ge 0\,.
\end{align}
When $\Pus{k} = 0$,
$\Ps{k}{\FD}{\Pbs{k}, \Pus{k}, \TSIR{}{}} = \Ps{k}{\HD}{\Pbs{k}, 0, \TSIR{}{}}$ 
and $\Ps{k}{\FD}{\Pus{k}, \Pbs{k}, \TSIR{}{}} = 0$, 
so from \eqref{eq:STPsFD} and \eqref{eq:SERay}, we have 
\begin{align}	
		\SE_k^{\text{o}}
	& = 
%		\TRate
%		% \sum_{k=1}^K
%		\LB{k}{}\!
%		\left( \sum_{t \in \SetI} \LB{t}{} \Bratio{tk}^{2/\PL{}}\right)\!
%		\Bigg\{
			\frac{
				1
			}{
				\sum_{i \in \SetI} \LB{i}{} \FnM{ik}{\FD}{\PL, \Pbs{k}}
			}
\end{align}
which is not affected by $\Pfd{k}$, resulting in 
$\frac{\partial \SE_k^{\text{o}}}{\partial \Pfd{k}}=0$ for $\Pus{k}=0$. 
Hence, we can see that for $0 \le \Pus{k} \le \Pbs{k}$,
$\frac{\partial \SE_k^{\text{o}}}{\partial \Pfd{k}}$ decreases as $\Pus{k}$ decreases and 
converges to $0$, i.e.,  
$\frac{\partial \SE_k^{\text{o}}}{\partial \Pfd{k}}\ge 0$.
%  for $0 \le \Pus{k} \le \Pbs{k}$. 
%
%
%Hence, $\Con{2}{\Pus{k}} \geq \Con{1}{\Pus{k}} \geq \Con{1}{\Pbs{k}}$ and we have  
%%
%% 
%\begin{align}
%	\frac{\partial \SE_k^{\text{o}}}{\partial \Pfd{k}}
%	& \geq
%		\frac{
%			%\left(
%			\Con{2}{\Pus{k}} - \Con{1}{\Pbs{k}}
%			%\right)
%%			\left(
%%				\Con{1}{\Pus{k}} \Pfd{k} + \Con{2}{\Pus{k}}
%%			\right)^2
%		}{
%%			\left(
%%				\Con{1}{\Pus{k}} \Pfd{k} + \Con{2}{\Pus{k}}
%%			\right)^2
%			\left(
%				\Con{1}{\Pbs{k}} \Pfd{k} + \Con{2}{\Pbs{k}}
%			\right)^2
%		}
%		\geq 0,\,\,\, \forall i \in \SetI\,.
%	\nonumber
%\end{align}
%%
%%
%% since $\Con{2}{\Pus{k}} \geq \Con{1}{\Pus{k}} \geq \Con{1}{\Pbs{k}}$. 
%Consequently, from \eqref{eq:SAEIC3}, we have
%%
%%
%\begin{align}% \nonumber	
%	\frac{\partial \SE}{\partial \Pfd{k}}
%	=
%		\frac{\TRate}{\Ws{}{}}
%		\sum_{k=1}^K 
%			\LB{k}{}\!
%			\left( \sum_{t \in \SetI} \LB{t}{} \Bratio{tk}^{2/\PL{}}\right)\!
%			\frac{\partial \SE_k^{\text{o}}}{\partial \Pfd{k}}
%	\ge 0,\, \forall k \in \SetI\,.
%\end{align}
%
%
Therefore, $\SE_k$ is an increasing function with $\Pfd{k},\,\forall k \in \SetI$ and 
the optimal $\Pfd{k}$ that maximizes $\SE_k$ is the maximum value of $\Pfd{k}$,
which is equal to $\PfdO{k} =1$.
\end{appendix}

% \bibliographystyle{IEEEtran}
%\bibliography{IEEEabrv,PCP}
%----- [Jemin]: Bibliography
\bibliographystyle{IEEEtran}
%\bibliography{bib/wgroup/StringDefinitions,bib/wgroup/IEEEabrv,wgroup/WGroup,bib/wgroup/BiblioCV,bib/mybib_Jemin_ver3,bib/PCP}
% 

\bibliography{/Users/jemin_lee/Documents/bib/wgroup/StringDefinitions,/Users/jemin_lee/Documents/bib/wgroup/IEEEabrv,/Users/jemin_lee/Documents/bib/wgroup/WGroup,/Users/jemin_lee/Documents/bib/wgroup/BiblioCV,/Users/jemin_lee/Documents/bib/mybib_Jemin_V3}

\begin{thebibliography}{10}
\providecommand{\url}[1]{#1}
\csname url@samestyle\endcsname
\providecommand{\newblock}{\relax}
\providecommand{\bibinfo}[2]{#2}
\providecommand{\BIBentrySTDinterwordspacing}{\spaceskip=0pt\relax}
\providecommand{\BIBentryALTinterwordstretchfactor}{4}
\providecommand{\BIBentryALTinterwordspacing}{\spaceskip=\fontdimen2\font plus
\BIBentryALTinterwordstretchfactor\fontdimen3\font minus
  \fontdimen4\font\relax}
\providecommand{\BIBforeignlanguage}[2]{{%
\expandafter\ifx\csname l@#1\endcsname\relax
\typeout{** WARNING: IEEEtran.bst: No hyphenation pattern has been}%
\typeout{** loaded for the language `#1'. Using the pattern for}%
\typeout{** the default language instead.}%
\else
\language=\csname l@#1\endcsname
\fi
#2}}
\providecommand{\BIBdecl}{\relax}
\BIBdecl

\bibitem{SabSchGuoBliRanWic:14}
A.~Sabharwal, P.~Schniter, D.~Guo, D.~W. Bliss, S.~Rangarajan, and R.~Wichman,
  ``In-band full-duplex wireless: Challenges and opportunities,'' \emph{{IEEE}
  J. Sel. Areas Commun.}, 2014, to appear.

\bibitem{SahSab:12}
A.~Sahai, G.~Patel, and A.~Sabharwal, ``Asynchronous full-duplex wireless,'' in
  \emph{Proc. IEEE Int. Conf. on Commun. Systems and Networks}, Bangalore, Jan.
  2010, pp. 1--9.

\bibitem{JuLimKimPooHon:12}
H.~Ju, S.~Lim, D.~Kim, H.~Poor, and D.~Hong, ``Full duplexity in
  beamforming-based multi-hop relay networks,'' \emph{{IEEE} J. Sel. Areas
  Commun.}, vol.~30, no.~8, pp. 1554--1565, Sep. 2012.

\bibitem{LiLiTeh:11}
Q.~Li, K.~Li, and K.~Teh, ``Achieving optimal diversity-multiplexing tradeoff
  for full-duplex {MIMO} multihop relay networks,'' \emph{{IEEE} Trans. Inf.
  Theory}, vol.~57, no.~1, pp. 303--316, Dec. 2011.

\bibitem{DayMarBliSch:12b}
B.~P. Day, A.~R. Margetts, D.~W. Bliss, and P.~Schniter, ``Full-duplex {MIMO}
  relaying: Achievable rates under limited dynamic range,'' \emph{{IEEE} J.
  Sel. Areas Commun.}, vol.~30, no.~8, pp. 1541--1553, Sep. 2012.

\bibitem{JuOhHon:09}
H.~Ju, E.~Oh, and D.~Hong, ``Catching resource-devouring worms in
  next-generation wireless relay systems: Two-way relay and full-duplex
  relay,'' \emph{{IEEE} Commun. Mag.}, vol.~47, no.~9, pp. 58--65, Oct. 2009.

\bibitem{DayMarBliSch:12a}
B.~Day, A.~Margetts, D.~Bliss, and P.~Schniter, ``Full-duplex bidirectional
  {MIMO}: Achievable rates under limited dynamic range,'' \emph{{IEEE} Trans.
  Signal Process.}, vol.~60, no.~7, pp. 3702--3713, Jul. 2012.

\bibitem{WeeCodLatEph:10}
P.~Weeraddana, M.~Codreanu, M.~Latva-aho, and A.~Ephremides, ``On the effect of
  self-interference cancelation in multihop wireless networks,'' \emph{EURASIP
  Journal on Wireless Communications and Networking}, vol. 2010, pp. 1--10,
  Oct. 2010.

\bibitem{BarRan:12}
S.~Barghi, A.~Khojastepour, K.~Sundaresan, and S.~Rangarajan, ``Characterizing
  the throughput gain of single cell {MIMO} wireless systems with full duplex
  radios,'' in \emph{Proc. IEEE Int. Symp. on Modeling and Optimization in
  Mobile, Ad Hoc and Wireless Networks}, Paderborn, Germany, May 2012, pp.
  68--74.

\bibitem{SahDigSab:13}
A.~Sahai, S.~Diggavi, and A.~Sabharwal, ``On degrees-of-freedom of full-duplex
  uplink/downlink channel,'' in \emph{Proc. IEEE Inf. Theory Workshop},
  Sevilla, Sep. 2013, pp. 1--5.

\bibitem{ZheKriLiPetOtt:13}
G.~Zheng, I.~Krikidis, J.~Li, A.~Petropulu, and B.~Ottersten, ``Improving
  physical layer secrecy using full-duplex jamming receivers,'' \emph{{IEEE}
  Trans. Signal Process.}, vol.~61, no.~20, pp. 4962--4974, Oct. 2013.

\bibitem{ZhaGuo:14}
L.~Zhang and D.~Guo, ``Virtual full duplex wireless broadcasting via compressed
  sensing,'' \emph{{IEEE/ACM} Trans. Netw.}, vol.~5, no.~5, pp. 1659--1671,
  Oct. 2014.

\bibitem{YamHanMurYos:11}
K.~Yamamoto, K.~Haneda, H.~Murata, and S.~Yoshida, ``Optimal transmission
  scheduling for a hybrid of full-and half-duplex relaying,'' \emph{{IEEE}
  Commun. Lett.}, vol.~15, no.~3, pp. 305--307, Mar. 2011.

\bibitem{NgLoSch:12}
D.~W.~K. Ng, E.~S. Lo, and R.~Schober, ``Dynamic resource allocation in
  {MIMO}-{OFDMA} systems with full-duplex and hybrid relaying,'' \emph{{IEEE}
  Trans. Commun.}, vol.~60, no.~5, pp. 1291--1304, May 2012.

\bibitem{RiiWerWic:11a}
T.~Riihonen, S.~Werner, and R.~Wichman, ``Hybrid full-duplex/half-duplex
  relaying with transmit power adaptation,'' \emph{{IEEE} Trans. Wireless
  Commun.}, vol.~10, no.~9, pp. 3074--3085, Sep. 2011.

\bibitem{DuaSab:10}
M.~Duarte and A.~Sabharwal, ``Full-duplex wireless communications using
  off-the-shelf radios: Feasibility and first results,'' in \emph{Proc.
  Asilomar Conf. on Signals, Systems, and Computers}, Pacific Grove, CA, Nov.
  2010, pp. 1558--1562.

\bibitem{SnoFulCha:11}
T.~Snow, C.~Fulton, and W.~Chappell, ``Transmit--receive duplexing using
  digital beamforming system to cancel self-interference,'' \emph{{IEEE} Trans.
  Microw. Theory Tech.}, vol.~59, no.~12, pp. 3494--3503, Dec. 2011.

\bibitem{LioVibColAth:10}
P.~Lioliou, M.~Viberg, M.~Coldrey, and F.~Athley, ``Self-interference
  suppression in full-duplex mimo relays,'' in \emph{Proc. Asilomar Conf. on
  Signals, Systems, and Computers}, Pacific Grove, CA, Nov 2010, pp. 658--662.

\bibitem{RiiWerWic:11}
T.~Riihonen, S.~Werner, and R.~Wichman, ``Mitigation of loopback
  self-interference in full-duplex {MIMO} relays,'' \emph{{IEEE} Trans. Signal
  Process.}, vol.~59, no.~12, pp. 5983--5993, Dec. 2011.

\bibitem{LeeSimChaKan:14}
L.~Anttila, D.~Korpi, E.~Antonio-Rodr\'{i}guez, R.~Wichman, and M.~Valkama,
  ``Modeling and efficient cancellation of nonlinear self-interference in
  {MIMO} full-duplex transceivers,'' 2014, available at
  \texttt{http://arxiv.org/abs/1406.0671}.

\bibitem{DuaDicSab:12}
M.~Duarte, C.~Dick, and A.~Sabharwal, ``Experiment-driven characterization of
  full-duplex wireless systems,'' \emph{{IEEE} Trans. Wireless Commun.},
  vol.~11, no.~12, pp. 4296--4307, Nov. 2012.

\bibitem{ChoJaiSriLevKat:10}
J.~I. Choi, M.~Jain, K.~Srinivasan, P.~Levis, and S.~Katti, ``Achieving single
  channel, full duplex wireless communication,'' in \emph{Proc. ACM Mobicom},
  Chicago, IL, Sep. 2010, pp. 1--12.

\bibitem{JaiChoKimBhaSir:11}
M.~Jain, J.~I. Choi, T.~Kim, D.~Bharadia, S.~Seth, K.~Srinivasan, P.~Levis,
  S.~Katti, and P.~Sinha, ``Practical, real-time, full duplex wireless,'' in
  \emph{Proc. ACM Mobicom}, Las Vegas, NV, Sep. 2011, pp. 301--312.

\bibitem{BhaMcmKat:13}
D.~Bharadia, E.~McMilin, and S.~Katti, ``Full duplex radios,'' in
  \emph{Proceedings of ACM SIGCOMM}, Hong Kong, Aug. 2013, pp. 375--386.

\bibitem{RiiWic:12}
T.~Riihonen and R.~Wichman, ``Analog and digital self-interference cancellation
  in full-duplex {MIMO}-{OFDM} transceivers with limited resolution in {A/D}
  conversion,'' in \emph{Proc. Asilomar Conf. on Signals, Systems, and
  Computers}, Pacific Grove, CA, Nov. 2012, pp. 45--49.

\bibitem{HuaMaLiaCir:13}
Y.~Hua, Y.~Ma, P.~Liang, and A.~Cirik, ``Breaking the barrier of transmission
  noise in full-duplex radio,'' in \emph{Proc. Military Commun. Conf.}, San
  Diego, CA, Nov. 2013, pp. 1558--1553.

\bibitem{CoxAck:13}
C.~Cox and E.~Ackerman, ``Demonstration of a single-aperture, full-duplex
  communication system,'' in \emph{{Proc. IEEE Radio and Wireless Symposium}},
  Austin, TX, Jan. 2013, pp. 148--150.

\bibitem{Kno:12}
M.~Knox, ``Single antenna full duplex communications using a common carrier,''
  in \emph{Proc. IEEE Wireless and Microwave Technology Conference}, Cocoa
  Beach, FL, Apr. 2012, pp. 1--6.

\bibitem{GoyLiuHuaPan:13}
S.~Goyal, P.~Liu, S.~Hua, and S.~Panwar, ``Analyzing a full-duplex cellular
  system,'' in \emph{Proc. Conf. on Inform. Sci. and Sys.}, Baltimore, MD, Mar.
  2013, pp. 1--6.

\bibitem{TonHae:14}
Z.~Tong and M.~Haenggi, ``Throughput analysis for wireless networks with
  full-duplex radios,'' Sep. 2014, pp. 1--5, submitted, available at
  \texttt{http://arxiv.org/abs/1409.7433}.

\bibitem{QueRocGuvKou:13}
T.~Q.~S. Quek, G.~de~la Roche, I.~Guvenc, and M.~Kountouris, \emph{Small Cell
  Networks: Deployment, {PHY} Techniques, and Resource Allocation}.\hskip 1em
  plus 0.5em minus 0.4em\relax Cambridge Univ Pr, 2013.

\bibitem{CheQueKou:12}
W.~C. Cheung, T.~Q.~S. Quek, and M.~Kountouris, ``Throughput optimization in
  two-tier femtocell networks,'' \emph{{IEEE} J. Sel. Areas Commun.}, vol.~30,
  no.~3, pp. 561--574, Apr. 2012.

\bibitem{DhiGanBacAnd:12}
H.~S. Dhillon, R.~K. Ganti, F.~Baccelli, and J.~G. Andrews, ``Modeling and
  analysis of {K}-tier downlink heterogeneous cellular networks,'' \emph{{IEEE}
  J. Sel. Areas Commun.}, vol.~30, no.~3, pp. 550--560, Apr. 2012.

\bibitem{NovDhiAnd:13}
T.~D. Novlan, H.~S. Dhillon, and J.~G. Andrews, ``Analytical modeling of uplink
  cellular networks,'' \emph{{IEEE} Trans. Wireless Commun.}, vol.~12, no.~6,
  pp. 2669--2679, Jun. 2013.

\bibitem{DhiGanAnd:13}
H.~Dhillon, R.~Ganti, and J.~Andrews, ``Load-aware modeling and analysis of
  heterogeneous cellular networks,'' \emph{{IEEE} Trans. Wireless Commun.},
  vol.~12, no.~4, pp. 1666--1677, Apr. 2013.

\bibitem{LopGuvRocKouQueZha:11}
D.~Lopez-Perez, I.~Guvenc, G.~de~la Roche, M.~Kountouris, T.~Q.~S. Quek, and
  J.~Zhang, ``Enhanced inter-cell interference coordination challenges in
  heterogeneous networks,'' \emph{{IEEE} Commun. Mag.}, vol.~18, no.~3, pp.
  22--30, Jun. 2011.

\bibitem{LeeAndHon:13}
J.~Lee, J.~G. Andrews, and D.~Hong, ``Spectrum-sharing transmission capacity
  with interference cancelation,'' \emph{{IEEE} Trans. Commun.}, vol.~61,
  no.~1, pp. 76--86, Jan. 2013.

\bibitem{HuaLauChe:09}
K.~Huang, V.~K.~N. Lau, and Y.~Chen, ``{Spectrum sharing between cellular and
  mobile ad hoc networks: transmission-capacity trade-off},'' \emph{{IEEE} J.
  Sel. Areas Commun.}, vol.~27, no.~7, pp. 1256--1267, Aug. 2009.

\bibitem{LeeAndHon:11}
J.~Lee, J.~G. Andrews, and D.~Hong, ``{Spectrum-sharing transmission
  capacity},'' \emph{{IEEE} Trans. Wireless Commun.}, vol.~10, no.~9, pp.
  3053--3063, Sep. 2011.

\bibitem{WilQueSluRab:13}
M.~Wildemeersch, T.~Q.~S. Quek, C.~H. Slump, and A.~Rabbachin, ``Cognitive
  small cell networks: Energy efficiency and trade-offs,'' \emph{{IEEE} Trans.
  Commun.}, vol.~61, no.~9, pp. 4016--4029, Sep. 2013.

\bibitem{SohQueKouShi:13}
Y.~S. Soh, T.~Q.~S. Quek, M.~Kountouris, and H.~Shin, ``Energy efficient
  heterogeneous cellular networks,'' \emph{{IEEE} J. Sel. Areas Commun.},
  vol.~31, no.~5, pp. 840--850, May 2013.

\bibitem{JoSanXiaAnd:12}
H.-S. Jo, Y.~J. Sang, P.~Xia, and J.~Andrews, ``Heterogeneous cellular networks
  with flexible cell association: A comprehensive downlink {SINR} analysis,''
  \emph{{IEEE} Trans. Wireless Commun.}, vol.~11, no.~10, pp. 3484--3495, Oct.
  2012.

\bibitem{SinDhiAnd:13}
S.~Singh, H.~S. Dhillon, and J.~G. Andrews, ``Offloading in heterogeneous
  networks: Modeling, analysis, and design insights,'' \emph{{IEEE} Trans.
  Wireless Commun.}, vol.~12, no.~5, pp. 2484--2497, May 2013.

\bibitem{SohQueKouCai:13}
S.~Soh, T.~Q.~S. Quek, M.~Kountouris, and G.~Caire, ``Cognitive hybrid division
  duplex for two-tier femtocell networks,'' \emph{{IEEE} Trans. Wireless
  Commun.}, vol.~12, no.~10, pp. 4852--4865, Oct. 2013.

\bibitem{WinPinShe:J09}
M.~Z. Win, P.~C. Pinto, and L.~A. Shepp, ``A mathematical theory of network
  interference and its applications,'' \emph{Proc. {IEEE}}, vol.~97, no.~2, pp.
  205--230, Feb. 2009.

\bibitem{PinWin:J10}
P.~C. Pinto and M.~Z. Win, ``Communication in a {Poisson} field of interferers
  -- {Part I}: Interference distribution and error probability,'' \emph{{IEEE}
  Trans. Wireless Commun.}, vol.~9, no.~7, pp. 2176--2186, Jul. 2010.

\bibitem{PinWin:J10a}
P.~C. Pinto and M.~Z. Win, ``Communication in a {Poisson} field of interferers -- {Part II}:
  Channel capacity and interference spectrum,'' \emph{{IEEE} Trans. Wireless
  Commun.}, vol.~9, no.~7, pp. 2187--2195, Jul. 2010.

\bibitem{LeeConRabWin:J13}
J.~Lee, A.~Conti, A.~Rabbachin, and M.~Z. Win, ``Distributed network secrecy,''
  \emph{{IEEE} J. Sel. Areas Commun.}, vol.~31, no.~9, pp. 1889--1900, Sep.
  2013.

\bibitem{HeaKouBai:13}
R.~Heath, M.~Kountouris, and T.~Bai, ``Modeling heterogeneous network
  interference using {Poisson} point processes,'' \emph{{IEEE} Trans. Signal
  Process.}, vol.~61, no.~16, pp. 4114--4126, Aug. 2013.

\bibitem{WinRabLeeCon:14}
M.~Z. Win, A.~Rabbachin, J.~Lee, and A.~Conti, ``Cognitive network secrecy with
  interference engineering,'' \emph{{IEEE} Netw.}, vol.~28, no.~5, pp. 86--90,
  Sep./Oct. 2014.

\bibitem{BanLeeKimHon:15}
J.~Bang, J.~Lee, S.~Kim, and D.~Hong, ``An efficient relay selection strategy
  for random cognitive relay networks,'' \emph{{IEEE} Trans. Wireless Commun.},
  2015, to appear.

\bibitem{AkoHea:13}
S.~Akoum and R.~W. Heath~Jr, ``Interference coordination: Random clustering and
  adaptive limited feedback,'' \emph{{IEEE} Trans. Signal Process.}, vol.~61,
  no.~7, pp. 1822--1834, Apr. 2013.

\bibitem{GarZhoShi:14}
V.~Garcia, Y.~Zhou, and J.~Shi, ``Coordinated multipoint transmission in dense
  cellular networks with user-centric adaptive clustering,'' \emph{{IEEE}
  Trans. Wireless Commun.}, vol.~13, no.~8, pp. 4297--4308, Aug. 2014.

\bibitem{DhiLiNugPiAnd:14}
H.~Dhillon, Y.~Li, P.~Nuggehalli, Z.~Pi, and J.~G. Andrews, ``Fundamentals of
  heterogeneous cellular networks with energy harvesting,'' \emph{{IEEE} Trans.
  Wireless Commun.}, vol.~13, no.~5, pp. 2782--2797, May 2014.

\bibitem{LimBenGhaLat:10}
C.~H.~M. De~Lima, M.~Bennis, K.~Ghaboosi, and M.~Latva-aho, ``Interference
  management for self-organized femtocells towards green networks,'' in
  \emph{Proc. IEEE Int. Symp. on Personal, Indoor and Mobile Radio Commun.},
  Instanbul, Turkey, Sep. 2010, pp. 352--356.

\bibitem{KimJuParHon:13}
D.~Kim, H.~Ju, S.~Park, and D.~Hong, ``Effects of channel estimation error on
  full-duplex two-way networks,'' \emph{{IEEE} Trans. Veh. Technol.}, vol.~62,
  no.~9, pp. 4666--4672, Nov. 2013.

\bibitem{CirRonHua:14}
A.~C. Cirik, Y.~Rong, and Y.~Hua, ``Achievable rates of full-duplex {MIMO}
  radios in fast fading channels with imperfect channel estimation,''
  \emph{{IEEE} Trans. Signal Process.}, vol.~62, no.~15, pp. 3874--3886, Aug.
  2014.

\bibitem{BolGreMeeTri:06}
G.~Bolch, S.~Greiner, H.~d. Meer, and K.~S. Trivedi, \emph{Queueing Networks
  and Markov Chains: Modeling and Performance Evaluation with Computer Science
  Applications}.\hskip 1em plus 0.5em minus 0.4em\relax Wiley-Blackwell, 2006.

\bibitem{GraRyz:B07}
I.~S. Gradshteyn and I.~M. Ryzhik, \emph{Table of Integrals, Series, and
  Products}, 7th~ed.\hskip 1em plus 0.5em minus 0.4em\relax San Diego, CA:
  Academic Press, Inc., 2007.

\bibitem{Kin:B93}
J.~F. Kingman, \emph{Poisson Processes}.\hskip 1em plus 0.5em minus 0.4em\relax
  Oxford University Press, 1993.

\end{thebibliography}

\begin{IEEEbiography}[{\includegraphics[width=1in,height=1.25in,clip,keepaspectratio]{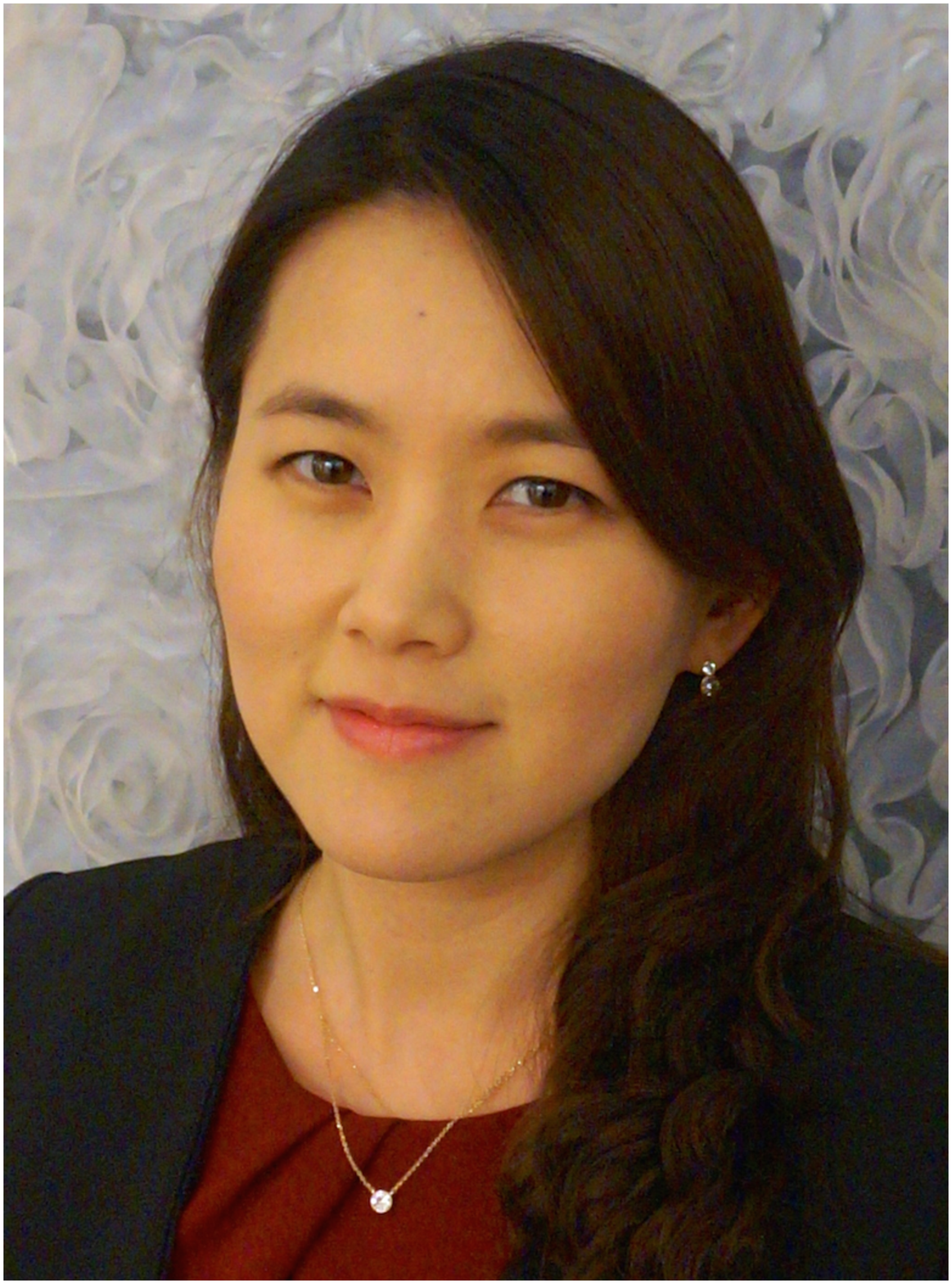}}]
{\bfseries Jemin Lee} (S'06-M'11)  is a Temasek Research Fellow %and a Principal Investigator 
at the Singapore University of Technology and Design (SUTD), Singapore.
%, and currently leading the Wireless Security and Communication Laboratory. 
She received the B.S. (with high honors), M.S., and Ph.D. degrees in Electrical and Electronic Engineering from Yonsei University, Seoul, Korea, in 2004, 2007, and 2010, respectively. She was a Postdoctoral Fellow at the Massachusetts Institute of Technology (MIT), Cambridge, MA from Oct. 2010 to Oct. 2013, and a Visiting Ph.D. Student at the University of Texas at Austin, Austin, TX from Dec. 2008 to Dec. 2009. Her current research interests include physical layer security, wireless security, heterogeneous networks, cognitive radio networks, and cooperative communications. 
%
% She has published several IEEE journals and conference proceedings in the area of wireless networks and patents related to cognitive radio networks and heterogeneous networks. 

Dr.~Lee is currently an Editor for the {\scshape IEEE Transactions on Wireless Communications} and the {\scshape IEEE Communications Letters}, 
and served as a Guest Editor of the Special Issue on Heterogeneous and Small Cell Networks for the {\scshape ELSEVIER Physical Communication} in 2013. 
She also served as a Co-Chair of the IEEE 2013 Globecom Workshop on Heterogeneous and Small Cell Networks, 
and Technical Program Committee Member for numerous IEEE conferences.  
She is currently a reviewer for several IEEE journals and has been recognized as an Exemplary Reviewer of {\scshape IEEE Communications Letters} and {\scshape IEEE Wireless Communication Letters} for recent several years. 
% She is currently a reviewer for several IEEE journals and has been recognized as an Exemplary Reviewer of {\scshape IEEE Communications Letters} in 2011 and 2012 and {\scshape IEEE Wireless Communication Letters} in 2012, 2013, and 2014. 
% 
She received the IEEE ComSoc Asia-Pacific Outstanding Young Researcher Award in 2014, the Temasek Research Fellowship in 2013, the Chun-Gang Outstanding Research Award in 2011, and the IEEE WCSP Best Paper Award in 2014.
\end{IEEEbiography}%\vfill

\begin{IEEEbiography}[{\includegraphics[width=1in,height=1.25in,clip,keepaspectratio]{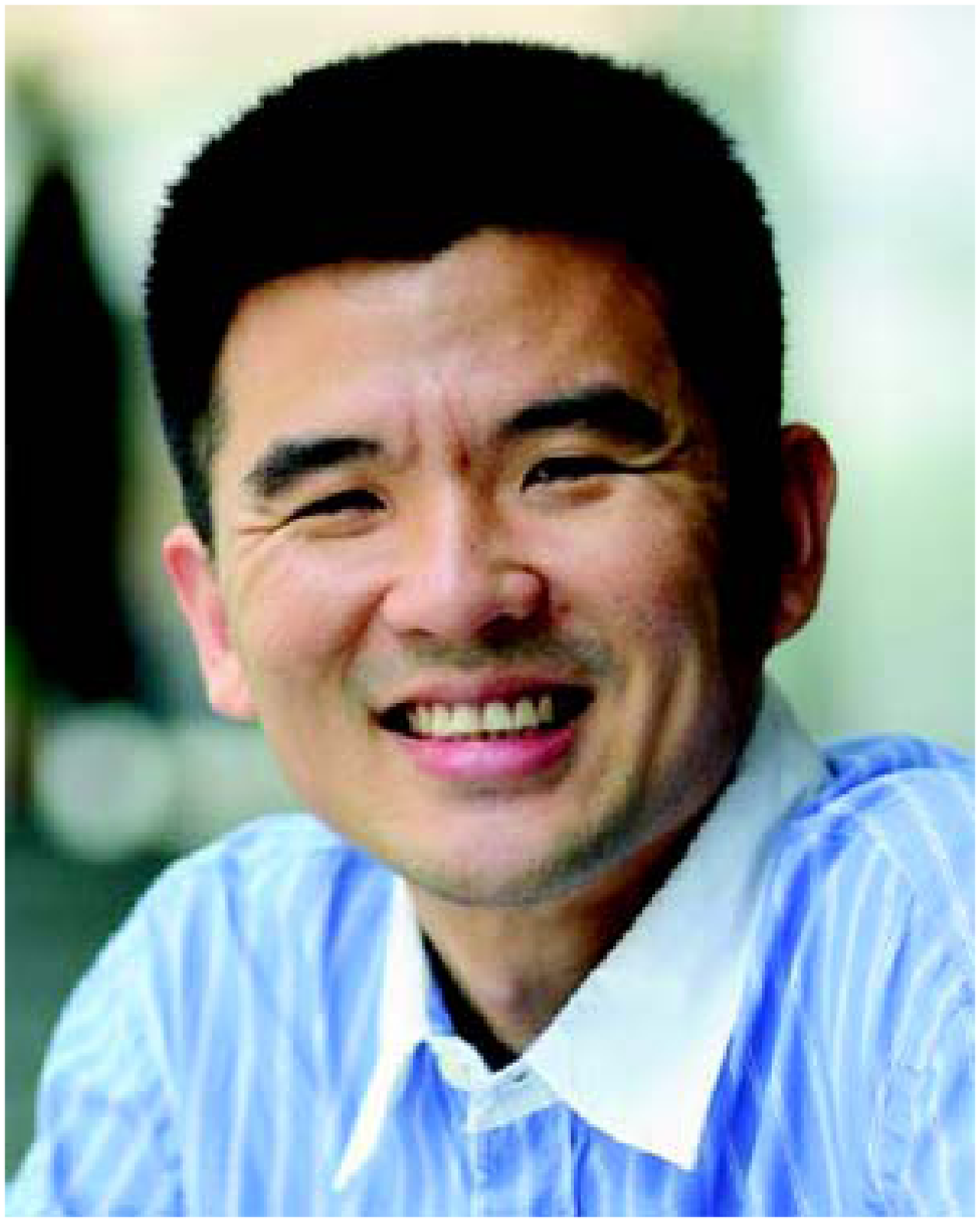}}]
{Tony Q.S. Quek}(S'98-M'08-SM'12) received the B.E.\ and M.E.\ degrees in Electrical and Electronics Engineering from Tokyo Institute of Technology, Tokyo, Japan, respectively. At Massachusetts Institute of Technology, he earned the Ph.D.\ in Electrical Engineering and Computer Science. Currently, he is an Assistant Professor with the Information Systems Technology and Design Pillar at Singapore University of Technology and Design (SUTD). He is also a Scientist with the Institute for Infocomm Research. His main research interests are the application of mathematical, optimization, and statistical theories to communication, networking, signal processing, and resource allocation problems. Specific current research topics include sensor networks, heterogeneous networks, green communications, smart grid, wireless security, compressed sensing, big data processing, and cognitive radio.

Dr.\ Quek has been actively involved in organizing and chairing sessions, and has served as a member of the Technical Program Committee as well as symposium chairs in a number of international conferences. He is serving as the PHY \& Fundamentals Track for IEEE WCNC in 2015, the Communication Theory Symposium for IEEE ICC in 2015, and the PHY \& Fundamentals Track for IEEE EuCNC in 2015. He is currently an Editor for the {\scshape IEEE Transactions on Communications}, the {\scshape IEEE Wireless Communications Letters}, and an Executive Editorial Committee Member for the {\scshape IEEE Transactions on Wireless Communications}. He was Guest Editor for the {\scshape IEEE Signal Processing Magazine} (Special Issue on Signal Processing for the 5G Revolution) in 2014, and the {\scshape IEEE Wireless Communications Magazine} (Special Issue on Heterogeneous Cloud Radio Access Networks) in 2015.

Dr.\ Quek was honored with the 2008 Philip Yeo Prize for Outstanding Achievement in Research, the IEEE Globecom 2010 Best Paper Award, the CAS Fellowship for Young International Scientists in 2011, the 2012 IEEE William R. Bennett Prize, the IEEE SPAWC 2013 Best Student Paper Award, and the IEEE WCSP 2014 Best Paper Award.
\end{IEEEbiography}
\end{document}